\documentclass[twocolumn]{aastex631} 

\usepackage{multirow}
\usepackage{MnSymbol}

\usepackage{makecell}
\usepackage{amsmath}
\usepackage{gensymb}
\usepackage[english]{babel}
\usepackage{blindtext}
\usepackage{CJKutf8}
\newcommand{\augustus}{\texttt{Augustus}}

\renewcommand{\arg}[1]{\! \left( #1 \right)}
\newcommand{\pcond}[2]{p \arg{ #1 \, | \, #2 }}

\newcommand{\params}{\ensuremath{\boldsymbol\theta}}
\newcommand{\eparams}{\ensuremath{\boldsymbol\phi}}

\newcommand{\mags}{\ensuremath{\mathbf{m}}}
\newcommand{\absmag}{\ensuremath{\mathbf{M}}}
\newcommand{\errors}{\ensuremath{\boldsymbol\sigma}}
\newcommand{\feh}{\ensuremath{{\rm [Fe/H]}}}
\newcommand{\afe}{\ensuremath{{\rm [\alpha/Fe]}}}
\newcommand{\rvec}{\ensuremath{\mathbf{R}}}

\newcommand{\brutus}{\textsc{brutus}}
\newcommand{\mist}{\texttt{MIST}}

\newcommand{\likelihood}{\ensuremath{\mathcal{L}}}
\newcommand{\prior}{\ensuremath{\pi}}

\shortauthors{Zucker, Saydjari, \& Speagle et al.}
\graphicspath{{./}{figures/}}

\begin{document}

\title{A Deep, High-Angular Resolution 3D Dust Map of the Southern Galactic Plane}

\correspondingauthor{Catherine Zucker}
\email{catherine.zucker@cfa.harvard.edu}

\author[0000-0002-2250-730X]{Catherine Zucker}
\affiliation{Center for Astrophysics $\mid$ Harvard \& Smithsonian, 
60 Garden St., Cambridge, MA, USA 02138}
\affiliation{Space Telescope Science Institute, 3700 San Martin Drive, Baltimore, MD 21218, USA}

\author[0000-0002-6561-9002]{Andrew K. Saydjari}
\altaffiliation{Equal Contribution}
\altaffiliation{Hubble Fellow}
\affiliation{Center for Astrophysics $\mid$ Harvard \& Smithsonian, 
60 Garden St., Cambridge, MA, USA 02138}
\affiliation{Department of Physics, Harvard University, 17 Oxford St., Cambridge, MA 02138, USA}
\affiliation{Department of Astrophysical Sciences, Princeton University,
Princeton, NJ 08544 USA}

\author[0000-0003-2573-9832]{Joshua S. Speagle (\begin{CJK*}{UTF8}{gbsn}沈佳士\ignorespacesafterend\end{CJK*})}
\altaffiliation{Equal Contribution}
\affiliation{Department of Statistical Sciences, University of Toronto, 9th Floor, Ontario Power Building, 700 University Ave, Toronto, ON M5G 1Z5, Canada}
\affiliation{David A. Dunlap Department of Astronomy \& Astrophysics, University of Toronto, 50 St George Street, Toronto, ON M5S 3H4, Canada}
\affiliation{Dunlap Institute for Astronomy \& Astrophysics, University of Toronto, 50 St George Street, Toronto, ON M5S 3H4, Canada}
\affiliation{Data Sciences Institute, University of Toronto, 17th Floor, Ontario Power Building, 700 University Ave, Toronto, ON M5G 1Z5, Canada} 

\author[0000-0002-3569-7421]{Edward F. Schlafly}
\affiliation{Space Telescope Science Institute, 3700 San Martin Drive, Baltimore, MD 21218, USA}

\author[0000-0001-5417-2260]{Gregory M. Green}
\affiliation{Max Planck Institute for Astronomy, K\"{o}nigstuhl 17, D-69117 Heidelberg, Germany}

\author[0000-0002-8109-2642]{Robert Benjamin}
\affiliation{Department of Physics, University of Wisconsin-Whitewater, 800 W Main St, Whitewater, WI 53190 USA}

\author[0000-0003-4797-7030]{Joshua Peek}
\affiliation{Space Telescope Science Institute, 3700 San Martin Drive, Baltimore, MD 21218, USA}

\author[0000-0003-3122-4894]{Gordian Edenhofer}
\affiliation{Max Planck Institute for Astrophysics, Karl-Schwarzschildstra\ss e 1, 85748 Garching, Germany}
\affiliation{Ludwig-Maximilians-Universit\"at, Geschwister-Scholl Platz 1, 80539 Munich, Germany}

\author[0000-0003-1312-0477]{Alyssa Goodman}
\affiliation{Center for Astrophysics $\mid$ Harvard \& Smithsonian, 
60 Garden St., Cambridge, MA, USA 02138}

 \author[0000-0002-0631-7514]{Michael A. Kuhn}
 \affiliation{Centre for Astrophysics Research, University of Hertfordshire, College Lane, Hatfield AL10 9AB, UK}

\author[0000-0003-2808-275X]{Douglas P. Finkbeiner}
\affiliation{Center for Astrophysics $\mid$ Harvard \& Smithsonian, 
60 Garden St., Cambridge, MA, USA 02138}
\affiliation{Department of Physics, Harvard University, 17 Oxford St., Cambridge, MA 02138, USA}



\begin{abstract}
We present a deep, high-angular resolution 3D dust map of the southern Galactic plane over $239^\circ < l < 6^\circ$ and $|b| < 10^\circ$ built on photometry from the DECaPS2 survey, in combination with photometry from VVV, 2MASS, and unWISE and parallaxes from Gaia DR3 where available. To construct the map, we first infer the distance, extinction, and stellar types of over 700 million stars using the {\brutus} stellar inference framework with a set of theoretical {\mist} stellar models. Our resultant 3D dust map has an angular resolution of $1 \arcmin$, roughly an order of magnitude finer than existing 3D dust maps and comparable to the angular resolution of the Herschel 2D dust emission maps. We detect complexes at the range of distances associated with the Sagittarius-Carina and Scutum-Centaurus arms in the fourth quadrant, as well as more distant structures out to a maximum reliable distance of $d \approx$ 10 kpc from the Sun. The map is sensitive up to a maximum extinction of roughly $A_V \approx 12$ mag. We publicly release both the stellar catalog and the 3D dust map, the latter of which can easily be queried via the Python package \texttt{dustmaps}. When combined with the existing \texttt{Bayestar19} 3D dust map of the northern sky, the DECaPS 3D dust map fills in the missing piece of the Galactic plane, enabling extinction corrections over the entire disk $|b| < 10^\circ$. Our map serves as a pathfinder for the future of 3D dust mapping in the era of LSST and Roman, targeting regimes accessible with deep optical and near-infrared photometry but often inaccessible with Gaia. 
\end{abstract}



\section{Introduction} \label{sec:intro}
The distribution of interstellar dust in the Milky Way has profound implications not only as a foreground contaminant for a broad range of astronomical observations, but also as a tracer of Galactic spiral structure and the sites of star formation within dense molecular clouds. 

Interstellar dust scatters and absorbs starlight at near-infrared, optical, and ultraviolet wavelengths, causing a reddening effect due to the preferential attenuation of higher energy photons. Near bright stars, interstellar dust clouds also reflect starlight, generating so called ``reflection nebulae" \citep{Sellgren_1984}. Due to the non-spherical nature of grains whose major axes align perpendicular to the magnetic field, interstellar dust polarizes starlight \citep{Han_2017}. Interstellar dust also re-radiates at mid- and far-infrared wavelengths, acting as a critical foreground for the Cosmic Microwave Background \citep{Finkbeiner_1999, Planck_2016}. 

Through its absorption, scattering, and re-processing of starlight, interstellar dust also plays a pivotal role in the physics and chemistry of the interstellar medium. For example, dust helps regulate the temperature of the interstellar medium, heating the gas via photoelectric heating \citep{Weingartner_2001}. Interstellar dust also catalyzes the formation of molecular hydrogen --- the key ingredient in molecular clouds --- and allows star formation to occur by removing the gravitational energy of collapsing clouds via far-infrared radiation \citep{Draine_2003}. 

Historically, most of our understanding of the distribution of Galactic dust has come from emission, either from associated \textsc{Hi} or from the dust itself. These maps are inherently two dimensional (2D) with varying angular resolution, where information on the structure of the dust as a function of distance has been projected onto the plane of the sky. A number of 2D dust maps have been built over the past half century. For example, \citet{Burstein_Heiles_1978} combined galaxy counts with \textsc{Hi} column density measurements derived from 21-cm spectral-line maps to probe integrated dust column density under the assumption that gas and dust are well-mixed \citep[see also][]{Burstein_Heiles_1982, Lenz_2017}. \citet{SFD} derived the dust column density from far-infrared emission at $100 \; \micron$ and $240 \; \micron$ using IRAS \citep{IRAS} and DIRBE data \citep{DIRBE}, calibrating the far-infrared flux to dust reddening using a sample of elliptical galaxies, and later SDSS photometry and spectra of 250,000 stars \citep{Schlafly_2011}, producing a map with an angular resolution of $6\arcmin$. 

More recently, \citet{Planck_2014} used a similar technique to \citet{SFD} for modeling far-infrared emission, combining the $100\,\micron$ IRAS data with Planck data between $353-857 \; \rm GHz$ to derive an all-sky 2D map of dust reddening at an angular resolution of $5\arcmin$.  Concurrently, \citet{Meisner_Finkbeiner_2015} produced all sky maps of diffuse Galactic dust emission based on WISE $12 \micron$ images at an angular resolution of $15\arcsec$. The WISE-based maps are roughly a factor of $4\times$ higher than the next highest angular resolution 2D dust emissions maps from \textit{Herschel}. Compared to WISE, the \textit{Herschel} maps achieve a lower $1\arcmin$ angular resolution \citep[$37\arcsec$ at $500\micron$;][]{Spire}, but suffer from less contamination from stars and galaxies. 

Alongside 2D emission-based maps, the recent rise in wide-field surveys has also enabled the construction of 2D dust maps based on photometry. These 2D dust maps primarily rely on measuring the infrared color excess toward red clump (RC) and red giant branch (RGB) stars, because these stellar populations act as so-called ``standard crayons," with very consistent intrinsic colors. By estimating the color excess to many standard crayons, these 2D star-based approaches can reconstruct the integrated extinction out to large distances, commensurate with the completeness limit of the underlying stellar populations. A number of such pioneering 2D-star-based map have been produced, particularly towards the inner galaxy.  For instance, \citet{Surot_2020} use RC and RGB stars detected in the VVV survey to map the color excess towards a 300 sq. deg. region of the bulge with very fine angular resolution ranging between $10 \arcsec$ and $2 \arcmin$ \citep[see also ][]{Gonzalez_2012, Alonso_Garcia_2017, Nataf_2013}. More recently, using intrinsic colors drawn from the StarHorse stellar parameter catalog \citep{Anders_2019} rather than purely RC and RGB stars, \citet{Zhang_Kainulainen_2022} build a 2D dust map over the full VVV survey area, achieving an angular resolution of $30\arcsec$ and sensitive to extinctions up to $A_V \approx 10-20$ mag. 

While these 2D emission and 2D star-based maps have been critical for correcting for the effects of dust obscuration primarily on extragalactic observations --- where the entire Milky Way is a foreground --- they are largely insufficient for corrections \emph{within} the Milky Way, or to probe the internal structure of the Milky Way's interstellar medium. For these applications, it is necessary to map the distribution of dust in three spatial dimensions (3D). 

3D dust mapping has flourished over roughly the past decade, adding the critical third dimension (distance) to our understanding of the interstellar dust distribution. Similar to the 2D star-based dust maps, 3D dust mapping relies on the principle that dust reddens stellar photometric colors, so these 3D maps are based on dust extinction, rather than dust emission. By modeling this cumulative reddening effect for stars at different distances along an individual line of sight in the Milky Way, it is possible to infer the differential reddening along that line of sight. By grouping hundreds of millions of stars sightline-by-sightline across the Galaxy, one can reconstruct the 3D distribution of dust. 

One of the primary challenges of early 3D dust mapping efforts is simultaneously inferring the type of the star (and thus its intrinsic colors), its distance, and reddening from photometry alone. Despite the challenges, several such 3D dust maps have been built solely on the optical and/or infrared photometry readily available in large-scale surveys. \citet{Marshall_2024} combined the near-infrared photometric colors of stars detected by 2MASS \citep{Skrutskie_2006} with models for stars' intrinsic colors and distances obtained from the Besançon Stellar Population Synthesis Model of the Galaxy \citep{Robin_2003} to infer the 3D distribution of reddening toward the inner Galactic plane. \citet{Sale_2014} used a hierarchical Bayesian model applied to photometry from the IPHAS survey \citep{IPHAS} to construct a 3D dust map towards the northern Galactic plane. \citet{Green_2015} probabilistically inferred the distances, reddenings, and stellar types of $\approx 800$ million stars using a combination of Pan-STARRS1 \citep{Chambers_2016} and 2MASS photometry to produce a 3D dust map over three-quarters of the northern sky \citep[the ``Bayestar" map; see also][]{Green_2014, Green_2018}.

By providing constraints on the distances of a billion stars --- independent of their colors --- the second data release of the Gaia mission ushered in a new era of 3D dust maps with unprecedented distance resolution. A number of efforts have combined the broadband photometric colors of stars with Gaia astrometry in pursuit of even more highly resolved maps with improved distance resolution. For example, \citet{Lallement_2019} applied a hierarchical inversion algorithm to a combination of Gaia and 2MASS photometry and Gaia parallax measurements to produce a 3D dust map out to 3 kpc from the Sun with a distance resolution (in Cartesian space) of $\rm \approx 25 \; pc$ \citep[see also][]{Lallement_2018, Lallement_2022, Vergely_2022}. Applying a Random Forest regression to a combination of Gaia, 2MASS and WISE photometry alongside Gaia parallax measurements, \citet{Chen_2019} infer the properties of $\approx 50$ million stars and construct a 3D dust map of the full Galactic plane out to $\rm \approx 5 \; kpc$ with an angular resolution of $6\arcmin$. \citet{Green_2019} incorporated Gaia parallaxes into their ``Bayestar" dust mapping pipeline and implemented a new spatial prior \citep[compared to][]{Green_2018} for correlating neighboring sightlines, resulting in a 3D dust map with four times finer distance resolution than their previous maps and a typical angular resolution of $\approx 7\arcmin$. Incorporating Gaia-informed estimates of stellar distances and reddenings \citep[obtained from the StarHorse catalog;][]{Anders_2019}, \citet{Leike_2020} combined Metric Gaussian Variational Inference \citep{Knollmuller_2019}, Gaussian Processes, and Information Field Theory to map the 3D distribution of dust at $\approx$ 1 pc Cartesian resolution out to $\rm \approx 400 \; pc$ from the Sun --- resolving the detailed internal structure of local molecular clouds for the first time \citep[see also][]{Leike_2019}. Using a new catalog of distance and extinction measurements from \citet{Zhang_2023} based on the Gaia $BP/RP$ spectra, \citet{Edenhofer_2023} built upon the methodology of \citet{Leike_2020} to produce a parsec-scale 3D dust map out to a distance of 1.25 kpc from the Sun at $14 \arcmin$ angular resolution.

While Gaia has been transformational to the field of 3D dust mapping, the gains have largely been limited to the solar neighborhood, within a few kiloparsecs of the Sun.  This limitation stems from the underlying requirement that stars must be detected \textit{behind} dense dust clouds in order for the cloud to be detected in 3D maps. If the dust column is so high -- as in the inner Galactic plane -- that it extinguishes the light from background stars entirely, the cloud goes undetected in 3D maps \citep{Green_2019,Zucker_Speagle_2019}. Because most existing maps rely on either easily extinguished optical photometry and astrometry from Gaia and/or shallow near-infrared photometry (from e.g. 2MASS), most structure in the inner Galactic plane at distances $\rm \gtrsim 2 \; kpc$ has largely gone unresolved in 3D. This structure includes some of the richest regions in the inner galaxy, including the Scutum-Centaurus arm, which hosts much of the Milky Way's most active star formation. 

In this work, we leverage two of the deepest infrared surveys toward the inner Galaxy --- the Dark Energy Camera Plane Survey 2 \citep[DECaPS2;][]{Saydjari_2023} and the Vista Variables in the Via Lactea Survey \citep[VVV;][]{Minniti_2010} --- to infer the distances, reddenings, and stellar types of hundreds of millions of stars. Incorporating Gaia parallax distances when available nearby, but relying on the deep infrared photometry in heavily dust-enshrouded regions at greater distances, we construct a 3D dust map of the southern Galactic plane. Our 3D dust map achieves three main goals:

\begin{enumerate}
    \item By incorporating deep infrared photometry into an improved stellar modeling pipeline, we resolve hitherto-undetected structure towards the inner Galactic plane at distances $\rm \gtrsim 2 \; kpc$.
    \item By leveraging the sheer stellar density of DECaPS2 ($\approx 800$ million stars over only 6\% of the sky), we produce the highest angular resolution 3D dust map ($\approx 1\arcmin$) to date, constituting almost an order of magnitude improvement in angular resolution over current 3D dust maps \citep{Green_2019,Chen_2019}.
    \item By targeting the portion the Galaxy unreachable by Pan-STARRS1 ($\delta < -30^\circ$), we can combine our map with the ``Bayestar19" 3D dust map \citep{Green_2019} to provide full coverage of the Galactic Plane at $|b| < 10^\circ$ and enable extinction corrections over the entire disk. 
\end{enumerate}

\begin{figure*}[]
\begin{center}
\includegraphics[width=0.85\textwidth]{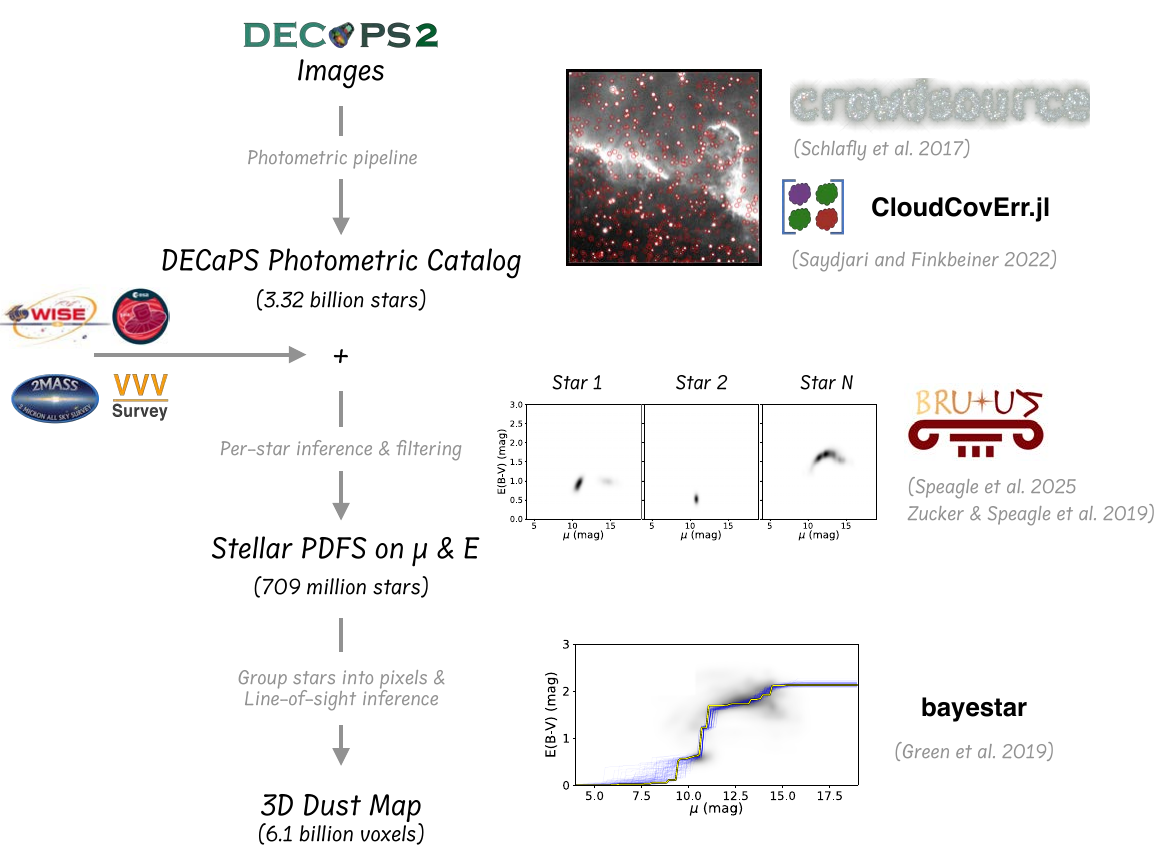}
\end{center}
\caption{Schematic overview of the process of turning images of the sky into a 3D map of dust. The DECaPS2 survey forms the foundation of our 3D dust map, whose photometric colors are combined with complementary photometric (VVV, 2MASS, unWISE) and astrometric surveys (Gaia) where available and fed into our stellar inference framework. We use the \texttt{brutus} stellar inference framework to infer the distance, extinction, and stellar type of hundreds of millions of stars. We then group stars into pixels and fit the set of distance-reddening measurements along the line of sight in each pixel to generate a 3D map of dust.} 
\label{fig:schematic}
\end{figure*}

In Figure \ref{fig:schematic} we provide a schematic overview of our pipeline. In \S \ref{sec:data} we describe the various datasets and the survey-level cuts applied to ensure data quality. In \S \ref{sec:compilation} we describe how the photometric and astrometric data from these surveys are combined into a single catalog. Leveraging the assembled catalog, we describe the methodology underpinning our stellar modeling and 3D dust mapping pipelines in \S \ref{sec:methods}. In \S \ref{sec:results} we present the 3D distribution of stars and dust in the southern Galactic plane. We discuss the implications of our map in \S \ref{sec:discussion} and compare with existing efforts. In \S \ref{sec:data_avail} we discuss the availability of the data generated and software produced under this work. Finally, we conclude in \S \ref{sec:conclusion}. For a summary of known artifacts in our 3D dust map, we refer readers to \S \ref{sec:artifacts} in the Appendix. In this work, our goal is to present the method of construction of the map, while future works will use this map to further investigate the spatial and dynamical organization of the interstellar medium in the inner Galaxy. 

\section{Data} \label{sec:data}

Our analysis is based on a combination of photometric and astrometric data from five surveys: 

\begin{itemize}
    \item The Dark Energy Camera Plane Survey 2\\ \citep[DECaPS2;][]{Saydjari_2023}
    \item The Vista Variables in the Via Lactea Survey \citep[VVV;][]{Minniti_2010}
    \item The Two Micron All Sky Survey \citep[2MASS;][]{Skrutskie_2006}
    \item The ``Unofficial" Wide-field Infrared Survey Catalog \citep[unWISE;][]{Schlafly_2019}
    \item The Third Data Release of the Gaia Mission \citep[Gaia DR3;][]{Gaia_DR3}
\end{itemize}

\subsection{DECaPS2} \label{subsec:decaps}

The second release of the Dark Energy Camera Plane Survey \citep[DECaPS2;][]{Saydjari_2023} is a deep five-band optical and near-infrared survey of the southern Galactic plane ($239^\circ < l < 6^\circ$, $|b| < 10^\circ$).\footnote{All the DECaPS2 photometry utilized here is available for download at \url{http://decaps.skymaps.info/release/data/files/DR2_REDUX/DATABASE/dbfits/} and accessible via NOIRLab's AstroDataLab} It builds upon the first data release \citep[DECaPS1;][]{Schlafly_2018} by improving the photometric reduction and extending the latitude range from $|b| < 4^\circ$ to $|b| < 10^\circ$, for a total of 2700 $\rm deg^{2}$ of sky coverage (6.5\%) with an average of three epochs per band. The DECaPS2 survey totals 3.32 billion objects and achieves average single-exposure AB magnitude depths of 23.5, 22.6, 22.1, 21.6, and 20.8 mag in the $g$, $r$, $i$, $z$, and $Y$ bands, respectively, though with considerable variation due to crowding. The filters span the wavelength range 400 to 1065 nm. We use the \texttt{CloudCovErr}-``corrected'' \texttt{crowdsource} photometry where the full background covariance is modeled, improving the uncertainty estimates in regions with structured backgrounds \citep[for details on the methods see][]{Schlafly_2021, Saydjari_2022}. On a band-by-band basis, we require an object to have an ``OK" detection in at least one epoch (\texttt{nmag\_cflux\_ok}$>0$), errors $<$ 0.1 mag, and \texttt{fracflux} $> 0.75$, indicating that at least 75\% of the PSF-weighted fraction of flux at the star's location is derived from itself (as opposed to neighboring sources). The DECaPS2 survey complements the Pan-STARRS1 survey, which covers three-quarters of the northern sky  ($\delta > -30^\circ$) at wavelengths similar to DECaPS2, which was previously used in the construction of the \texttt{Bayestar19} 3D dust map \citep{Green_2019}. 

\subsection{VVV} \label{subsec:vvv}

The VISTA Variables in the Via Lactea (VVV) survey \citep{Minniti_2010} is a near-infrared survey targeting 562 $\rm deg^{2}$ of the bulge ($-10^\circ < l < 10^\circ$, $-10^\circ < b < 5^\circ$) and southern Galactic plane ($-65^\circ < l < -10^\circ$, $-2^\circ < b < 2^\circ$). The original PSF reduction of the VVV survey by \citet{Alonso_Garcia_2018} used DoPhot \citep{DoPhot_1993} to derive a catalog of 846 million sources with coverage in the $Z$, $Y$, $J$, $H$, and $K_s$ filters, spanning $0.84-2.5 \; \mu m$. We utilize a more recent PSF reduction by \citet{Zhang_2019}, which applies the DaoPHOT \citep{DaoPHOT_1987} to the $J$, $H$, and $K_s$ bands only to obtain a 926 million source catalog that goes roughly one magnitude deeper than \citet{Alonso_Garcia_2018}. The \citet{Zhang_2019} reduction achieves $5\sigma$ limiting magnitudes of 20.8, 19.5, and 18.7 mag in the $J$, $H$, and $Ks$ bands, respectively, in the Vega system. \citet{Zhang_2019} flag sources as spurious detections by sigma-clipping outlying sources using the magnitude-error relations in different band combinations.

The DECaPS2 error modeling implemented by \citet{Saydjari_2023} debiased and improved uncertainty estimates for photometry on structured backgrounds using the \texttt{CloudCovErr} algorithm. Injection tests performed on every DECam exposure in the survey indicate that the photometric error estimates are correct except in regions of the most extreme crowding. Without marginalizing over structured backgrounds, uncertainty can often be underestimated by a factor of two or more. Therefore, we adopt a more stringent error cut for VVV, 2MASS, and unWISE since their uncertainties may be underestimated. On a band-by-band basis, we require that no source is flagged as spurious in any magnitude-error relation using that band and that the error is $<0.05 \; \rm mag$. 

\subsection{2MASS} \label{subsec:2mass}

The Two Micron All Sky Survey \citep[2MASS;][]{Skrutskie_2006} is an all sky infrared survey in the $J$, $H$, and $K$ bands, achieving a $10\sigma$ point-source depth of 15.8, 15.1, and 14.3 mag, respectively, in the Vega system. The point-source catalog contains a total of 470 million objects, of which roughly 340 million sources are considered good quality. On a band-by-band basis, we only utilize detection that meet the 2MASS ``high-reliability" criteria\footnote{For a description of the high-reliability criteria, see the \href{https://irsa.ipac.caltech.edu/data/2MASS/docs/releases/allsky/doc/sec1\_6b.html}{2MASS All Sky Data Release Explanatory Supplement}.} with errors $<0.05$ mag. We also exclude sources that are flagged as having possible contamination from nearby bright point sources (requiring \texttt{cc\_flg}$==0$) and galaxies (requiring \texttt{gal\_contam}$==0$). 

\subsection{unWISE} \label{subsec:unwise}
The ``Unofficial" Wide-field Infrared Survey Explorer \citep[unWISE;][]{Schlafly_2019} catalog analyzes the unblurred ``unWISE" coadds \citep{Lang_2014} derived from the WISE images to detect two billions sources over the full sky in the W1 and W2 bands at $3.4 \; \mu \rm m$ and $4.5 \; \mu \rm m$, respectively. Compared to the official ALLWISE catalog \citep{Cutri_2013}, the unWISE catalog is based on deeper imaging and uses the \texttt{crowdsource} algorithm \citep{Schlafly_2021} to improve modeling of crowded regions. The unWISE catalog extends 0.7 magnitudes deeper than ALLWISE, achieving a $5\sigma$ point-source depth of $\approx$ 20.7 and 20.0 mag for W1 and W2 in the AB system. On a band-by-band basis, we require that no flags are set (\texttt{flags\_unwise}$==0$), errors $<$ 0.05 mag, and \texttt{fracflux} $> 0.85$, indicating that at least 85\% of the PSF-weighted fraction of flux at the star's location is derived from itself rather than contamination from adjacent sources.

\subsection{Gaia DR3} \label{subsec:gaia}
Gaia \citep{Gaia_Mission} is an all-sky optical survey providing astrometry (parallaxes and proper motions) and photometry (in the $G$, $BP$, and $RP$ bands) for over a billion stars. Given the breadth of the Gaia passbands ($330 - 1050$ nm in $G$, $330-680$ nm in $BP$ and $630-1050$ nm in $RP$) in comparison to the DECaPS2 filters covering a similar wavelength range, we only utilize the Gaia astrometry (specifically the Gaia parallax measurements) in this work. We leverage the parallax measurements from the third data release \citep[Gaia DR3;][]{Gaia_DR3}, which provides a median parallax uncertainty of $\rm 0.02 - 0.03 \; mas$ for $G = 9-14$ mag and $\rm 0.5 \; mas$ uncertainty for $G=20$ mag \citep{Gaia_EDR3_Astrometry}. Following \citet{Fabricius_2021}, 
we exclude stars with renormalized unit weight error \texttt{ruwe} $>$ 1.4. 

\begin{figure}[t]
\includegraphics[width=0.45\textwidth]{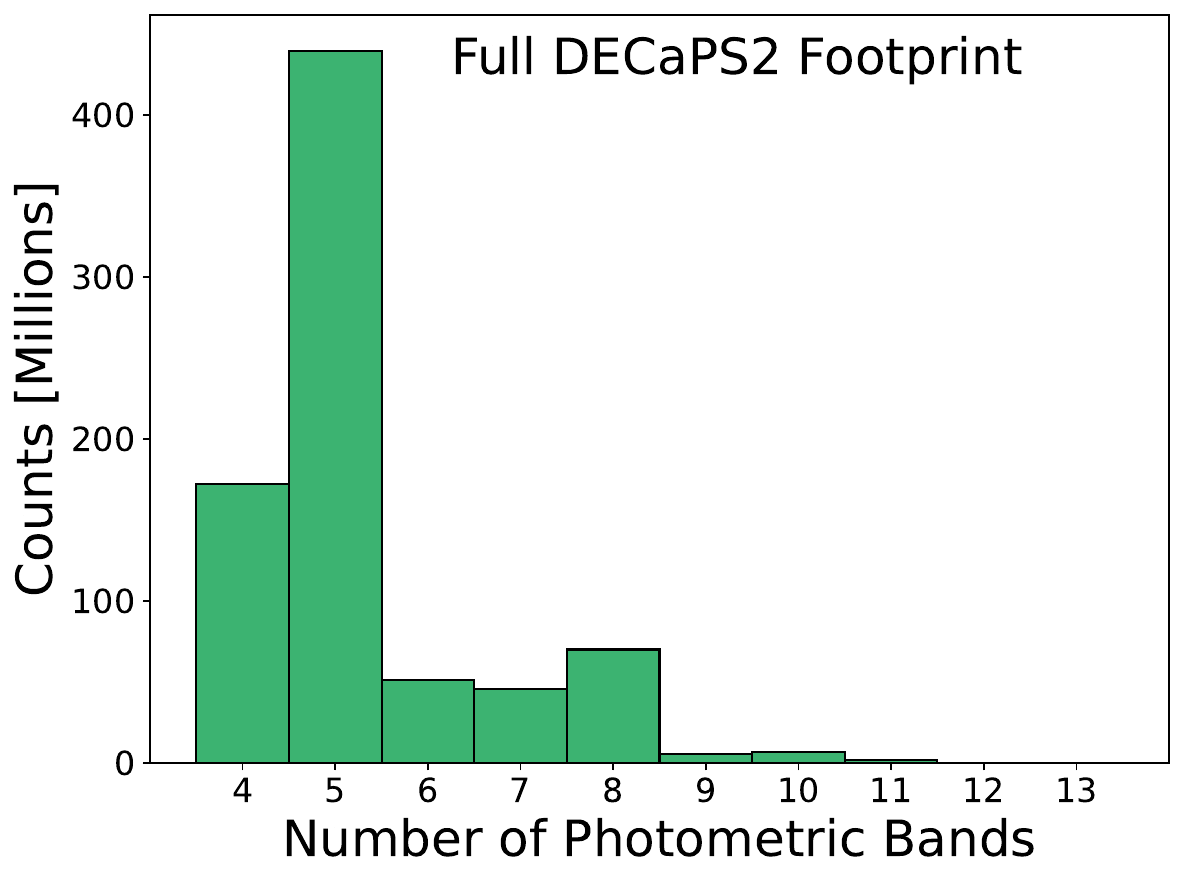}
\caption{Number of photometric bands per star incorporated into the stellar modeling, tabulated across all stars in the full DECaPS2 footprint. The number of photometric detections ranges from four (the minimum) to thirteen (the maximum), with an average of five bands incorporated per star. } 
\label{fig:ndet_per_star}
\end{figure}

\begin{figure*}
\begin{center}
\includegraphics[width=\textwidth]{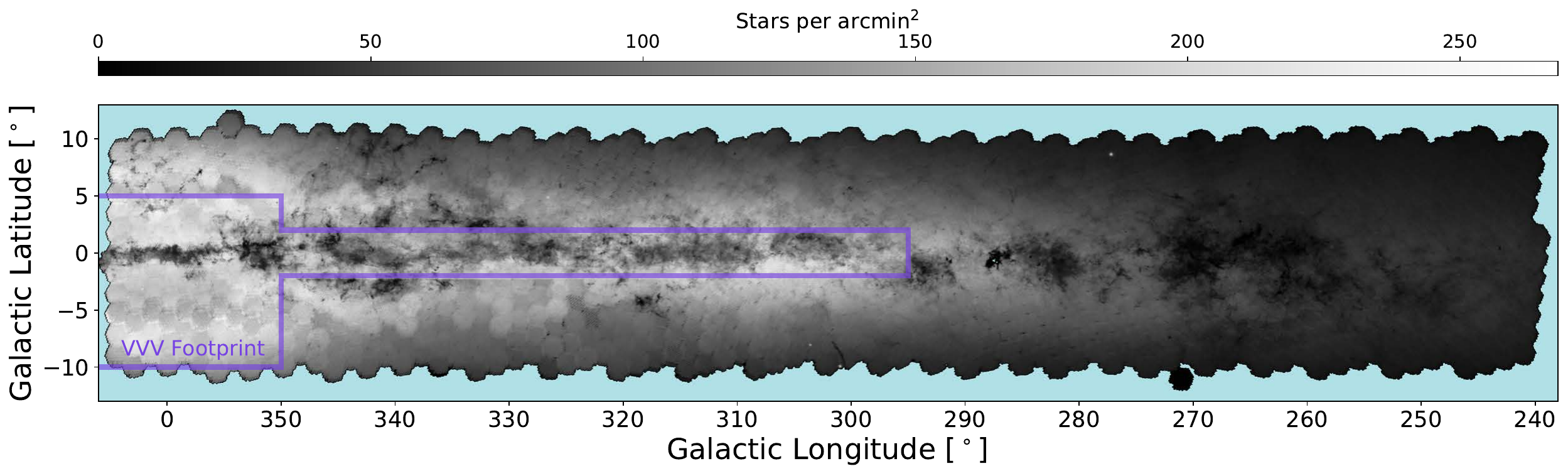}
\end{center}
\caption{Distribution of source density on the plane of the sky for stars we model in \S \ref{subsec:perstar_inference}, the majority of which will be used in the line-of-sight dust reconstruction. The VVV footprint (purple polygon) shows a modest increase in source density in comparison to the rest of the DECaPS2 footprint. The median source density is 70 stars per arcmin$^2$, with an interquartile range of $37-125$ stars per arcmin$^2$ .} 
\label{fig:source_density}
\end{figure*}

\begin{figure*}[t]
\begin{center}
\includegraphics[width=\textwidth]{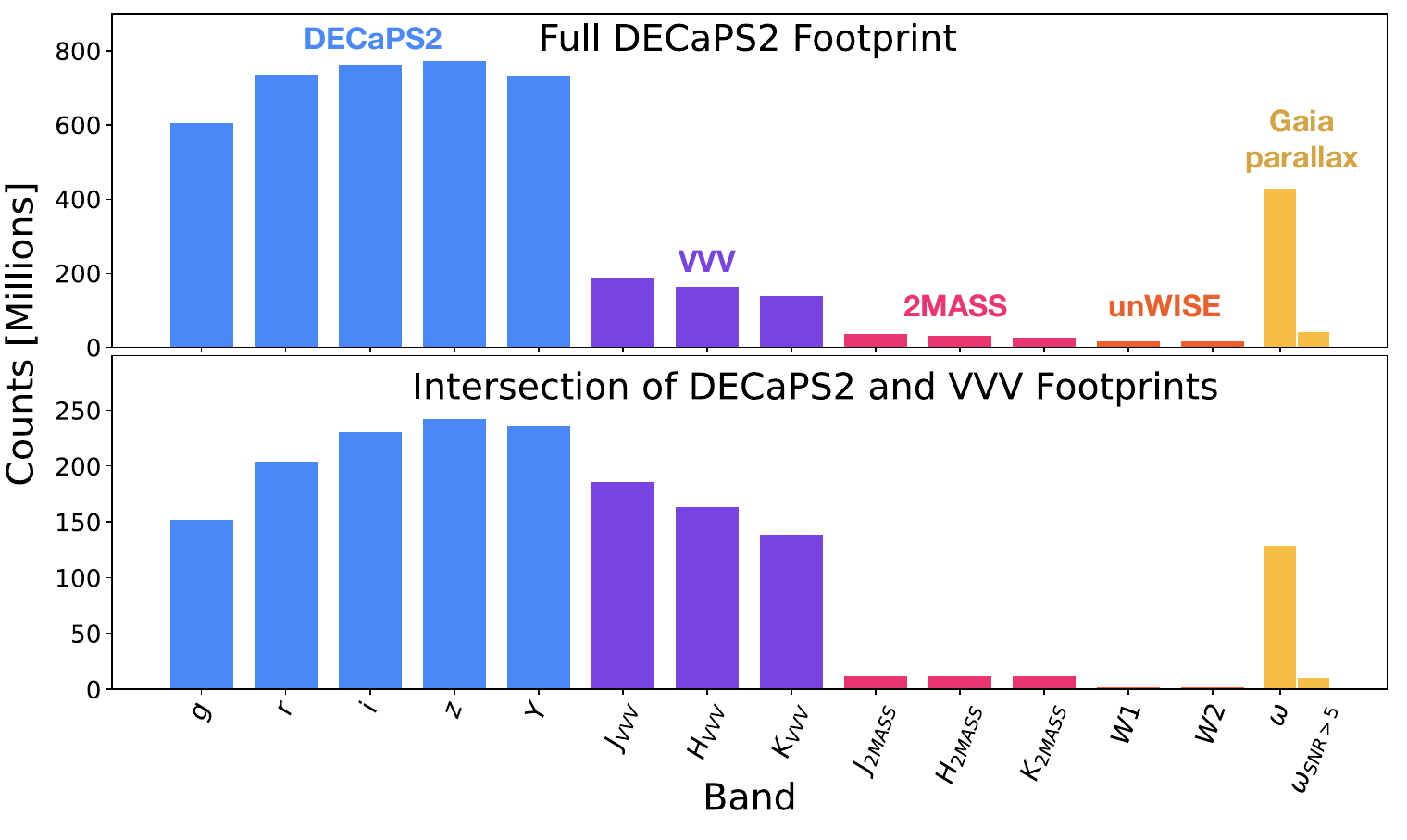}
\end{center}
\caption{Breakdown of the number of stars detected in each photometric band (shown in blue for DECaPS2, purple for VVV, pink for 2MASS, orange for unWISE) and with an available Gaia parallax measurement (shown in yellow). For Gaia, we further subdivide the stars into all those with a Gaia parallax detection and only those with a signal-to-noise ratio on the parallax detection $>5$. The top-panel shows the breakdown of band coverage for the entire DECaPS2 footprint, while the bottom panel shows the breakdown for the subset of the DECaPS2 footprint that overlaps with the VVV footprint (see purple polygon in Figure \ref{fig:source_density}).} 
\label{fig:coverage_hist}
\end{figure*}

\section{Assembling a Final Catalog} \label{sec:compilation}

To assemble a final catalog, we crossmatch the DECaPS2, VVV, 2MASS, unWISE, and Gaia DR3 surveys using the Large Survey Database (LSD) architecture \citep{Juric_2012}. DECaPS2 serves as the primary catalog for the astrometry, meaning we find photometry and astrometry for sources in the other catalogs that crossmatch to detections in DECaPS2, adopting a crossmatch radius of $0.5\arcsec$. After applying the band-by-band survey-level photometric quality cuts described in \S \ref{subsec:decaps}-\S\ref{subsec:unwise}, we require that the star be detected in at least four bands ($g,r,i,z,Y,J,H,K,W1,W2$), at least one of which must be a DECaPS2 band. To ensure continuity across the VVV survey boundary, we do consider sources that have detections in both VVV and 2MASS, but do not double count the $J$, $H$, and $K$ detections for the purposes of imposing the four-band minimum. For example, if a star is detected in 2MASS $J$, $H$, $K$ and VVV $J$, $H$, $K$ alongside a DECaPS $Y$ band detection, this star would be labeled as a four-band detection when filtering, but both the 2MASS and VVV data would be incorporated in the fit. We do not require a Gaia parallax measurement but include these distance constraints in the catalog when available. In Figure \ref{fig:ndet_per_star}, we break down the number of photometric bands per star incorporated into the stellar modeling, tabulated across the full DECaPS2 footprint. The number of bands ranges from four (the minimum) to thirteen (the full filter coverage, consisting of $g$, $r$, $i$, $z$, $Y$, $J_{VVV}$, $H_{VVV}$, $K_{VVV}$, $J_{2MASS}$, $H_{2MASS}$, $K_{2MASS}$, $W1$, $W2$), with an average of five bands per star. 

Our resulting catalog contains 793 million sources. Despite targeting only one third of the Galactic plane, our source count is just shy of the 799 million stars used to reconstruct the \texttt{Bayestar19} 3D dust map (covering the remaining two thirds of the plane and the rest of the northern sky), underlining the increased source density in the DECaPS2 footprint (Figure \ref{fig:source_density}).  In Figure \ref{fig:coverage_hist}, we break down the number of stars detected per band.  There are between $605-772$ million detections in any individual DECaPS2 $grizY$ band. Roughly $20\%$ of the sample have detections in a VVV band ($138-186$ million detections per band), compared to roughly 3\% for 2MASS ($25-35$ million detections per band). Less than 2\% of stars are detected in unWISE (15 million detections per band). In contrast, roughly half of the sample (427 million stars) have a Gaia parallax measurement, though only 5\% of stars (40 million) have a parallax that alone will strongly constrain the star's distance, (signal-to-noise ratio $>5$). In many cases the low SNR parallaxes are still useful for breaking the dwarf-giant degeneracy.  Considering only the region of the sky where there is overlap between the VVV and DECaPS2 footprints (purple polygon in Figure \ref{fig:source_density}, encompassing part of the bulge and the plane and the southern plane $|b| < 2^\circ$), roughly 75\% of stars have both DECaPS2 and VVV detections, underlining the importance of the deeper NIR photometry of VVV compared to 2MASS in dust-enshrouded regions.

\section{Methods} \label{sec:methods}

Here we outline our methodology for mapping the three-dimensional distribution of dust in the southern Galactic plane. As described in \S \ref{sec:perstar}, we start by inferring the stellar parameters, dust extinction, and distance on a star-by-star basis using our assembled photometric and astrometric catalog from \S \ref{sec:compilation}. We then group the stars into discrete pixels on the celestial sphere and use the per-star distance and extinction estimates within each pixel to model the distribution of dust as a function of distance along the line of sight, as described in \S \ref{sec:los}. The per-star and line-of-sight inference required a significant number of CPU hours, and the computational resources required to generate these data products are described in Appendix \S\ref{sec:compute_deets}.

\subsection{Stellar Modeling} \label{sec:perstar}

To perform the stellar modeling, we leverage the open-source Python package {\brutus} \citep{Speagle_2023a}. {\brutus} derives estimates of extrinsic stellar parameters $\eparams$ (distance, extinction, total-to-selective extinction ratio) over a grid of intrinsic stellar parameters $\params$ (e.g. initial mass).  The {\brutus} pipeline has already successfully been applied to estimate distances, extinctions, and stellar properties of 170 million stars at high galactic latitude (the {\augustus} catalog) as presented in \citet{Speagle_2023b}. In \S \ref{subsec:perstar_inference}, we briefly summarize the core inference framework from \citet{Speagle_2023a} and highlight specific adaptions implemented in this work with respect to \citet{Speagle_2023b}. In \S \ref{subsec:perstar_setup} we describe the {\brutus} setup (version \texttt{v0.8.3}) used to perform the inference on the assembled catalog from \S \ref{sec:compilation}. 

\subsubsection{Statistical Framework} \label{subsec:perstar_inference}

We assume that the observed magnitudes of a star, denoted by $\mags \equiv \{ m_i \}_{i=1\ldots b}$, measured across a set of $b$ photometric passbands, can be modeled as
\begin{equation} \label{eq:model_mag}
    \mags_{\params, \eparams} 
    \equiv \absmag_{\params} + \mu + A_V \times 
    \left(\rvec + R_V \times \rvec'\right)
\end{equation}

\noindent where
\begin{itemize}
 \item $\absmag_{\params}$ is the set of absolute magnitudes for a star as a function of its intrinsic stellar parameters {\params}
 
 \item $\mu$ is the distance modulus

 \item $A_V$ is the dust extinction, measured as the total attenuation in magnitudes in the $V$ band
 
 \item $\rvec$ is the reddening vector
 
 \item $R_V$ is the total-to-selective extinction ratio, measured as the attenuation in the $V$ band $A_V$ relative to the color excess in the B and V bands, $E_{B-V}$ ($R_V = \frac{A_V}{E_{B-V}}$)
 
 \item $\rvec'$ is the differential reddening vector that sets the shape of the reddening vector $\rvec'$.
 
\end{itemize}

Following \citet{Speagle_2023b}, to parameterize {\params}, we utilize the MESA Isochrone and Stellar Tracks (\mist) models \citep{Choi_2016}, which are a set of theoretical isochrones that relate intrinsic stellar evolutionary parameters $\params$ to surface level parameters $\params_\star$ via stellar atmospheric models \citep[see][]{Cargile_2020}. The intrinsic stellar parameters $\params$ include the initial mass $M_{\rm init}$, the initial metallicity $\feh_{\rm init}$, and the Equivalent Evolutionary Phase ${\rm EEP}$, which breaks down the various stages of stellar evolution into an equal number of steps to ensure efficient sampling of phases even when stars are rapidly evolving on the post main-sequence. The surface level parameters $\params_\star$ include the effective temperature $T_{\rm eff}$, the surface gravity $\log g$, the surface metallicity $\feh_{\rm surf}$, and the surface $\alpha$-abundance enhancement $\afe_{\rm surf}$. As in \citet{Speagle_2023b}, $\afe_{\rm surf} = 0$ by default since the current models assume solar-scaled abundance patterns. However, we hope to explore the effect of $\alpha$-abundance enhancements --- and particularly any potential degeneracies between $\afe_{\rm surf}$ and extinction --- in future work. 

We utilize Version 1.2 of the non-rotating {\mist} models. We do not model the secondary mass fraction and assume all stars are single stars. The exact grid of models for the stellar evolutionary parameters {\params} utilized in this work (\texttt{grid\_mist\_v10}) is summarized in Table \ref{tab:mist_grid} and available for download via the {\brutus} GitHub page. Notably, we extend the mass grid down to $0.2 \; \rm M_\odot$ from the minimum mass of $0.5 \; \rm M_\odot$ considered in \citet{Speagle_2023b} to account for the abundance of M-dwarfs now being detected at greater distances in the DECaPS2 survey. Note that we do not include any models on the pre-MS (${\rm EEP} <  202$) or beyond the start of the thermally pulsing asymptotic giant branch (${\rm EEP} > 808$).

\begin{deluxetable}{lcc}
\tablecolumns{3}
\tablecaption{Parameter Grid for the {\mist} models (\texttt{grid\_mist\_v10}). 
\label{tab:mist_grid}}
\tablehead{Minimum & Maximum & Spacing}
\startdata
\cutinhead{\textbf{Initial Mass} ($M_{\rm init}$)}
$0.2\,\rm M_\odot$ & $2.0\,\rm M_\odot$ & $0.025\,\rm M_\odot$ \\
\cutinhead{\textbf{Initial Metallicity} ($\feh_{\rm init}$)}
$-3.0$ & $-2.0$ & $0.10$ \\
$-2.0$ & $0.45$ & $0.05$ \\
\cutinhead{\textbf{Equivalent Evolutionary Point} (${\rm EEP}$)}
$202$ & $454$ & $6$ \\
$454$ & $808$ & $2$ \\
\enddata
\end{deluxetable}

The quantities $\rvec$ and $\rvec'$ are fixed to the extinction curve of \citet{Schlafly_2016} and published filter curves for the different bandpasses\footnote{We document the extinction curve script originally published in \citet{Schlafly_2016} on Zenodo (see \href{https://zenodo.org/records/16813633}{doi:10.5281/zenodo.16813633})}.  Unlike \citet{Speagle_2023b}, we assume that $\rvec$ and $\rvec'$ 
are the same for all models, such that $\rvec_{\params} = \rvec$ and $\rvec'_{\params} = \rvec'$. Therefore, the extinction curve does not change as a function of spectral type. The gray component of the extinction curve is not measured in \citet{Schlafly_2016}, and instead relies on \citet{Indebetouw_2005}, which finds $A_H/A_K = 1.55$.  The components of $\rvec$ are computed as derivatives of $A_{542}$ at $A_{542} = 2.3~\mathrm{mag}$ for a 4500~K solar metallicity giant, where $A_{542}$ is the extinction at 542~nm.  The values of $\rvec'$ give the derivative of the extinction at $R_V = 3.32$ for each filter for the same dust column and solar spectrum. We re-normalize the extinction curve to scale with $A_V$, and the extinction coefficients normalized to $A_V=1 \; \rm mag$ for the mean $R_V = 3.32$ are summarized in Table \ref{tab:extcurve}. Note that the extinction coefficient for the WISE W2 band, $A_{W2}$ is larger than typically reported in the literature, on par with $A_J$ \citep[see e.g.][]{Wang_2019}. The W2 band is the only infrared band (and the only band other than $g$ band) to have a negative derivative $R'$, causing this value to be slightly larger than literature values when we re-normalized the extinction curve to scale with $A_V$ (see Eqn. \ref{eq:model_mag}). Less than 2\% of stars have detections in W2 (and the extinction behaves as expected when scaling with $A_V$ in Eqn. \ref{eq:model_mag}) so this effect should be negligible on the resulting stellar inference.

This model for the extinction curve is an approximation, assuming that a particular parcel of dust will change the magnitude of any observed star in exactly the same way.  This model would be correct if the bandpasses were infinitely narrow or the extinction curve constant in wavelength over a bandpass.  In fact variation of the extinction curve over the bandpass introduces small dependences of $\rvec$ and $\rvec'$ in Table~\ref{tab:extcurve} on extinction and stellar type.  Here we neglect that variation, which is quite small for the bandpasses considered in this work but is much larger for, for example, the very broad Gaia $G$, $BP$ and $RP$ bands.  This model has the advantage that it is simpler and can be implemented more efficiently. While we technically allow the extinction curve to vary from star to star (by sampling in $R_V$), any inference surrounding the dust extinction curve is highly prior driven, as we discuss further in Appendix \S\ref{sec:rv}. 

\begin{deluxetable}{ccc}
\tablecolumns{3}
\tablecaption{Adopted Reddening Vector 
\label{tab:extcurve}}
\setlength{\tabcolsep}{14pt}
\tablehead{Filter & \rvec & $\rvec'$}
\startdata
   DECam $g$ & 1.3285 & -0.0169 \\
   DECam $r$ & 0.7000 &  0.0631 \\
   DECam $i$ & 0.3529 &  0.0982 \\
   DECam $z$ & 0.1814 &  0.0989 \\
   DECam $Y$ & 0.1297 &  0.0933 \\
   \hline
   VISTA $J$ & 0.0823 &  0.0609 \\
   VISTA $H$ & 0.0492 &  0.0371 \\
   VISTA $K$ & 0.0318 &  0.0241 \\
   \hline
   2MASS $J$ & 0.0835 &  0.0624 \\
   2MASS $H$ & 0.0489 &  0.0370 \\
   2MASS $K$ & 0.0315 &  0.0239 \\
   \hline
   WISE $W1$ & 0.0484 &  0.0052 \\
   WISE $W2$ & 0.0746 & -0.0080
\enddata
\tablecomments{Parameterization of the mean reddening vector $\rvec$ (corresponding to $R_V = 3.32$) and the differential reddening vector $\rvec'$ in this work (see Eqn. \ref{eq:model_mag}). The extinction coefficients have been normalized to $A_V = 1 \; \rm mag$ with the gray component fixed assuming $\frac{A_H}{A_K} = 1.55$ following \citet{Indebetouw_2005}.}
\end{deluxetable}

Based on our model, the posterior probability $P(\params, \eparams | \; \hat{m}, \hat{\varpi})$ that the set of intrinsic stellar parameters $\params$ and extrinsic parameters $\eparams$  is consistent with the observed magnitudes $\hat{m}$, measured Gaia parallax $\hat{\varpi}$, and prior information on $\params$ and $\eparams$ is given by Bayes' Theorem: 

\begin{align}
    P(\params, \eparams | \; \hat{m}, \hat{\varpi}) &\propto
    P(\hat{m}, \hat{\varpi} | \; \params, \eparams) P(\params, \eparams)
    \nonumber \\
    &\boxed{\equiv 
    \likelihood_{\rm phot}(\params, \eparams) 
    \likelihood_{\rm astr}(\eparams)
    \prior(\params, \eparams)
    }
\end{align}

where $P(\params, \eparams | \; \hat{m}, \hat{\varpi})$
is the posterior probability for $\params$ and $\eparams$,
$P(\hat{m}, \hat{\varpi} | \; \params, \eparams)
\equiv \likelihood_{\rm phot}(\params, \eparams) \likelihood_{\rm astr}(\eparams)$
is the likelihood --- broken into the
photometric $\likelihood_{\rm phot}(\params, \eparams)$ 
and astrometric likelihood $\likelihood_{\rm astr}(\eparams)$ ---
and $P(\params, \eparams) \equiv \prior(\params, \eparams)$
is the galactic prior.

Our photometric likelihood $\likelihood_{\rm phot}(\eparams)$ compares the set of observed magnitudes {\mags} with the predicted magnitudes $\mags_{\params, \eparams}$ given the associated observational errors $\hat{\errors}$. We assume $\likelihood_{\rm phot}(\params, \eparams)$ to be independent and roughly Gaussian in the measured magnitude in each band $b$ such that: 

\begin{equation}
    \likelihood_{\rm phot}(\params, \eparams)
    \equiv \prod_{i=1}^{b} \frac{1}{\sqrt{2\pi\hat{\sigma}_i^2}}
    \exp\left[-\frac{1}{2}\frac{\left(m_i(\params, \eparams) 
    - \hat{m}_i\right)^2}{\hat{\sigma}_i^2}\right]
\end{equation}

Likewise, the astrometric likelihood $\likelihood_{\rm astr}(\eparams)$ compares the predicted parallax $\varpi(\eparams)$ with the observed parallax $\hat{\varpi}$ given the associated parallax error $\hat{\sigma}_\varpi$. We also assume $\likelihood_{\rm astr}(\eparams)$ to be Gaussian such that: 

\begin{equation}
    \likelihood_{\rm astr}(\eparams) \equiv
    \frac{1}{\sqrt{2\pi\hat{\sigma}_{\varpi}^2}}
    \exp\left[-\frac{1}{2}\frac{\left(\varpi(\eparams) 
    - \hat{\varpi}\right)^2}{\hat{\sigma}_\varpi^2}\right]
\end{equation}

The Galactic prior $\prior(\params, \eparams)$ encompasses prior beliefs on the 3D distribution and properties of stars and dust in the Milky Way. The prior is broken up into several components that are assumed to be independent: 

\begin{align}
    \pi(\params, \eparams) \propto 
    &\:\:\underbrace{\pi(M_{\rm init})}_{\rm IMF} \nonumber \\
    &\times \underbrace{\pi(d | \ell, b)}_{\rm 3D\:number} \nonumber \\
    &\times \underbrace{\pi(\feh_{\rm init} | d, \ell, b)}_{\rm 3D\:metallicity}
    \nonumber \\
    &\times \underbrace{\pi(t_{\rm age} | d, \ell, b)}_{\rm 3D\:age} \\
    &\times \underbrace{\pi(A_V | d, \ell, b)}_{\rm 3D\:extinction} \nonumber \\
    &\times \underbrace{\pi(R_V)}_{\rm Dust\: extinction \: curve} \nonumber
\end{align}
Here $d$ is the distance to the star and $(\ell, b)$ are the star's Galactic coordinates. We adopt the same default priors for the initial mass function (IMF), metallicity, age, and dust extinction curve as described in Appendix \S A of \citet{Speagle_2023a}. Briefly, we assume that the initial masses of stars follow a broken power law following \citet{Kroupa_2001}. For the 3D number density prior, we adopt a three component thin disk, thick disk and halo model on stellar distance, metallicity, and age informed by previous studies \citep{Bland_Hawthorn_2016, Anders_2019, Xue_2015}. Our 3D number density prior is conceptually similar to \citet{Bailer_Jones_2018}, except they adopt a Galactic model based on a Gaia mock catalog from \citep{Rybizki_2018}. Following \citet{Schlafly_2016}, for the prior on the dust extinction curve we assume a mean $R_V$ of $\mu_{R_V}=3.32$ and standard deviation of $\sigma_{R_V}=0.18$. This is a very tight prior on $R_V$, so any significant variations we may see in $R_V$ are likely due to our fit compensating for limitations in other areas (e.g. the stellar models) rather than capturing meaningful variations in the dust extinction curve.  Unlike \citet{Speagle_2023b}, who adopt a 3D extinction prior based on the 3D dust map of \citet{Green_2019}, we instead place a flat prior on extinction over the range $A_V  = 0-24 \; \rm mag$. The range for the flat prior was chosen to match that reddening range \citet{Green_2019}, who sampled for the stellar reddening between $0-7$ mag.

\subsubsection{Application to Data} \label{subsec:perstar_setup}

We apply the {\brutus} pipeline to all 793 million stars in the assembled catalog described in \S \ref{sec:compilation}.  As discussed in \citet{Speagle_2023a,Speagle_2023b}, there are systematic offsets between the MIST models and the underlying photometric data that are not fully captured by the photometric errors. We thus apply both a zeropoint correction and an error floor to the photometry. 

Full details on both the zeropoint correction and the error floor are summarized in Table 1 of \citet{Speagle_2023b}. Briefly, the photometric zeropoint corrections range from $1\%-4\%$ on a band-by-band basis, and are multiplied to the observed flux densities after transforming from magnitude space to flux density space. The error floors are added in quadrature to the reported observational uncertainties on the magnitudes and range from $0.02$ mag for DECaPS2 up to $0.04$ mag for unWISE. 

In addition to the photometric corrections, we also apply a correction to the Gaia astrometry. Specifically, we implement a parallax zero point correction \citep[as described in][]{Lindegren_2021_ZP} dependent on the star's magnitude, color, and ecliptic latitude. To apply the correction, we utilize the \href{https://gitlab.com/icc-ub/public/gaiadr3_zeropoint}{\texttt{gaiadr3-zeropoint}} package.

The \texttt{brutus} configuration used to generate our stellar modeling results, including all photometric and astrometric corrections, is available online via Zenodo (\href{https://doi.org/10.5281/zenodo.14915000}{doi:10.5281/zenodo.14915000}).

We generate one thousand random samples from the posterior to construct a 2D binned posterior on distance and reddening for each star, which will be used to fit the line-of-sight 3D dust model as described in \S \ref{sec:los}. To generate the 2D posteriors, we use the built-in \texttt{bin\_pdfs\_distred} function in {\brutus}, evaluating the posterior density of each star on a grid with distance modulus spanning $\mu=4-19$ mag in steps of $0.125$ mag and reddening spanning $E = 0-7$ mag in steps of 0.01 mag. We convert from extinction to reddening using the $R_V$ samples. Following \citet{Green_2019}, we smooth the posteriors, adopting a Gaussian kernel with a standard deviation equal to 1\% of the total range.  

In addition to the 2D distance-reddening posteriors, we use the one thousand random samples to compute the 2.5th, 16th, 50th, 84th, and 97.5th percentiles (median, $1\sigma$, and $2\sigma$ errors) of the intrinsic and extrinsic stellar parameters. After saving the marginalized distance-extinction 2D posteriors and the percentiles, we thin the samples and save a subset of random samples to enable other science cases beyond those described in this work. 

\subsection{3D Dust Modeling} \label{sec:los}

In this section, we describe our approach to utilizing the posterior density estimates of distance and reddening for individual stars generated as part of the modeling in \S \ref{subsec:perstar_setup} to constrain the line-of-sight distribution of dust in the Milky Way. In \S \ref{subsec:filtering} we explain the quality cuts imposed on individual stellar posteriors and, in \S \ref{subsec:pixelization}, how we group stars into individual pixels on the celestial sphere to fit for the extinction as a function of distance along each line of sight. Finally, in \S \ref{subsec:bayestarlos} we discuss the inference framework for the line-of-sight fits, and in \S \ref{subsec:infill} we describe how we use data-driven Gaussian process priors to infill missing pixels. 

\subsubsection{Filtering Per Star Posteriors} \label{subsec:filtering}
To fit the line-of-sight dust distribution, we partition the sky using a HEALPix pixelization \citep{Gorski_2005} with $N_{\rm side}=8192$ and a nested ordering scheme. Before grouping stars by HEALPix pixel, we filter out stars with unreliable modeling. First, we remove stars with poor best-fit chi-square ($\chi^2$) values. Because $\chi^2$ depends on the number of bands detected per source, which varies from four to thirteen bands, we remove all stars for which:  
\begin{equation} \label{eq:filtering}
P(\chi^2_{n_{bands}} > \chi^2_{\rm best}) < 0.01
\end{equation}
\noindent with $n_{bands}$ being the number of detected bands. Eqn. \ref{eq:filtering} is designed to filter stars that fail to achieve even a single reasonable fit. The term $P(\chi^2_{n_{bands}} > \chi^2_{\rm best})$ denotes the $\chi^2$ survival function, or the probability of observing a $\chi^2$ value larger than $\chi^2_{\rm best}$ assuming the number of degrees of freedom is equal to the number of detected photometric bands. By filtering at the  $P(\chi^2_{n_{bands}} > \chi^2_{\rm best}) < 0.01$ threshold, we remove stars whose best-fit model is statistically inconsistent with the data at the 99\% confidence level, roughly equivalent to removing outliers at the $2\sigma-3\sigma$ level.

Next, we impose a cut to remove low-mass M-dwarfs from the sample. The theoretical {\mist} isochrones are known to exhibit large systematic biases in predicted colors at masses below $M_{\rm init} < 0.5 \; \rm M_\odot$ \citep[see e.g.][]{Choi_2016, Speagle_2023a, Speagle_2023b}. Thus, we remove all stars where the 50th percentile of the $M_{\rm init}$ samples is less than $0.5 \; \rm M_\odot$. As noted in \S \ref{subsec:perstar_inference} we extend the initial mass grid down to $M_{\rm init} = 0.2 \; \rm M_\odot$. Although stars with masses between $M_{\rm init} = 0.2 - 0.5 \; \rm M_\odot$ are ultimately filtered out, extending the mass grid down to lower masses improves the line-of-sight fits, since these stars are more effectively excluded rather than potentially being mismodeled as higher mass solutions due to known degeneracies between, for instance, dwarf and giant \citep[see e.g.][]{Green_2014}.

Next, we impose a cut on the maximum distance to a star, removing all stars where the 2.5th percentile of the distances samples ($2\sigma$ below the median distance) is greater than 30 kpc. This cut roughly corresponds to the edge of the stellar disk on the far side of the Milky Way and removes a very small fraction ($< 1\%$) of spurious fits, which largely manifest in nebulous and crowded regions near $b=0^\circ$, likely due to source blending. 

Finally, we apply a cut to remove a small fraction of stars displaying a prominent very nearby ($\mu \lesssim 8$ mag), high reddening ($E_{B-V} \gtrsim$ 3 mag) mode. This mode appears predominantly within a few degrees of the Galactic center and does not correspond with any nearby known dense clouds. We explored several pathways to explain the existence of this mode, ruling out any obvious issues in our pipeline such as poor stellar model coverage, the quality of the stellar photometry in the most crowded area of the sky, or the more extreme variation in the extinction law towards the Galactic center. While we could not definitively identify the root cause of this issue, we know the modeling of this mode is nonphysical based on the absence of such extreme, nearby dust in high-Cartesian-resolution 3D dust maps of the solar neighborhood \citep[see e.g.][]{Edenhofer_2023,Leike_2020}. To address this issue, we remove stars where a majority of the probability in their 2D binned posterior on distance and reddening is inconsistent with the \citet{Edenhofer_2023} 3D dust map. \citet{Edenhofer_2023} provide twelve samples of the extinction density at parsec-scale resolution out to 1.25 kpc from the Sun. For each $\mu$ bin in our binned posteriors, we determine the maximum extinction density of the \citet{Edenhofer_2023} map (over the entire sky and all twelve samples) at that distance, considering only distance bins out to the limit of the \citet{Edenhofer_2023} map ($\mu=10.4375$ mag).\footnote{We convert from the unitless extinction of the \citet{Edenhofer_2023} map to $E_{B-V}$ using the published extinction coefficients from \citet{Zhang_2023}.} We then sum over all probability within $\mu = 10.4325$ mag and above the max reddening from \citet{Edenhofer_2023} at each distance, removing stars where $> 50\%$ of their probability lies in this unphysical regime. For the remaining stars, we set this unphysical portion of the binned posterior equal to zero, and renormalize each posterior before running the line-of-sight fits. 

Of the 793 million stars in the catalog, 84 million are removed via this filtering scheme, leaving 709 million stars to constrain the line-of-sight dust distribution. We hereafter refer to these 709 million stars as the ``high-quality" stellar sample. Over the DECaPS2 survey footprint, we sample the 3D dust distribution in roughly 51 million $N_{side} = 8192$ pixels and 120 distance bins (over 6.1 billion total voxels) using this high-quality stellar catalog.

\subsubsection{Pixelization} \label{subsec:pixelization}

The angular resolution of extinction-based 3D dust maps depends on both the stellar density and chosen pixelization. If there are insufficient reliable posteriors for stars in a given resolution element, then the line-of-sight reddening for that pixel will be poorly constrained. Previous works have assigned stars to pixels with a binary weight function, including stars that fall within a HEALPix pixel and excluding them from all other pixels \citep[e.g.][]{Green_2019}.

One of the primary reasons for the high angular resolution of our map is our use of the DECaPS2 catalog, which goes deeper than Pan-STARRS1 and carefully deblends crowded fields to obtain an extremely high stellar density catalog. The other key to our high angular resolution is methodological, choosing to use a Gaussian weighting function when assigning stars to pixels, rather than a binary one. Specifically, before a star's posterior contributes to the likelihood for a given HEALPix pixel (see \S \ref{sec:los}, Eq. \ref{eqn:los-posterior-final}), its posterior is multiplied by a Gaussian weighting function on the great circle distance between the star and the HEALPix pixel center, with FWHM $= 2.5$ pixels ($1\arcmin.07$ at $N_{\rm side}=8192$, which has $26\arcsec$ pixels) so it is well sampled. We truncate this weighting function at 2$\times$ FWHM (5 pixels, $4.7\sigma$, $2\arcmin.15$ at $N_{\rm side}=8192$) for computational efficiency, because stars far in the tails of the weighting function do not change the line-of-sight inference significantly. Note that we implement this Gaussian weighting scheme \textit{not} due to uncertainty on the star's sky coordinates, but rather because it acts as a form of regularization that correlates the inference across neighboring pixels

While this weighting scheme can increase the accessible angular resolution with a given stellar density, the maximum angular resolution is still limited by the stellar density of the photometric catalogs used. We find that $\sim1\%$ of pixels have too few stars ($< 10$) for a reliable line-of-sight inference when using a HEALPix pixelization with $N_{side} = 8192$. For comparison, this angular resolution is $5\times$ higher than the Planck integrated reddening maps \citep{Planck_2014}.

One added benefit of our Gaussian weighting scheme is that, in the limit of uniform stellar density, the dust map presented here has a well-defined PSF\footnote{We caution that a pixel on the edge of a dense cloud or filament may have a very non-uniform angular distribution of stars. In this case, the PSF is still somewhat ambiguous.} with FWHM $=$ 1'.07. This facilitates comparisons with (N)IR emission-based dust maps and is in contrast to many other 3D dust maps (where angular resolution varies across the sky and as a function of distance), despite most astronomers expecting image data to have a well-defined PSF. Previous 3D dust mapping works have imposed correlations between independent pixels after the fact \citep{Green_2019} or imposed a spatial correlation kernel prior at inference time \citep{Leike_2019,Leike_2020,Edenhofer_2023}, but neither approach yields a well-defined PSF on the sky.

\subsubsection{Line-of-Sight Inference} \label{subsec:bayestarlos}

For each $N_{\rm side}=8192$ pixel, we split the line of sight up into discrete distance bins, evenly spaced in distance modulus from $\mu = 4$ mag to 19 mag, with a bin spacing of 0.125 mag. We model the dust distribution as a step function, where the parameter $\alpha$ denotes the increases in reddening in the set of distance bins, discretized as integer multiples of 0.01 mag. Recall that we pre-computed the joint posterior density on distance modulus $\mu$ and reddening $E$ for individual stars in \S \ref{subsec:perstar_setup}, which we now denote by $\tilde{p} \arg{\mu, \, E }$.

Following \citet{Green_2019}, the posterior probability of $\alpha$ is given by: 

\begin{align}
    \pcond{\alpha}{\!\left\{ \mu, \, E \right\}}
    &\propto
    p \arg{ \alpha}
    \prod_{i \, \in \, \mathrm{stars}}
    \int
    \tilde{p}_i \arg{\mu_i, \, E \arg{\alpha, \mu_i}}
    \, \mathrm{d}\mu_i
    \, .
    \label{eqn:los-posterior-final}
\end{align}

$p(\alpha)$ is a prior that encompasses our expectations on the differential reddening. Following \citet{Green_2019}, we place a log-normal prior on the increase in reddening in each distance bin, adopting a smooth model of the distribution of dust in the Galaxy based on \citet{Drimmel_Spergel_2001}. The term $\int \tilde{p}_i \arg{\mu_i, \, E \arg{\alpha, \mu_i}}  \, \mathrm{d}\mu_i$ is our likelihood function for the $i\rm th$ star in the pixel, and is equivalent to taking the line-integral over $\mu$ through the star's pre-computed distance-extinction posterior following the distance-reddening curve defined by $\alpha$. The total likelihood is the product of the individual stellar likelihoods over all stars. 

We sample for $\alpha$ using a custom C++ implementation\footnote{\href{https://github.com/andrew-saydjari/bayestar.git}{https://github.com/andrew-saydjari/bayestar.git} v0.2.0} of parallel tempered Markov Chain Monte Carlo \citep[MCMC, ][]{Speagle_2019}, which fits a model in which line-of-sight extinction is discretized at the 0.01\,mag level \citep{Green_2019}. We begin with a randomized initial extinction profile for each line of sight. In each distance bin, the initial guess in the jump in extinction is drawn from a chi-square distribution (with one degree of freedom) and the total extinction (in the large-distance limit) is normalized to equal ($\pm$20\% of) the 90th percentile of the extinctions of the individual stars along the line of sight. 

We use chains at four different temperatures (with inverse temperatures of $0.9^0$, 0.9$^1$, $0.9^2$ and $0.9^3$). Each chain is generated using Metropolis-Hastings sampling, with a mix of four proposal types: 33.3\% ``shift'' steps (moving differential extinction up or down by 0.01\,mag in a given distance bin), 16.7\% ``absolute'' steps (setting the total extinction at a given distance to a random value between the extinctions of the neighboring distance bins), 33.3\% ``swap'' steps (swapping the differential extinction at two randomly chosen distances), and 16.7\% ``swap neighbor'' steps (swapping the differential extinction at two neighboring distances). We sample the ensemble of temperature chains in rounds of updates. During each update round, we first take 2400 (2400, 4800, 4800) steps in each temperature chain. Then, we propose a ``temperature-swap'' step, which swaps the state of two randomly selected temperature chains. We perform a total of 2500 (5000, 5000, 10000) rounds of updates. The first 20\% of update rounds are discarded as burn-in.  

Convergence was assessed through the autocorrelation times of three variables: the prior, the likelihood, and the strength of the first principal component of the parameter vector in the MCMC chain. We require that none of the autocorrelation times exceed 5\% of the total number of MCMC steps. Pixels that failed to converge were re-run a maximum of three times per setting, under each of four increasingly expensive MCMC settings (settings listed parenthetically). Empirically, we found most pixels that could ``converge'' under a given configuration would do so within three attempts, so we repeat computationally cheaper sampling configurations before changing configurations. Even if pixels did not ``converge'' under this autocorrelation metric, the result of the final (lowest temperature) line-of-sight sampling is reported in our map, though it is flagged by an accompanying bitmask.

\subsubsection{Infilling} \label{subsec:infill}
A small fraction ($\sim1\%$) of our pixels have too few stars for a reliable line-of-sight fit ($< 10$ stars). These pixels occur most frequently at the edges of the DECaPS2 imaging footprint or for lines of sight with sufficient extinction to significantly suppress detection of stars behind the dust. One could simply excise such pixels from the map or fill them with an average of neighboring pixels, but both of these options are unsatisfying.  We do have \emph{some} information about the dust extinction in these pixels, based on their neighbors and the spatial correlation of dust density. We can use the extinction in neighboring pixels to make an informed prediction that gives a mean and variance for the extinction in the missing pixels, which is often called ``inpainting'' or ``infilling.'' While principled infilling is better than returning no value for the map at those locations, the infilled values will be biased low because we include no prior to account for the fact that the stars/pixels we are missing are preferentially more extinguished.  Such pixels are flagged in the bitmask, so the user can decide whether to ignore them, or use them as a ``best guess.''

We perform this infilling with a method akin to conditional Gaussian process regression, using correlations only within a single distance slice (i.e. the cumulative reddening integrated to that slice). However, because these infills are conditioned on the pixels that are \emph{not} missing, which \emph{are} correlated along the distance axis, the infills will be reasonably correlated along the distance axis even though we have not imposed a prior on correlations in that dimension. 

We adapt the \cite{Saydjari_2022} method, designed for Cartesian pixel images, to work on ring-ordered HEALPix images far from the poles. The restriction on ordering allows a simple translation operator to be defined, which is necessary to obtain local samples of images the same size as the region-of-interest. We use these samples to learn a completely data-driven model of the covariance of pixels in the region-of-interest. This translation operator is only valid at low latitude because HEALPix rings above $|b| \gtrapprox 41.8\degree$ have a decreasing number of pixels toward the poles.  Fortunately our map at $|b| < 10^\circ$ is well within the region of validity. 

First, we do a preliminary moving median infill, replacing missing pixels with a rough guess so that every sample the size of the region of interest can be used to build a model of the pixel-pixel correlations. The rough first pass uses the median of missing pixels in an 11 pixel-wide region, as long as more than 5\% of the pixels are not missing. This infilling is repeated seven times to completely fill larger holes in the map.

Then, we loop over pixels missing from the original map, infill all missing pixels in a region 21 pixels wide around a given missing pixel, and repeat until no missing pixels remain. We use samples of this 21-pixel-wide region that are translates of the original region within a larger 51-pixel-wide region to learn a local mean and pixel-pixel covariance matrix. This allows us to predict missing pixels in the region of interest conditioned on the non-missing pixels. The prediction gives a posterior on the missing pixels, so we obtain not only a mean, but also realistic samples of the missing pixels with noise properties and fluctuations consistent with the local non-missing pixels. We stably seed the random number generator used for the draws to ensure draw consistency between different distance bins. 

To ensure the self-consistency of the infill, the infills for subsequent missing pixels are conditioned on previously infilled pixels. This is performed independently for each infill draw (and the mean). Thus, while each infill is guaranteed to make use of the same pixels in their conditional prediction, the values of those pixels can differ if they were infilled in a previous step in the loop. This self-consistency comes at the cost of introducing a (slight) dependence on the infill order. Given this dependence, we make the infills at least deterministic by infilling pixels in order of their (great-circle) distance from the approximate center of our survey footprint ($\ell = 301\degree, b = 0\degree$). This iterative infill, conditioned on past infills could extend arbitrarily far from the survey edge, but we limit it by requiring all pixels in the training samples are not missing after the preliminary (moving median) infill, which sets the boundary implicitly.

We provide the mean infill for the mean map and a random draw of the infill for the samples of the map, alongside the code used to generate the infill, in \S \ref{sec:data_avail}. Pixels that were infilled are identified in the associated bitmask.

\section{Results} \label{sec:results}
This work presents two main data products: a star catalog of distance, reddening, and other parameters of 709 million stars, and the 3-D dust map derived from them. 
\begin{figure}[!h]
\begin{center}
\includegraphics[width=0.5\textwidth]{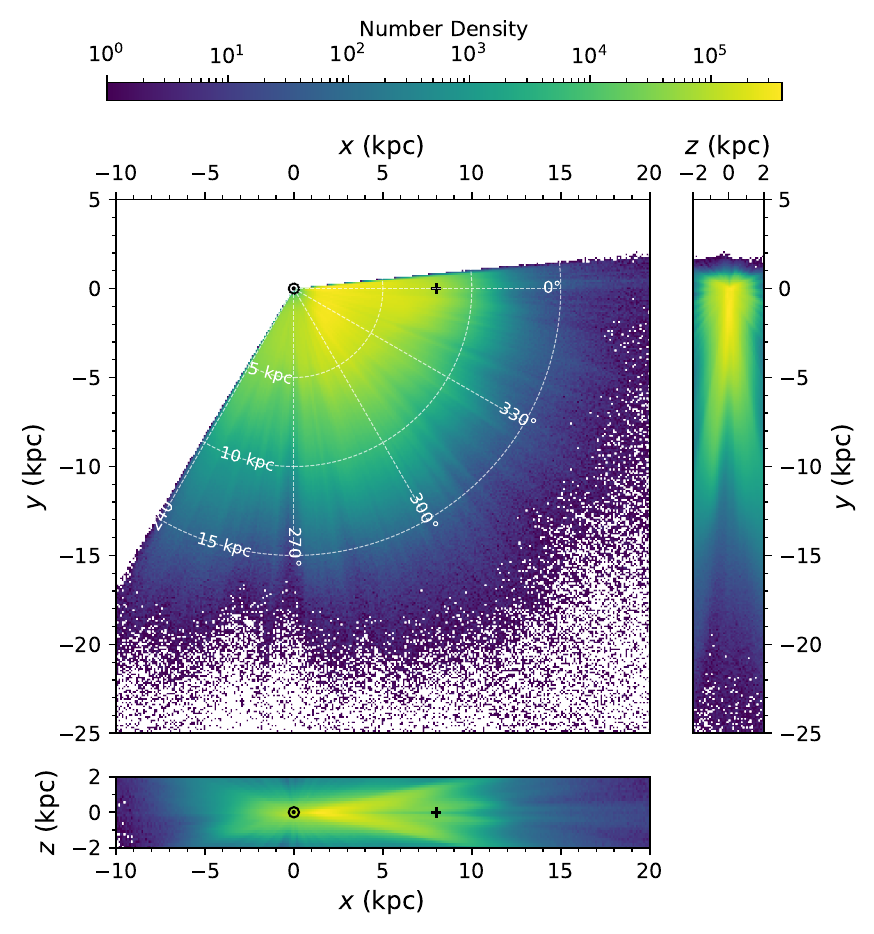}
\end{center}
\caption{Spatial distribution of stars over the full catalog. Each panel shows a projection of the stellar density in Heliocentric Galactic Cartesian coordinates. The Galactic center is marked with a $\mathbf{\plus}$ symbol and the Sun with a $\mathbf{\odot}$ symbol in the XY and XZ projections. We detect stars out to distances of $d=15$ kpc and beyond, but with an underdensity of stars near the midplane ($z=0$ pc) due to high levels of dust extinction.} 
\label{fig:spatial_distribution}
\end{figure}

\begin{figure*}
\begin{center}
\includegraphics[width=\textwidth]{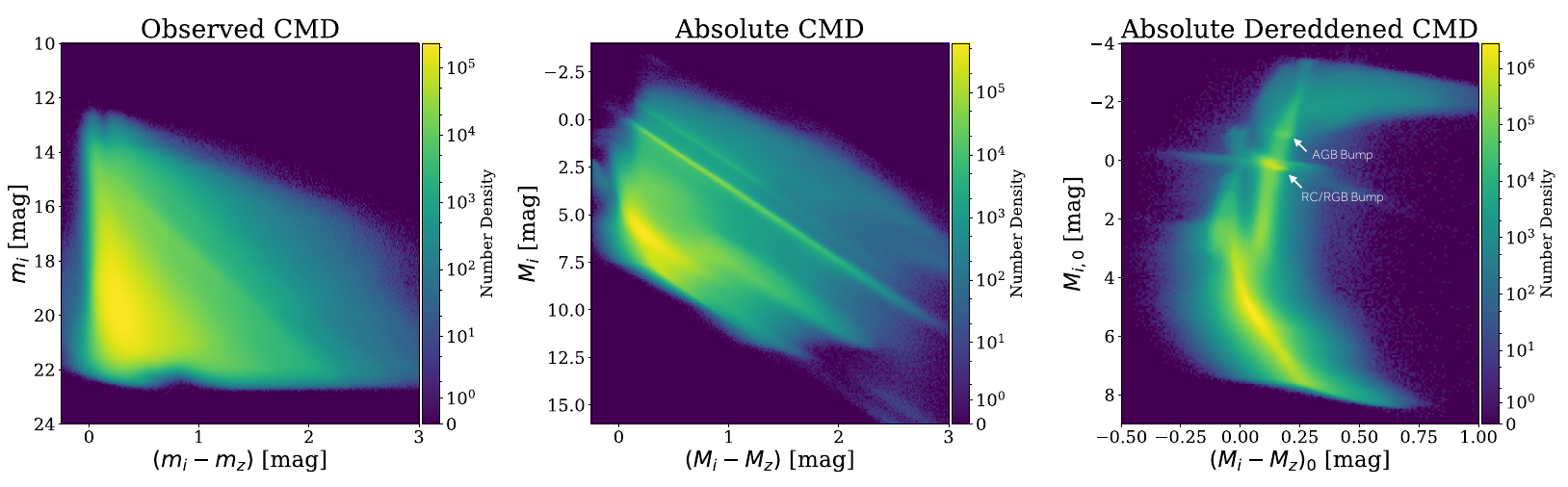}
\end{center}
\caption{Comparison between the observed CMD (apparent magnitude versus apparent color), the absolute CMD (absolute magnitude vs. absolute color) and the absolute, dereddened CMD (dereddened absolute magnitude versus intrinsic color) for DECaPS $i$ vs $i-z$. We compute the CMDs over the full high-quality stellar sample that have detections in both $i$ and $z$ bands, totaling roughly 670 million stars. In the absolute, dereddened CMD, we see clear overdensities corresponding to the asymptotic giant branch (the ``AGB bump") and the red clump and red giant branches (the ``RC/RGB bump"), highlighting the strength of our modeling in the post-main sequence regime.} 
\label{fig:CMD}
\end{figure*}

\subsection{Stellar Catalog}
We release the stellar catalog of distance, extinction, total-to-selective extinction ratio ($R_v$), and stellar type, including the median, $1\sigma$, and $2\sigma$ errors (expressed by the 2.5th, 16th, 50th, 84th, 97.5th percentiles of the posterior samples). We also release five random samples from the posterior. Our catalog includes all 709 million high-quality sources used in the line-of-sight dust reconstruction, selected from the 793 million stars over which we performed the stellar inference. The table (\texttt{decaps\_dr2.stellar\_inference}) is also available to query via NOIRLab's AstroDataLab\footnote{\url{https://datalab.noirlab.edu/data-explorer?showTable=decaps\_dr2.stellar\_inference}} and accessible via TAP-accessible clients, including the \texttt{astroquery} Python package. 

In Figure \ref{fig:spatial_distribution}, we show XY, XZ, and YZ projections of the spatial distribution of stars in the catalog in a Heliocentric Galactic Cartesian frame. The underdensity of stars along the $z=0 \; \rm pc$ axis is due to the high levels of dust extinction in the midplane, while the dearth of stars in the immediate solar vicinity ($d \lesssim 1 \; \rm kpc$) is due to both the latitude limits of the survey and the M-dwarf cut we impose, given the unreliability of the {\mist} models in the low-mass regime. We also detect a noticeable drop-off in stellar density around $l \approx 300 ^\circ$, which aligns with the edge of the VVV footprint. 

In Figure \ref{fig:CMD}, we show a comparison between an observed color magnitude diagram (CMD), an absolute CMD (corrected for distance), and an absolute, dereddened CMD (corrected for distance and extinction) in the DECaPS bands $i$ versus $i - z$. We compute the CMD for high-quality stars that have detections in both $i$ and $z$ bands, totaling roughly 670 million stars. We derive the dereddened CMD using the median (50th percentile) of the distances and extinction samples. As expected, we see a pronounced tightening of the CMD around the main sequence and giant branches, with noticeable overdensities corresponding to a blended red clump (RC) and red giant branch (RGB) bump, as well as a bump in the asymptotic giant branch (AGB), as marked in Figure \ref{fig:CMD}. 

\begin{figure}[!h]
\begin{center}
\includegraphics[width=0.5\textwidth]{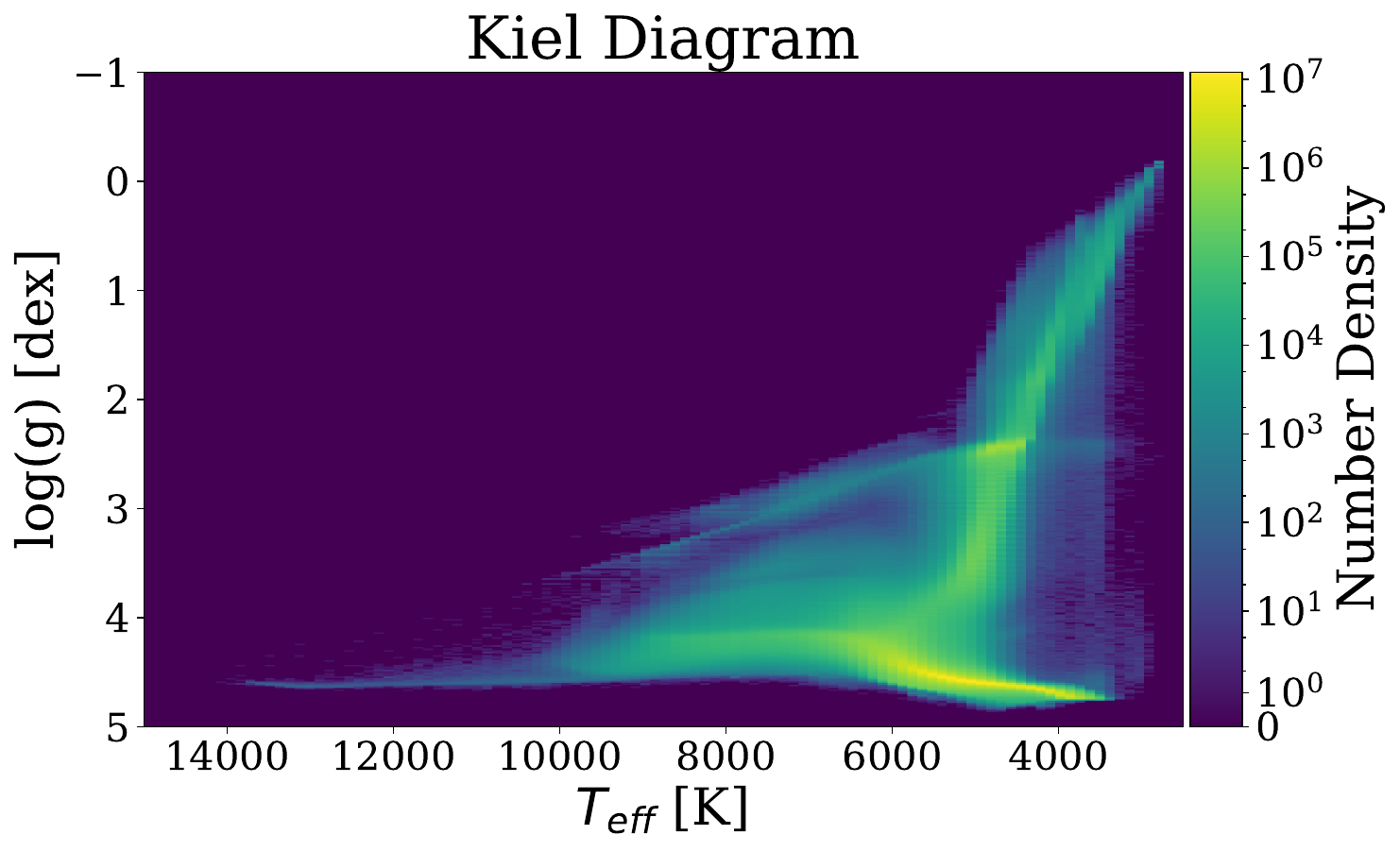}
\end{center}
\caption{Kiel Diagram, showing the distribution of two of the inferred stellar parameters, surface gravity, $\log(g)$, as a function of their effective temperature, $T_{\rm eff}$. We compute the Kiel Diagram across the full sample used in the line-of-sight dust reconstruction, totaling 709 million stars.} 
\label{fig:kiel}
\end{figure}

In Figure \ref{fig:kiel} we show a Kiel Diagram, which displays the surface gravity, $\log(g)$, versus effective temperature, $T_{\rm eff}$. Like the CMDs, we compute the Kiel Diagram over the full catalog, and take the median of the surface gravity and effective temperature samples. As expected, we see a clear bifurcation in the Kiel diagram between the main sequence ($\log(g) \lesssim 4$) and the post-main sequence ($\log(g) \gtrsim$ 4). Roughly 12\% of the sample (89 million stars) has evolved off the main sequence, corresponding to $\rm EEP > 454$ in the context of the MIST models. This large fraction of bright, midplane giants is critical for constraining the 3D distribution of dust at large distances, as we will show in \S \ref{sec:dustresults}. Beyond the Kiel diagram shown in Figure \ref{fig:kiel}, we are also able to broadly reproduce general trends in the age-metallicity relation. However, our inferred ages and metallicities are highly prior dominated and affected by our choice of binning when constructing the model grid described in \S \ref{subsec:perstar_inference}. Therefore, unlike our estimates of distance and extinction, we caution future users to not overinterpret the stellar type parameters given that a majority of stars are fit with only five bands of photometry (see Figure \ref{fig:ndet_per_star}). 

Finally, in Appendix \S \ref{sec:gaia_G}, we show that we can recover the Gaia \textit{G} band photometry. We infer the intrinsic type, distances, and extinction of stars based on other broadband photometry from a combination of DECaPS2, VVV, 2MASS, and unWISE and do not incorporate Gaia photometry (only the astrometry) into our model. However, using these inferred stellar parameters, we can predict the G band photometry and compare this prediction to the observed G band photometry for stars detected in Gaia. While we find overall good agreement across all signal-to-noise regimes, we systematically underpredict the Gaia G band magnitudes by $\approx 0.05$ mag, which we discuss more in Appendix \S \ref{sec:gaia_G}.

\subsection{3D Dust Data Product} \label{sec:dustresults}
We produce a 3D dust map with an angular resolution of FWHM=$1\arcmin$ (HEALPix $N_{side} = 8192$ pixels of $0.43'$) sampled in 120 logarithmic distance bins evenly-spaced between 63 pc and 63 kpc. Our map is probabilistic, and we generate 100 samples of the line-of-sight reddening (see \S \ref{subsec:bayestarlos}). Our primary data product is the mean line-of-sight reddening in each distance bin. Our map is released in units of $E(B-V)$ in magnitudes. $E(B-V)$ is derived from the underlying stellar inference on $A_V$ and $R_V$, where $E(B-V) = \frac{A_V}{R_V}$, with a mean of $R_V = 3.32$ based on our prior on the variation in the extinction curve from \citet{Schlafly_2016}. For each pixel of the map, we provide an estimate of the minimum and maximum reliable distance, which we discuss more in \S \ref{sec:reliable}. Given that we sample over 51 million HEALPix pixels on the plane of the sky and 120 distance bins along the line of sight, our primary data product constitutes the reddening inferred in over 6.1 billion voxels. 

Given our high angular resolution, recall that some pixels lack the minimum number of stars ($\geq10$ stars) requisite to perform the line-of-sight inference and these pixels were infilled (at each distance slice) following the procedure described in \S \ref{subsec:infill}. We flag these pixels and also release a corresponding HEALPix bitmask of the infilled pixels. 

The map can be downloaded at the Harvard Dataverse (see \href{https://doi.org/10.7910/DVN/J9JCKO}{doi:10.7910/DVN/J9JCKO}). The map can also be easily accessed via the \texttt{dustmaps} Python package \citep{dustmaps}, which provides a standard interface to query this map --- as well as a host of other 3D and 2D reddening maps --- over a user-defined set of sky coordinates and distances. 

\begin{figure*}
\begin{center}
\includegraphics[width=0.75\textwidth]{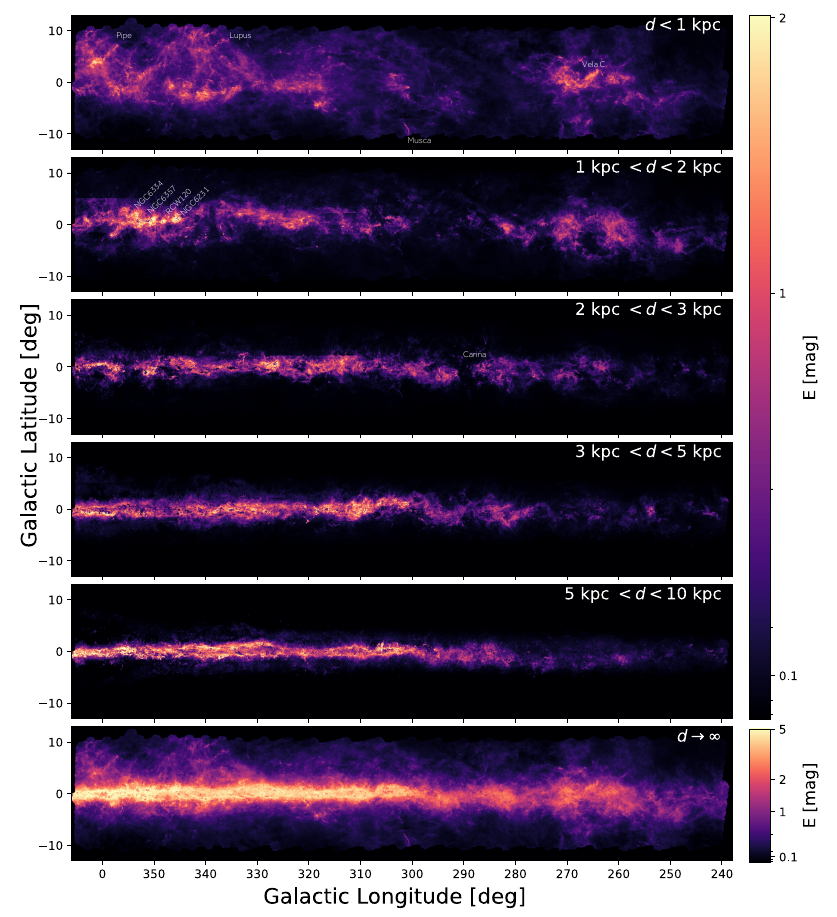}
\end{center}
\caption{Plane of sky reddening in $E_{B-V}$ of the DECaPS 3D dust map, integrated over discrete distance intervals.} 
\label{fig:integrated_sky}
\end{figure*}

In Figure \ref{fig:integrated_sky}, we show the plane of sky reddening integrated over discrete distance ranges ($d < 1 \; \rm kpc$, $1 \; {\rm kpc} < d < 2 \; \rm kpc$, $2 \; {\rm kpc} < d < 3 \; \rm kpc$, $3 \; {\rm kpc} < d < 5 \; \rm kpc$, $5 \; {\rm kpc} < d < 10 \; \rm kpc$, and $d \rightarrow \infty$). We see a wealth of structure over the distance range probed by our map. Within 1 kpc we detect nearby molecular clouds including Vela C, the Pipe Nebula, and Lupus. Beyond 1 kpc, we detect massive star forming regions like NGC6334 and RCW120. We also detect a diversity of feedback-driven bubbles, including several $\rm H{I}$ shells cataloged by \citet{Ehlerov_2013}. 

Like every data product, our map contains artifacts that users should be aware of, many of which only occur in specific regions of the sky and/or manifest at specific distances. These artifacts may be less obvious to the reader, since our maps are not produced the same way typical astronomical images are produced. We show and discuss these artifacts in more detail in Appendix \S \ref{sec:artifacts}, but note that the boundary of the VVV survey is visible near $b=\pm 2^\circ$ primarily between $d = 2-3$ kpc across part of the fourth quadrant in Figure \ref{fig:integrated_sky}. This artifact stems from the fact that VVV goes three magnitudes deeper in $J$,$H$, and $K$ compared to 2MASS. Since highly reddened stars needed to be detected behind clouds for the cloud to be detected with 3D dust mapping, the vastly different depths translate to some dense clouds being detected in VVV within $|b| < 2^\circ$ that are largely undetected and/or placed at a different distance at higher latitudes where only 2MASS data existed at the time of the map's generation.\footnote{After generating our 3D dust map, the VVVx survey \citep{Saito_2024} has become publicly available, which extends the latitude coverage of the original VVV survey that we utilize here. Utilizing the VVVx survey in future 3D dust maps likely shift this artifact to higher latitudes.}

\begin{figure*}[]
\begin{center}
\includegraphics[width=0.75\textwidth]{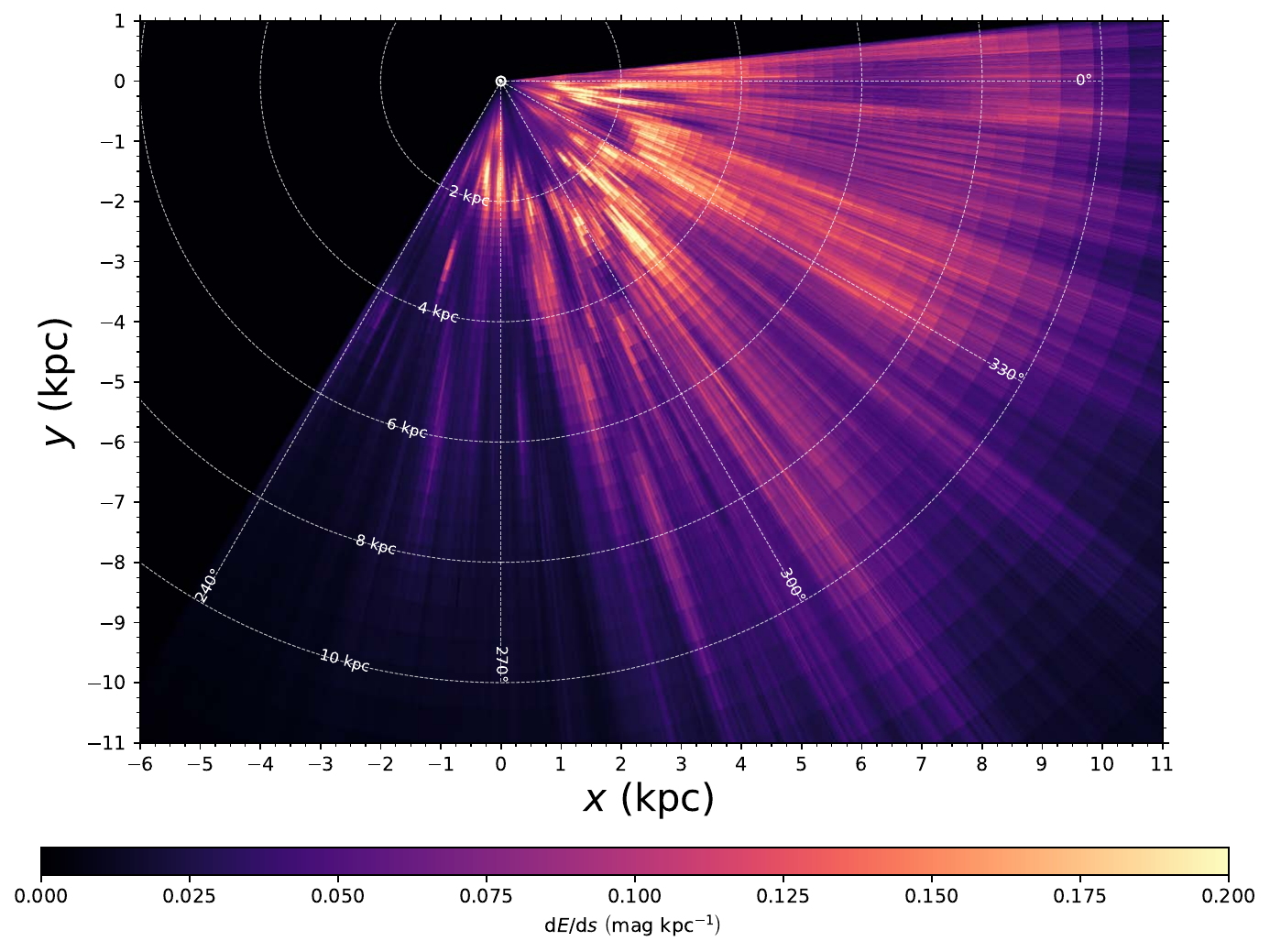}
\end{center}
\caption{Top-down view of DECaPS 3D dust map, integrated over $z=\pm 300 \; \rm pc$. The Sun is marked with a $\odot$ symbol at (x,y) = (0,0) and the Galactic center is to the right.} 
\label{fig:birdseye}
\end{figure*}

In Figure \ref{fig:birdseye}, we show a bird's-eye view of the southern Galactic plane, integrated over $z = \pm 300$ pc. Despite evidence of fingers of god --- owing to our angular resolution being finer than our distance resolution --- we detect a significant amount of dense structure in the fourth quadrant between $d=2-4$ kpc previously unresolved in purely Gaia-based maps. Particularly at $l< 300^\circ$ we also detect discrete dust complexes between $d = 5 - 10$ kpc. We observe several kpc-sized voids in this top-down view, akin to those seen in nearby face-on galaxies with JWST \citep{Lee_2022}. 

\section{Discussion} \label{sec:discussion}
Here we highlight the two greatest strengths of the map in the context of existing approaches: its high-angular resolution (\S \ref{subsec:herschel}) and its depth (\S \ref{subsec:depth}). In \S \ref{subsec:depth}, we also combine the DECaPS 3D dust map with the complementary Bayestar19 3D dust map in the northern sky to provide complete coverage of the plane $|b| < 10^\circ$. The combination of the map's angular resolution and its ability to resolve structure at large distances provides an important proof of concept for the future of 3D dust mapping in the era of LSST and Roman (see \S \ref{subsec:future}). 

\begin{figure*}[h!]
\begin{center}
\includegraphics[width=1.0\textwidth]{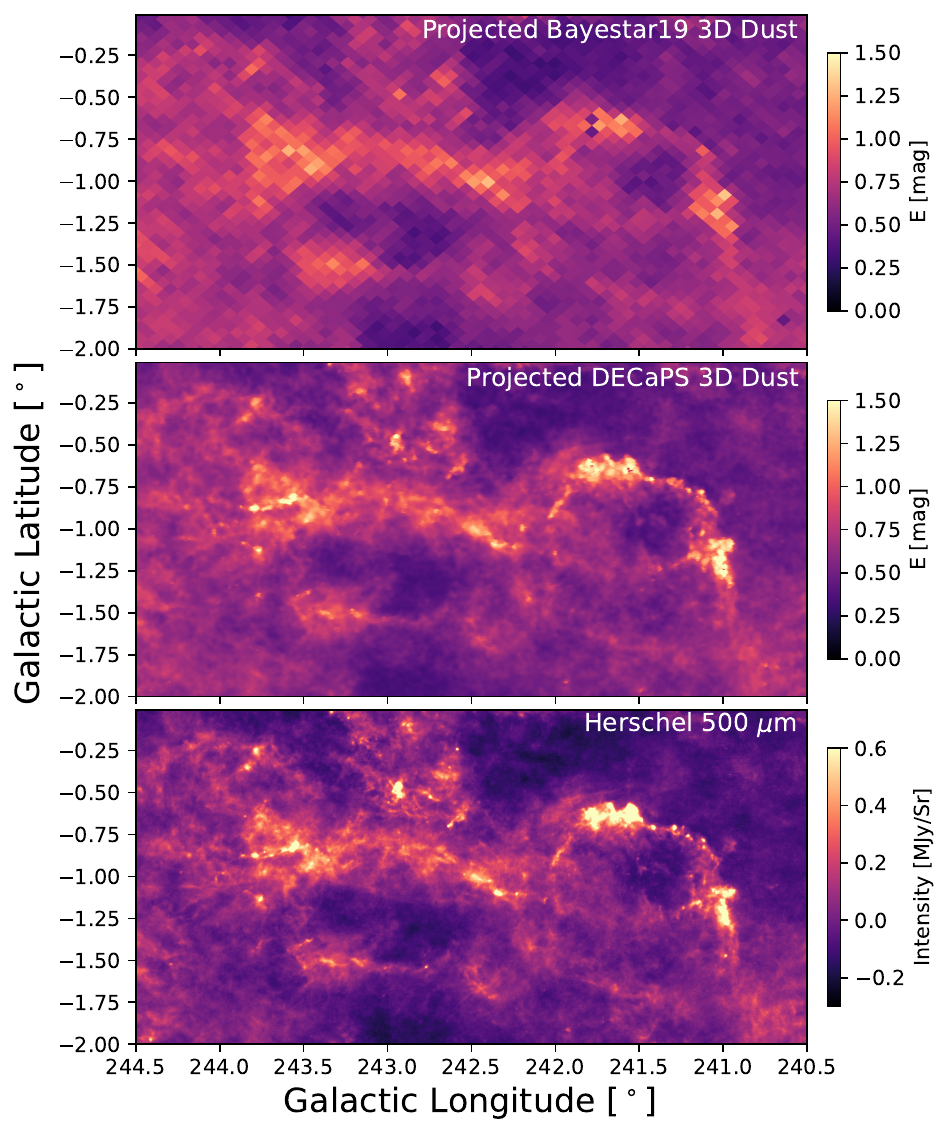}
\end{center}
\caption{Plane of sky comparison of a dense filament in the Galactic plane. The top two panels show the projected Bayestar19 3D dust map and the DECaPS 3D dust reddening maps, integrated out to a distance of 7 kpc. The bottom panel shows the 2D Herschel 500 $\mu m$ emission map, which is by definition integrated to infinity.} 
\label{fig:herschel_comp}
\end{figure*}

\subsection{Herschel-Resolution 3D Dust Mapping} \label{subsec:herschel}

As detailed in \S \ref{subsec:pixelization}, the angular resolution of 3D dust maps is directly tied to the density of stars on the plane of the sky and the chosen pixelization. Thanks to the increase in plane-of-sky stellar density made possible by the DECaPS2 survey, the data support a pixelization at HEALPix $N_{\rm side}=8192$, with an angular resolution defined by a $1'$ Gaussian PSF ($2.5\times$ the $0.43'$ pixel scale at HEALPix $N_{\rm side}=8192$) across the full footprint. 

In Figure \ref{fig:herschel_comp}, we show a cutout of our $1\arcmin$ resolution map alongside a cutout from the \texttt{Bayestar19} 3D dust map \citep{Green_2019} and the Herschel far-infrared emission map at $\rm 500 \; \mu m$ \citep{Molinari_2016} towards a dense filament in the Galactic plane centered at $(l,b) = (242.5^\circ, -1^\circ)$. Both the Bayestar19 3D dust map and the DECaPS 3D dust map have been integrated out to a distance of 7 kpc, roughly the maximum reliable distance of the DECaPS map towards the filamentary dust structure in this region of the sky. While the angular resolution varies across the Bayestar19 3D dust map, \citet{Green_2019} estimate a typical resolution of 6.8' (see their \S 2.3.1). For comparison, the Herschel emission map has an angular resolution of $37\arcsec$ at $500 \; \mu \rm m$. 

Figure \ref{fig:herschel_comp} clearly highlights the strength of our map's angular resolution and sensitivity: we are able to capture a majority of the rich filamentary structure seen with Herschel in dense regions. Our map constitutes the highest angular resolution 3D dust map available today, achieving $7\times$ higher angular resolution than Bayestar19 and an angular resolution within a factor of two of the Herschel $500 \mu \rm m$ emission maps. However, unlike Herschel, we have distance information and can place the filament in 3D within its broader Galactic environment given the significant depth of our 3D dust map, as discussed in \S \ref{subsec:depth}. While Figure \ref{fig:herschel_comp} demonstates the strength of our map, we also caution that compared to 2D approaches, we still underpredict the total amount of reddening in very dense regions, particularly toward the Galactic center. In Appendix \S \ref{sec:artifacts}, we discuss the limitation of our map towards the Galactic center, by comparing our projected 3D dust map with the 2D reddening map of \citet{Surot_2020}, which relies solely on infrared VVV photometry. 

\subsection{Depth} \label{subsec:depth}
Alongside possessing an angular resolution comparable to the Herschel $500 \; \mu m$ emission maps, our approach resolves structure substantially deeper into the Galactic plane than most existing 3D maps. Unlike the angular resolution of 3D dust maps, which is tied to the density of stars on the plane of the sky (\S \ref{subsec:herschel}), the depth of 3D dust maps depends on the ability to resolve a sufficient number of highly reddened stars at large distances. If reddened stars are not detected background to a dense cloud, that cloud will not appear in the resulting map, even if the cloud nominally lies within the distance range targeted in the reconstruction. In contrast to most existing approaches, 76\% of our stellar sample (541 million stars) lies at $d>3$ kpc, 43\% (303 million stars) at $d > 5 \; \rm kpc$, and 17\% (119 million stars) at $d > 7\; \rm kpc$ underlining the ability of our map to resolve structure significantly beyond the solar neighborhood.  

We can characterize the depth of our maps in the context of two regimes: those that require Gaia parallax measurements and those that do not. In contrast to Gaia-reliant approaches, our inference pipeline is flexible enough to infer stellar distances in the absence of parallaxes, allowing us to probe much larger volumes of the Galaxy. And in comparison to maps that also do not depend on Gaia, we incorporate deeper photometry and more sophisticated modeling of highly reddened post-main sequence stars, allowing us to penetrate further into the midplane. Our pipeline's flexibility to incorporate and model deeper photometry in the absence of Gaia translates to a data product capable of resolving dense clouds up to a depth of roughly $d=10$ kpc. However, the exact depth varies across the sky, particularly as a function of latitude, and is estimated in part by our maximum reliable distance calculation (see Appendix \S \ref{sec:reliable}). 

\begin{figure}
\includegraphics[width=0.5\textwidth]{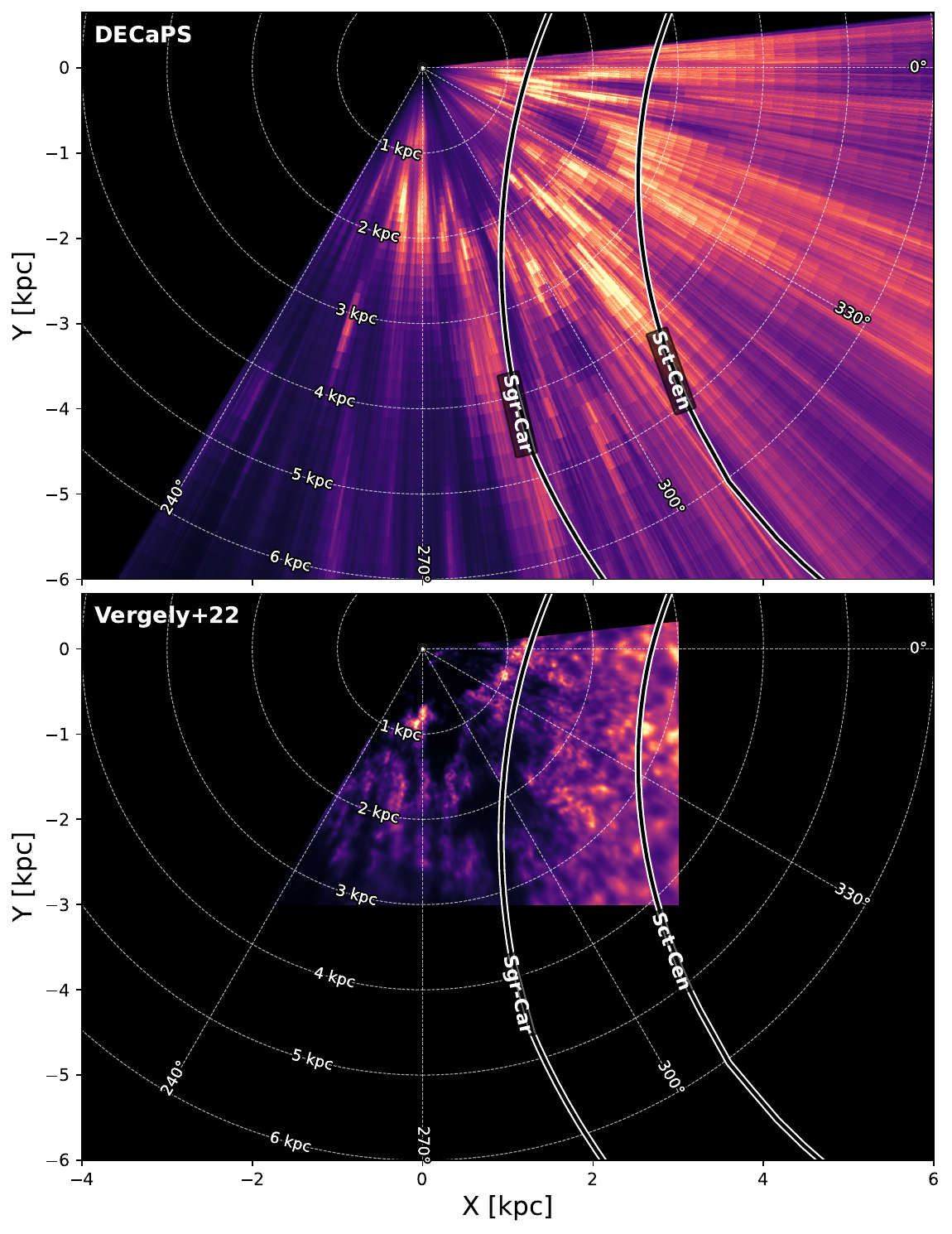}
\caption{Top-down view of DECaPS 3D dust map (top) alongside the \citet{Vergely_2022} map (bottom, restricted to the DECaPS footprint), each integrated over $z=\pm 400 \; \rm pc$. The \citet{Vergely_2022} map has been restricted to $239^\circ < l < 6 ^\circ$ to facilitate comparison with DECaPS over the same longitude range. The Sun is marked with a $\odot$ symbol at (x,y) = (0,0). Both are displayed on the same relative extinction scale and saturate at the 99.95\% percentile of the respective reconstruction. Spiral arm models from \citet{Reid_2019} for the Sagittarius-Carina (Sgr-Car) and Scutum-Centaurus (Sct-Cen) arms are overlaid. Within $d \lesssim 5$ kpc, we find that a majority of the massive cloud complexes lie in the interarm region, deviating from log-spiral fits in this region of the Galaxy. Click \href{https://faun.rc.fas.harvard.edu/czucker/Paper\_Figures/DECaPS\_Vergely\_Comparison.html}{here} for an interactive version of this figure that allows you to flash back/forth between the panels.} 
\label{fig:vergely_comp}
\end{figure}

\subsubsection{Comparisons within the Solar Neighborhood} \label{subsec:vergely}
Most modern 3D dust maps \citep{Lallement_2019,Vergely_2022,Chen_2019,Dharmawardena_2024} rely on higher-quality Gaia parallax measurements to constrain stellar distances, a key ingredient for 3D dust mapping alongside stellar extinction estimates. As an optical instrument, Gaia has a limiting magnitude of $G\approx21$ mag for stars with a full astrometric solution \citep{GaiaDR3Cat}. Maps that require Gaia detections typically lose a sufficient number of background sources at $d \gtrapprox  2\; \rm kpc$ in the inner Galaxy due to the high levels of dust extinction. Therefore most Gaia-based maps focus on the solar neighborhood \citep[$d<2-3$ kpc;][]{Edenhofer_2023, Dharmawardena_2024, Lallement_2019, Leike_2020} and caution that the detection of structure is inhomogeneous and sparsely sampled at the edges of their maps.

\begin{figure}[]
\begin{center}
\includegraphics[width=0.5\textwidth]{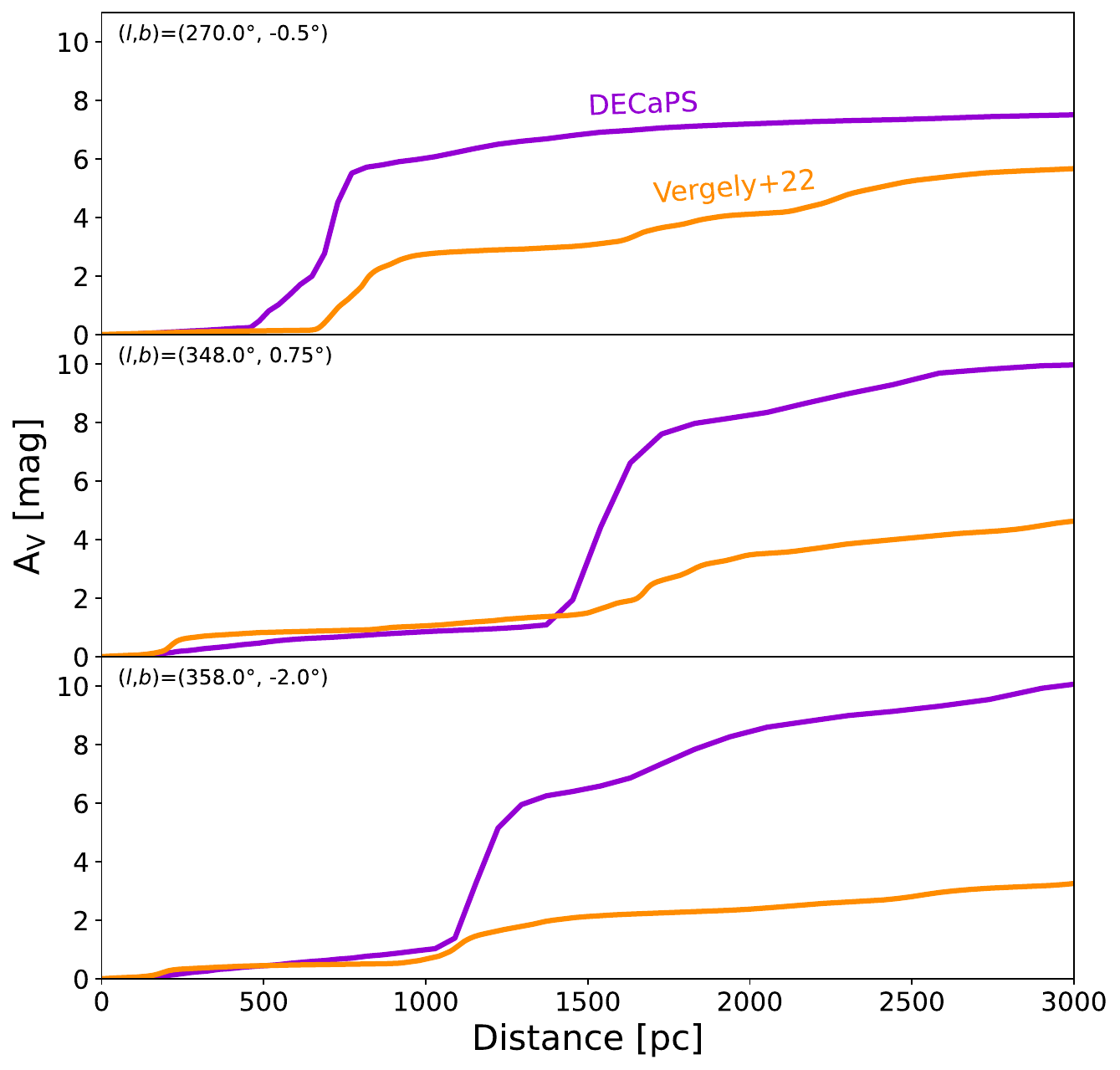}
\end{center}
\caption{Comparison of the cumulative extinction as a function of distance between the \citet{Vergely_2022} map (orange profiles) and the DECaPS map (purple profiles) towards three lines of sight in the Galactic plane. The DECaPS map recovers significantly more extinction per cloud (particularly beyond 1 kpc) compared to \citet{Vergely_2022}.} 
\label{fig:extinction_profiles}
\end{figure}

In Figure \ref{fig:vergely_comp}, we show a comparison between the Gaia-based 3D dust map from \citet{Vergely_2022} (extending out $\pm 3$ kpc in $\rm X$ and $\rm Y$ with a Cartesian resolution of 10 pc) alongside the DECaPS 3D dust map. In comparison to the \citet{Vergely_2022} map shown in Figure \ref{fig:vergely_comp}, our pipeline does \textit{not} require a Gaia parallax, and 46\% of stars in our sample (366 million stars) are not detected in Gaia and modeled using photometry alone.  Thanks to the flexibility of our pipeline to model stars in the absence of parallaxes, Figure \ref{fig:vergely_comp} illustrates the vastly different scale probed by the DECaPS map. Even focusing on the region of overlap, Figure \ref{fig:vergely_comp}  shows that we also detect more clouds between $d=2-3 \; \rm kpc$ and recover more extinction per cloud, indicating that we are detecting more highly reddened stars background to each cloud complex. We further emphasize this point in Figure \ref{fig:extinction_profiles}, where we compare the extinction profiles (the cumulative extinction as a function of distance) between \citet{Vergely_2022} and our DECaPS map along three lines of sight in the Galactic plane out to a distance of 3 kpc from the Sun. While the maps largely agree at low extinction within $d \lesssim 1\; \rm kpc$ and $A_V \lesssim 2 \; \rm mag$, we find significant differences at larger distances, with DECaPS detecting a factor of a few times more extinction per cloud compared to \citet{Vergely_2022}.\footnote{The \citet{Vergely_2022} map is provided in units of $A_{550 \rm nm}$, equivalent to extinction in the $V$ band. We query the \citet{Vergely_2022} extinction profiles using the \texttt{G-TOMO} tool at \href{https://explore-platform.eu}{https://explore-platform.eu}. For this comparison, we convert our DECaPS 3D dust map from $E(B-V)$ to $A_V$ assuming an $R_V = 3.32$.}

Therefore, while Gaia-based solar neighborhood 3D dust maps achieve a higher distance resolution in the nearest 2 kpc, the DECaPS map recovers more dense clouds in the southern Galactic plane at $d \gtrsim 2  \rm \; kpc$, where most of the star formation is occurring along the Sagittarius-Carina and Scutum-Centaurus arms \citep[see e.g. the distribution of YSOs in the southern Galactic plane from][]{Kuhn_2021}. 

In Figure \ref{fig:vergely_comp} we overlay the traces for the Sagittarius-Carina and Scutum-Centaurus arms, constrained by maser parallax measurements towards high-mass star-forming regions from \citet{Reid_2019}. We find that while we recover large cloud complexes at the correct \textit{range} of distances, most of the dust lies in the \textit{interarm} region (between the near Sagittarius-Carina and near Scutum-Centaurus arms, at distances $d \lesssim 4$ kpc) as defined by \citet{Reid_2019}. Given the dearth of maser measurements in the fourth quadrant, we anticipate substantial revision may be needed to spiral arm models in the fourth quadrant, which can be informed by a combination of this 3D dust map, future maser parallax measurements, and complementary spatial-dynamical constraints on the Milky Way's ISM \citep[e.g. Diffuse Intersteller Bands (DIBS);][]{Saydjari_2023_DIBs, KTII} in the southern Galactic plane. 

\begin{figure*}[]
\begin{center}
\includegraphics[width=1.\textwidth]{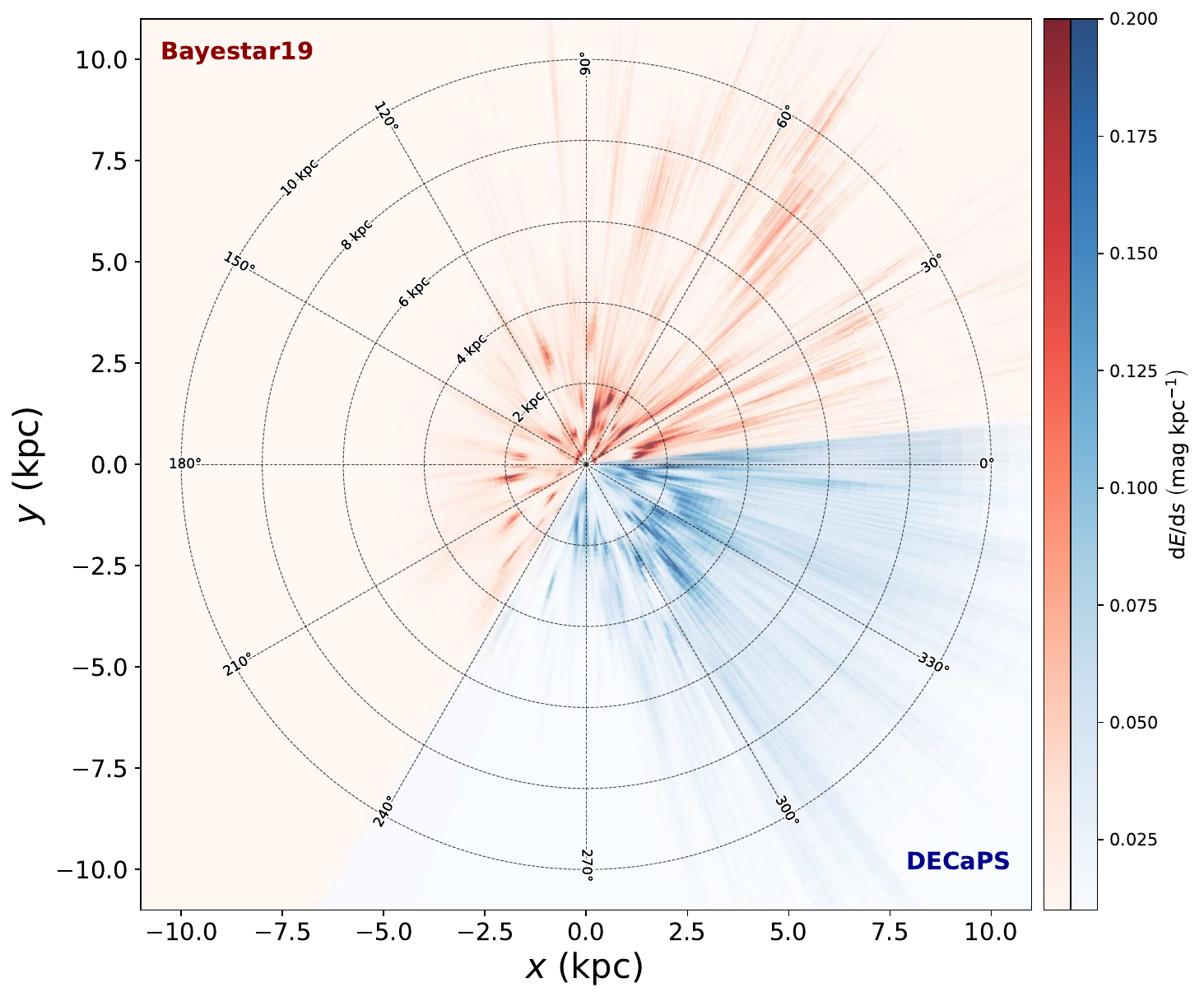}
\end{center}
\caption{Top down view of the Bayestar19 3D dust map \citep[red][]{Green_2019} combined with this DECaPS 3D Dust map (blue), both integrated over $|z| < 100$ pc and limited to $|b| < 10^\circ$.} 
\label{fig:full_plane_comp}
\end{figure*}

\subsubsection{Comparisons beyond the Solar Neighborhood}
As established in \S \ref{subsec:vergely}, maps that do not depend on Gaia to infer stellar distances are able to probe much larger volumes of the Galaxy. Several such 3D dust maps currently exist in the literature, and we focus here on comparisons with \citet{Green_2019} and \citet{Marshall_2024}, the latter of which is an update to the original 3D dust mapping work from \citet{Marshall_2006}. \citet{Rezaei_2024} also reconstructs the distribution of dust out to $d= 10 \; \rm kpc$. However, we forgo detailed comparisons with \citet{Rezaei_2024} since they rely on APOGEE spectroscopy for their 3D dust reconstruction. APOGEE DR16 is so sparsely sampled in the southern Galactic plane that no stars are incorporated in much of the fourth quadrant (see their Figure 2) and only 44,000 stars are included in the sample over the remainder of the plane. 

The Bayestar19 3D dust map \citep{Green_2019} is a complementary map to the DECaPS map and reconstructs the 3D dust distribution over three-quarters of sky in the north, primarily based on Pan-STARRS1 photometry for 799 million stars. While \citet{Green_2019} use an empirical set of stellar models rather than the theoretical MIST models we adopt here, the procedure for determining stellar distances is similar to the methodology outlined in \S \ref{subsec:perstar_inference} --- measured parallaxes and their corresponding uncertainties are incorporated into the per-star modeling as an additional likelihood where available; otherwise, stellar posteriors on distance and extinction are inferred using a minimum of four photometric passbands, as in this work. In Figure \ref{fig:full_plane_comp}, we combine the Bayestar19 3D dust map with the DECaPS 3D dust map to provide complete $360^\circ$ coverage of the Galactic plane. The combination of the two maps provides complete coverage of the plane $|b| < 10^\circ$. In Figure \ref{fig:full_plane_comp}, we show a top down view of the plane  integrated over $|z| < 100 \; \rm pc$ to highlight differences in the maps near the midplane.\footnote{We mask out the Bayestar19 3D dust map at latitudes $|b| > 10^\circ$ before integrating over $z$ to facilitate a one-to-one comparison to the DECaPS map, which only spans $|b| < 10^\circ$.} Close to $z=0 \; \rm pc$ we find that the DECaPS map detects more dense clouds at distances between $d=2-6 \; \rm kpc$ compared to Bayestar19. 

We attribute the greater depth of the DECaPS map to two causes. First, we incorporate significantly deeper infrared photometric data, allowing us to detect more stars background to dense clouds at larger distances. \citet{Green_2019} combine PS1 photometry with 2MASS photometry, but due to its shallow depth, only 10-20\% of PS1 sources also have 2MASS detections. While we also utilize 2MASS data, we primarily rely on VVV photometry in the infrared near the midplane, which is three magnitudes deeper than 2MASS and provides complementary detections to DECaPS for a majority of sources in the VVV footprint. In addition to the incorporation of deeper infrared data, we also employ more sophisticated modeling of post-main sequence stars, allowing us to detect highly reddened giants at larger distances. \citet{Green_2019} derive empirical stellar templates based on observations of uniformly old stellar populations towards globular clusters with SDSS and only approximate the morphology of the giant branch \citep[see discussion in][]{Green_2014}. In contrast, the MIST models encompasses a broader range of evolved evolutionary phases (up to the beginning of the thermally-pulsing ABG phase) and extend to higher initial masses \citep[for a comparison between the Bayestar19 and MIST post-main sequence models, see Figure 8 in][]{Speagle_2023b}

\begin{figure}[]
\begin{center}
\includegraphics[width=0.5\textwidth]{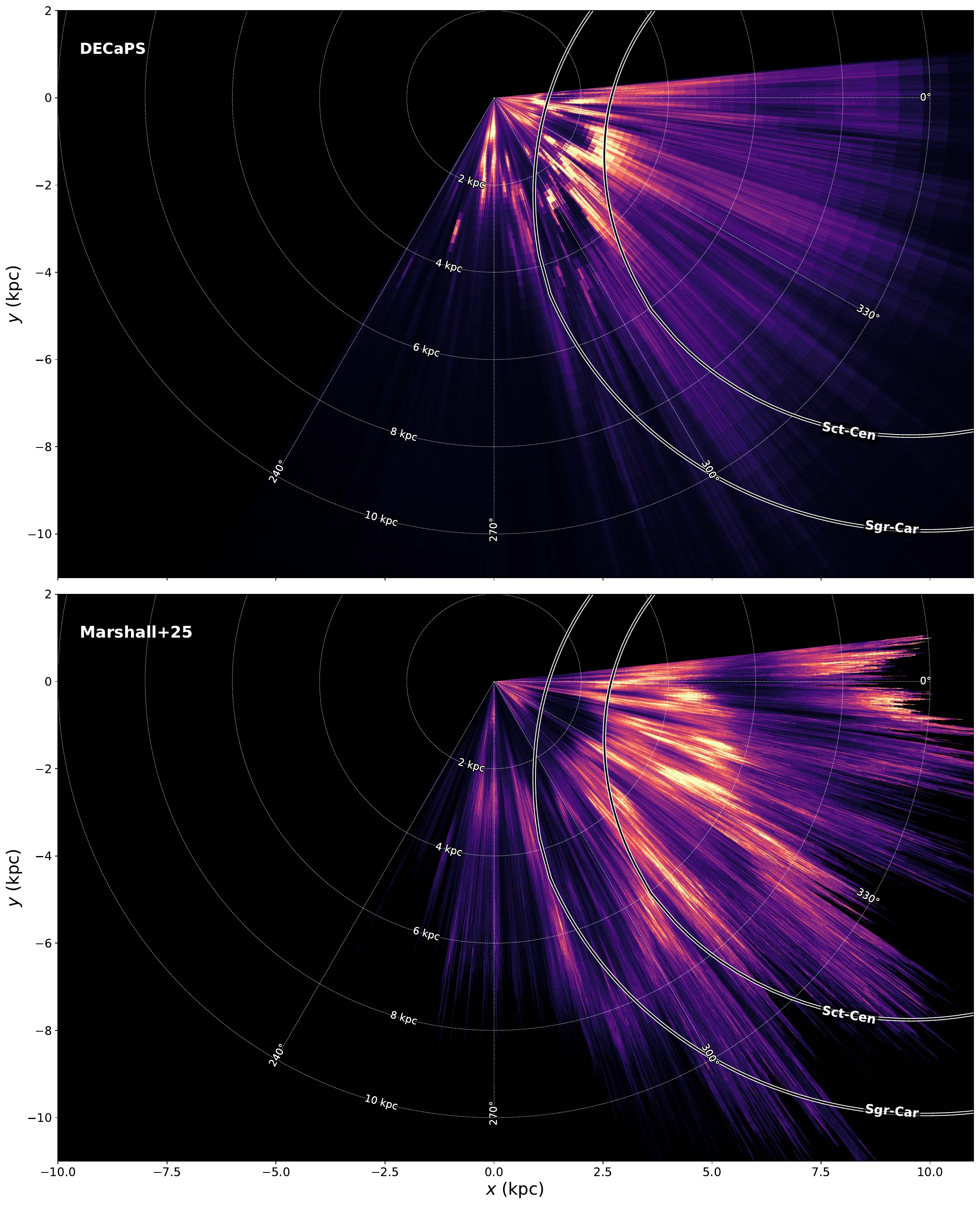}
\end{center}
\caption{Comparison of the DECaPS dust map (top panel) and the \citet{Marshall_2024} 3D dust map (bottom panel, restricted to the DECaPS footprint). To better highlight structures over a narrow latitude range, we show the average differential extinction, rather than the integral (as in Figure \ref{fig:vergely_comp}). Both are displayed on the same relative extinction scale and saturate at the 99.5\% percentile of the respective reconstruction. Arm models from \citet{Reid_2019} for the Sagittarius-Carina (Sgr-Car) and Scutum-Centaurus (Sct-Cen) arms are overlaid. Click \href{https://faun.rc.fas.harvard.edu/czucker/Paper\_Figures/DECaPS\_Marshall\_Comparison.html}{here} for an interactive version of this figure that allows you to flash back/forth between the panels.} 
\label{fig:marshall_comp}
\end{figure}

In Figure \ref{fig:marshall_comp}, we compare with the \citet{Marshall_2024} 3D dust map, which is the most similar to the DECaPS 3D dust map in terms of depth. \citet{Marshall_2024} adopts a 3D stellar distribution based on the Besançon Galaxy model \citep[see][]{Robin_2003,Czekaj_2014}. The Besançon model provides the intrinsic (unreddened) colors of simulated stars in the 2MASS bands,  which can be compared with the observed reddened 2MASS photometric colors to infer the near-infrared color excess along each line of sight.  Our map is qualitatively similar to the \citet{Marshall_2024} reconstruction, and we place several dense clouds in the third and fourth quadrant at similar distances. In some cases, we find the same features, but systematically offset in distance (see e.g. dense clouds between $l=240^\circ-270^\circ$), which may be the result of systematic uncertainties in the Besançon model used in the reconstruction of \citet{Marshall_2024}. As raised in \S \ref{subsec:vergely}, we argue that there is scant delineation between the dust associated with the Scutum-Centaurus and Sagittarius-Carina arms within $d \lesssim 5 \; \rm kpc$ in the DECaPS dust map. However, \citet{Marshall_2024} argues for more clearly defined arms and finds evidence of the Centaurus and Carina tangencies in the fourth quadrant, particularly on scales spanning  $10 \; \rm kpc$ in the bird's eye view in Figure \ref{fig:marshall_comp}, where \citet{Marshall_2024} is more sensitive to large-scale structure. While beyond the scope of this work, a more detailed comparison of the nature of spiral structure in the fourth quadrant is worthy of follow-up investigation. For now, see the interactive version of Figure \ref{fig:marshall_comp} to flash back and forth between the \citet{Marshall_2024} and DECaPS maps on the same grid.

\subsection{The Future Era of LSST and Roman} \label{subsec:future}
The future of 3D dust mapping lies not in the era of Gaia but in the era of LSST and Roman. The fourth and fifth data releases of Gaia circa 2026 and 2030 will herald even more precise astrometry, with parallaxes expected to improve by factors of $1.4\times$ and $1.9\times$, respectively \citep{Gaia_DR3}. However, future Gaia data releases will not substantially improve the limiting magnitude of the survey ($G\approx \rm 21\; mag$), so the number of DECaPS stars that have an accompanying parallax detection in Gaia DR4 and Gaia DR5 will not substantially increase. Therefore, studies that rely on Gaia alone will never be able to push significantly beyond the solar neighborhood, either now or in the future. 

Recall from \S \ref{subsec:depth} that 46\% of stars in our sample (366 million stars) do not have a Gaia parallax measurement: their stellar type, distance, and extinction are inferred using photometry alone. The future era of LSST and Roman will be an era of even deeper photometry than utilized here, and we need to be prepared to meet it. Expected to start science operations in 2025, LSST will target the southern Galactic plane between $\lambda = 320–1050 \rm nm$ in the $u$, $g$, $r$, $i$, $z$, and $y$ bands. While the filter balance and footprint are still being finalized, the LSST strategy will consist of a high-visit region near the bulge and a thick strip of the Galactic plane in the fourth quadrant as part of its Galactic Plane ``Wide Fast Deep" survey (WFD), alongside a low-visit ``Dusty Plane" survey that fills in the remainder of the third quadrant \citep{lsst_recommendations_2024}. LSST is expected to produce a photometric catalog of roughly 20 billion stars over its ten year duration, with a typical $5\sigma$ single-visit point-source depth in $r$ of 24.5 mag in the AB system \citep{Ivezic_2019}. 

Complementing LSST, Roman will be targeting the Galactic plane in the first two years of its mission as part of its General Astrophysics Survey program \citep{Sanderson_2024}. Launching by mid-2027, Roman's Wide-Field Instrument (WFI) has eight science filters spanning $0.48 - 2.3 \; \micron$ \citep{nasa_wfi_technical_2024}, where the longest wavelength F213 filter will prove critical for peering through highly extinguished regions in the plane. Like LSST, the filter balance and footprint for the Roman Galactic Plane Survey are still being defined, but is expected to reach a $5\sigma$ point-source depth of 25.4 mag in F146 in the AB system, assuming a typical 57 s integration \citep{nasa_wfi_technical_2024}. Roman has the potential to provide multi-band photometric imaging for tens of billions of stars \citep{Paladini_2023}. Therefore, either alone or in combination, LSST and Roman will produce photometry for at least an order of magnitude more stars in the Galactic plane than currently available. 

To match the potential of these next-generation photometric surveys, we anticipate three required methodological improvements. First, we need to employ faster inference frameworks. Second, we need to develop more accurate templates for the intrinsic colors of stars that capture both low initial stellar masses \textit{and} a broad range of post-main sequence evolutionary phases. And third, we need to implement more nuanced modeling of the extinction curve. 

The stellar inference pipeline we present here is effective but slow --- on average, {\brutus} took 4 s/star to generate our stellar catalog (see \S \ref{sec:compute_deets}).  Machine learning approaches are more easily scalable and offer an alternative to our brute-force Bayesian inference pipeline, producing outputs in a fraction of a second per star once the model is trained. For example, \citet{Zhang_2023} developed a data-driven model to infer stellar atmospheric parameters, distances, and extinctions for stars in the Gaia XP catalog \citep[see also data-driven models from][]{Green_2021}. The model was trained on the 1\% of Gaia XP stars with stellar atmospheric labels ($T_{\rm eff}$, log(g) and [Fe/H]) from the LAMOST survey. Training took 25 hours on one GPU node and was used to infer the parameters for all 220 million Gaia XP stars in roughly 36 hours on 2-4 GPU nodes (Private Communication; Zhang \& Green 2024), making this type of approach very computationally efficient in the era of LSST and Roman. 

One potential challenge of data-driven approaches is that the quality of the inference is predicated upon capturing a broad range of stellar types in the spectroscopic training dataset: if a certain type is not represented in training, those stars will be mismodeled in the final catalog. Two of the most critical stellar types to capture in the era of LSST and Roman are stars with low initial stellar masses (e.g. M-dwarfs) and those that have evolved off the main sequence (e.g. red giant branch, horizontal branch, and asymptotic giant branch stars). Roman will detect low-mass dwarfs out to a few kiloparsecs, representing a critical foreground population for anchoring the distribution of dust at larger distances. Likewise, strong coverage of the post-main sequence will allow future dust maps to model highly reddened giants at distances far beyond the distances probed in this work, enabling the construction of 3D dust maps beyond the Galactic center. The data-driven models from \citet{Zhang_2023} have poor coverage of low-mass M-dwarf solutions, as well as sparser coverage of post-main sequence evolution than the MIST models we employ here. Therefore, incorporating new and upcoming spectroscopic surveys like SDSS-V's Milky Way Mapper \citep{Almeida_2023} will prove critical for accurately modeling these stellar types in the context of future data-driven machine learning approaches.  

Finally, future 3D dust mapping frameworks will need to implement more sophisticated modeling of the 3D variation in the extinction curve. Improved modeling of the intrinsic colors of stars (across a broader range of spectral types) will only go so far if the modeling of the extinction curve is too simplistic. Recall from \S \ref{subsec:perstar_inference} that the shape of the extinction curve is parameterized by the total-to-selective extinction ratio, $R_V$. In this work, we adopt a tight prior on $R_V$, with a mean $R_V=3.32$ and a standard deviation of $\sigma_{{R_V}} = 0.18$ based on \citet{Schlafly_2016}.  We also assume that the extinction curve is independent of stellar type. However, the dust extinction curve depends on the underlying stellar spectrum and is known to vary substantially throughout the Milky Way \citep{Zhang_2024}. Failure to capture these effects could, for example, create systematics in the modeling of $A_V$ that would be nearly indistinguishable from changes in a star's effective temperature $T_{\rm eff}$. SDSS-V's Milky Way Mapper is currently targeting a much larger number of stars at high extinction. These measurements will prove critical for studying variations in the extinction curve across the Galaxy at large, which can be folded into future inference frameworks in the era of LSST and Roman. 

\section{Code and Data Availability} \label{sec:data_avail}
The software to reproduce the per-star and line-of-sight inference is publicly available on Zenodo (\href{https://doi.org/10.5281/zenodo.16813633}{doi:10.5281/zenodo.16813633}). The stellar inference pipeline is based on the publicly available {\brutus} software package (v0.8.3) which is publicly available on Zenodo (\href{https://doi.org/10.5281/zenodo.14915000}{doi:10.5281/zenodo.14915000})

For the per-star inference, we release the 2.5th, 16th, 50th, 84th, and 97.5th percentiles of the samples for the stellar properties, distance, extinction, and the total-to-selective extinction ratio of each star in our high-quality catalog (709 million stars; see \S \ref{subsec:filtering}), which have been computed from all 1000 posterior samples prior to thinning. We also release five random samples of distance, extinction, total-to-selective extinction ratio, and model indices, the latter of which can be translated to samples of stellar type given the underlying {\brutus} model grid. In addition to being available on the Dataverse (\href{https://doi.org/10.7910/DVN/K88GFI}{doi:10.7910/DVN/K88GFI}), the stellar catalog is archived in the \href{https://datalab.noirlab.edu/data-explorer?showTable=decaps\_dr2.stellar\_inference}{AstroDataLab} (\texttt{decaps\_dr2.stellar\_inference}) so it is accessible via TAP-accessible clients, including the \texttt{astroquery} Python package.

For the line-of-sight inference, we release the mean 3D reddening map (computed from all 100 samples) and five random samples (in units of $E_{B-V}$ in mags) as the core data product, which is publicly available for download on the Dataverse (\href{https://doi.org/10.7910/DVN/J9JCKO}{doi:10.7910/DVN/J9JCKO}). $E_{B-V}$ is derived from the underlying stellar inference on $A_V$ and $R_V$, where $E(B-V) = \frac{A_V}{R_V}$, with a mean of $R_V = 3.32$ based on our prior on the variation in the extinction curve from \citet{Schlafly_2016}. This core data product is released in HEALPix $N_{\rm side} = 8192$ format, where we provide the reddening in 120 logarithmically-spaced distance bins. Recall from \S \ref{subsec:infill} that we infill roughly 1\% of pixels in our footprint due to an insufficient number of stars. All five released samples have been infilled, alongside the mean map. We release a number of quality flags, including whether the line-of-sight fit converged in a given pixel, whether the pixel was infilled, and the minimum and maximum reliable distance modulus in that pixel (see \S \ref{sec:reliable}). We also provide the number of stars used to inform the line of sight fit for each $N_{\rm side} = 8192$ pixel. 

In addition to being available on the Dataverse, the map is also queryable via the Python package \texttt{dustmaps} using the \texttt{DECaPSQuery} class, which is the recommended way of performing extinction corrections. We also provide the functionality to query with memory mapping (\texttt{DECaPSQueryLite}) so the entire map does not have to be read into memory. Via \texttt{dustmaps}, users can also combine the DECaPS query with the existing Bayestar query to perform extinction corrections across the entire disk $|b| < 10^\circ$, and an example is provided online in the \texttt{dustmaps} documentation.

\section{Conclusion} \label{sec:conclusion}

We present a deep, high-angular-resolution 3D dust map over the range $239^\circ < l < 6^\circ$ and $|b| < 10^\circ$. We start by inferring the distance, extinction, and stellar type of almost one billion stars using optical and infrared photometry from the DECaPS2, VVV, 2MASS, and unWISE surveys. Unlike most solar neighborhood based maps, we do not require Gaia parallax measurements, though we incorporate these distance constraints when available (roughly half the sample). We then group the stars into pixels, and fit the distance and extinction measurements of stars in each pixel to infer the distribution of dust along 51 million lines of sight in the southern Galactic plane, totalling over 6 billion voxels. Our main conclusions are as follows: 

\begin{itemize}
\item Thanks to the increased stellar density provided by
the DECaPS2 survey, we produce a 3D dust map with an angular resolution of FWHM= $1\arcmin$, which is roughly an order of magnitude finer than any existing 3D dust map and on par with the angular resolution of the Herschel 2D dust emission maps. 

\item The flexibility of our pipeline to model stars using photometry alone allows us to probe significantly deeper into the Galactic plane than most current maps, targeting regions inaccessible to Gaia. We resolve cloud complexes towards the nearby Sagittarius-Carina and Scutum-Centaurus arm as well as more distant complexes lying between $6-10 \; \rm kpc$ from the Sun. However, we find that most of the cloud complexes lie in the interarm regions (between Sagittarius-Carina and Scutum-Centaurus), suggesting the need for substantial revision to spiral arm models in the fourth quadrant, where constraints are currently limited by a lack of maser parallax measurements. 

\item Our map fills in the one-third of the Galactic plane absent from the Bayestar19 3D dust map \citep{Green_2019}. By combining Bayestar19 with our new map, we enable extinction corrections over the entire disk within $|b| < 10^\circ$. 

\end{itemize}

Our map serves as a valuable proof of concept for the future of 3D dust mapping in the era of LSST and Roman, which will provide deep optical and infrared photometry for tens of billions of stars.  In this work, we focused on pushing the angular resolution frontier, in contrast to, for example, the \citet{Edenhofer_2023} map, which pushed the limits of distance resolution. As we transition from voxel counts of several billion in this current work to the trillion that may be possible with LSST and Roman, simply storing and handling maps of this size will become challenging, and we may need to move beyond voxelization. Improved storage and data handling --- alongside the development of faster stellar inference frameworks, improved modeling of low-mass and post-main sequence stars, and more sophisticated treatments of the dust extinction curve --- will prove critical for fully realizing the potential of 3D dust mapping in the coming decade.


\section*{Acknowledgments}
CZ acknowledges insightful discussions with Doug Marshall, Mark Reid, Tom Dame, Michael Rugel, Philipp Frank, and Torsten En\ss{}lin. CZ thanks Doug Marshall for facilitating the comparison with his 3D dust map in Figure \ref{fig:marshall_comp}. CZ, DPF, and AG acknowledge support by NASA ADAP grant 80NSSC21K0634 “Knitting Together the Milky Way: An Integrated Model of the Galaxy’s Stars, Gas, and Dust.” AKS acknowledges support by a National Science Foundation Graduate Research Fellowship (DGE-1745303) and that support for this work was provided by NASA through the NASA Hubble Fellowship grant HST-HF2-51564.001-A awarded by the Space Telescope Science Institute, which is operated by the Association of Universities for Research in Astronomy, Inc., for NASA, under contract NAS5-26555. AKS acknowledges Sophia S\'{a}nchez-Maes for helpful discussions and much support. JSS was supported by funding from the Dunlap Institute, an NSERC Banting Postdoctoral Fellowship, NSERC Discovery Grant RGPIN-2023-04849, and a University of Toronto Connaught New Researcher Award. JSS would also like to thank Rebecca Bleich for not stealing too many french fries during the process of preparing and revising this manuscript. The authors thank the anonymous referee for their prompt and constructive review of our manuscript that improved the quality of this work. 

This work was supported by the National Science Foundation under Cooperative Agreement PHY-2019786 (The NSF AI Institute for Artificial Intelligence and Fundamental Interactions). The authors acknowledge Interstellar Institute's program ``With Two Eyes'' and the Paris-Saclay University's Institut Pascal for hosting discussions that nourished the development of the ideas behind this work.

This research has used data, tools or materials developed as part of the EXPLORE project that has received funding from the European Union’s Horizon 2020 research and innovation programme under grant agreement No 101004214.

A portion of this work was enabled by the FASRC Cannon cluster supported by the FAS Division of Science Research Computing Group at Harvard University. 


\software{\texttt{astropy} \citep{astropy}, \texttt{brutus} \citep{Speagle_2023a}, \texttt{glue} \citep{glueviz}, \texttt{numpy} \citep{numpy}, \texttt{scipy} \citep{scipy}, \texttt{healpy} \citep{healpy_one,healpy_two}, \texttt{dustmaps} \citep{dustmaps}}

\bibliography{main}{}

\begin{thebibliography}{}
\expandafter\ifx\csname natexlab\endcsname\relax\def\natexlab#1{#1}\fi
\providecommand{\url}[1]{\href{#1}{#1}}
\providecommand{\dodoi}[1]{doi:~\href{http://doi.org/#1}{\nolinkurl{#1}}}
\providecommand{\doeprint}[1]{\href{http://ascl.net/#1}{\nolinkurl{http://ascl.net/#1}}}
\providecommand{\doarXiv}[1]{\href{https://arxiv.org/abs/#1}{\nolinkurl{https://arxiv.org/abs/#1}}}

\bibitem[{{Almeida} {et~al.}(2023){Almeida}, {Anderson},
  {Argudo-Fern{\'a}ndez}, {Badenes}, {Barger}, {Barrera-Ballesteros}, {Bender},
  {Benitez}, {Besser}, {Bird}, {Bizyaev}, {Blanton}, {Bochanski}, {Bovy},
  {Brandt}, {Brownstein}, {Buchner}, {Bulbul}, {Burchett}, {Cano D{\'\i}az},
  {Carlberg}, {Casey}, {Chandra}, {Cherinka}, {Chiappini}, {Coker}, {Comparat},
  {Conroy}, {Contardo}, {Cortes}, {Covey}, {Crane}, {Cunha}, {Dabbieri},
  {Davidson}, {Davis}, {de Andrade Queiroz}, {De Lee}, {M{\'e}ndez Delgado},
  {Demasi}, {Di Mille}, {Donor}, {Dow}, {Dwelly}, {Eracleous}, {Eriksen},
  {Fan}, {Farr}, {Frederick}, {Fries}, {Frinchaboy}, {G{\"a}nsicke}, {Ge},
  {Gonz{\'a}lez {\'A}vila}, {Grabowski}, {Grier}, {Guiglion}, {Gupta}, {Hall},
  {Hawkins}, {Hayes}, {Hermes}, {Hern{\'a}ndez-Garc{\'\i}a}, {Hogg},
  {Holtzman}, {Ibarra-Medel}, {Ji}, {Jofre}, {Johnson}, {Jones}, {Kinemuchi},
  {Kluge}, {Koekemoer}, {Kollmeier}, {Kounkel}, {Krishnarao}, {Krumpe},
  {Lacerna}, {Lago}, {Laporte}, {Liu}, {Liu}, {Liu}, {Lopes}, {Macktoobian},
  {Majewski}, {Malanushenko}, {Maoz}, {Masseron}, {Masters}, {Matijevic},
  {McBride}, {Medan}, {Merloni}, {Morrison}, {Myers}, {M{\'e}sz{\'a}ros},
  {Negrete}, {Nidever}, {Nitschelm}, {Oravetz}, {Oravetz}, {Pan}, {Peng},
  {Pinsonneault}, {Pogge}, {Qiu}, {Ramirez}, {Rix}, {Fern{\'a}ndez Rosso},
  {Runnoe}, {Salvato}, {Sanchez}, {Santana}, {Saydjari}, {Sayres},
  {Schlaufman}, {Schneider}, {Schwope}, {Serna}, {Shen}, {Sobeck}, {Song},
  {Souto}, {Spoo}, {Stassun}, {Steinmetz}, {Straumit}, {Stringfellow},
  {S{\'a}nchez-Gallego}, {Taghizadeh-Popp}, {Tayar}, {Thakar}, {Tissera},
  {Tkachenko}, {Hernandez Toledo}, {Trakhtenbrot}, {Fern{\'a}ndez-Trincado},
  {Troup}, {Trump}, {Tuttle}, {Ulloa}, {Vazquez-Mata}, {Vera Alfaro},
  {Villanova}, {Wachter}, {Weijmans}, {Wheeler}, {Wilson}, {Wojno}, {Wolf},
  {Xue}, {Ybarra}, {Zari}, \& {Zasowski}}]{Almeida_2023}
{Almeida}, A., {Anderson}, S.~F., {Argudo-Fern{\'a}ndez}, M., {et~al.} 2023,
  \apjs, 267, 44, \dodoi{10.3847/1538-4365/acda98}

\bibitem[{{Alonso-Garc{\'\i}a} {et~al.}(2017){Alonso-Garc{\'\i}a}, {Minniti},
  {Catelan}, {Contreras Ramos}, {Gonzalez}, {Hempel}, {Lucas}, {Saito},
  {Valenti}, \& {Zoccali}}]{Alonso_Garcia_2017}
{Alonso-Garc{\'\i}a}, J., {Minniti}, D., {Catelan}, M., {et~al.} 2017, \apjl,
  849, L13, \dodoi{10.3847/2041-8213/aa92c3}

\bibitem[{{Alonso-Garc{\'\i}a} {et~al.}(2018){Alonso-Garc{\'\i}a}, {Saito},
  {Hempel}, {Minniti}, {Pullen}, {Catelan}, {Ramos}, {Cross}, {Gonzalez},
  {Lucas}, {Palma}, {Valenti}, \& {Zoccali}}]{Alonso_Garcia_2018}
{Alonso-Garc{\'\i}a}, J., {Saito}, R.~K., {Hempel}, M., {et~al.} 2018, \aap,
  619, A4, \dodoi{10.1051/0004-6361/201833432}

\bibitem[{{Anders} {et~al.}(2019){Anders}, {Khalatyan}, {Chiappini}, {Queiroz},
  {Santiago}, {Jordi}, {Girardi}, {Brown}, {Matijevi{\v{c}}}, {Monari},
  {Cantat-Gaudin}, {Weiler}, {Khan}, {Miglio}, {Carrillo}, {Romero-G{\'o}mez},
  {Minchev}, {de Jong}, {Antoja}, {Ramos}, {Steinmetz}, \&
  {Enke}}]{Anders_2019}
{Anders}, F., {Khalatyan}, A., {Chiappini}, C., {et~al.} 2019, \aap, 628, A94,
  \dodoi{10.1051/0004-6361/201935765}

\bibitem[{{Astropy Collaboration} {et~al.}(2022){Astropy Collaboration},
  {Price-Whelan}, {Lim}, {Earl}, {Starkman}, {Bradley}, {Shupe}, {Patil},
  {Corrales}, {Brasseur}, {N{\"o}the}, {Donath}, {Tollerud}, {Morris},
  {Ginsburg}, {Vaher}, {Weaver}, {Tocknell}, {Jamieson}, {van Kerkwijk},
  {Robitaille}, {Merry}, {Bachetti}, {G{\"u}nther}, {Aldcroft},
  {Alvarado-Montes}, {Archibald}, {B{\'o}di}, {Bapat}, {Barentsen},
  {Baz{\'a}n}, {Biswas}, {Boquien}, {Burke}, {Cara}, {Cara}, {Conroy},
  {Conseil}, {Craig}, {Cross}, {Cruz}, {D'Eugenio}, {Dencheva}, {Devillepoix},
  {Dietrich}, {Eigenbrot}, {Erben}, {Ferreira}, {Foreman-Mackey}, {Fox},
  {Freij}, {Garg}, {Geda}, {Glattly}, {Gondhalekar}, {Gordon}, {Grant},
  {Greenfield}, {Groener}, {Guest}, {Gurovich}, {Handberg}, {Hart},
  {Hatfield-Dodds}, {Homeier}, {Hosseinzadeh}, {Jenness}, {Jones}, {Joseph},
  {Kalmbach}, {Karamehmetoglu}, {Ka{\l}uszy{\'n}ski}, {Kelley}, {Kern},
  {Kerzendorf}, {Koch}, {Kulumani}, {Lee}, {Ly}, {Ma}, {MacBride}, {Maljaars},
  {Muna}, {Murphy}, {Norman}, {O'Steen}, {Oman}, {Pacifici}, {Pascual},
  {Pascual-Granado}, {Patil}, {Perren}, {Pickering}, {Rastogi}, {Roulston},
  {Ryan}, {Rykoff}, {Sabater}, {Sakurikar}, {Salgado}, {Sanghi}, {Saunders},
  {Savchenko}, {Schwardt}, {Seifert-Eckert}, {Shih}, {Jain}, {Shukla}, {Sick},
  {Simpson}, {Singanamalla}, {Singer}, {Singhal}, {Sinha}, {Sip{\H{o}}cz},
  {Spitler}, {Stansby}, {Streicher}, {{\v{S}}umak}, {Swinbank}, {Taranu},
  {Tewary}, {Tremblay}, {de Val-Borro}, {Van Kooten}, {Vasovi{\'c}}, {Verma},
  {de Miranda Cardoso}, {Williams}, {Wilson}, {Winkel}, {Wood-Vasey}, {Xue},
  {Yoachim}, {Zhang}, {Zonca}, \& {Astropy Project Contributors}}]{astropy}
{Astropy Collaboration}, {Price-Whelan}, A.~M., {Lim}, P.~L., {et~al.} 2022,
  \apj, 935, 167, \dodoi{10.3847/1538-4357/ac7c74}

\bibitem[{{Bailer-Jones} {et~al.}(2018){Bailer-Jones}, {Rybizki}, {Fouesneau},
  {Mantelet}, \& {Andrae}}]{Bailer_Jones_2018}
{Bailer-Jones}, C.~A.~L., {Rybizki}, J., {Fouesneau}, M., {Mantelet}, G., \&
  {Andrae}, R. 2018, \aj, 156, 58, \dodoi{10.3847/1538-3881/aacb21}

\bibitem[{{Bland-Hawthorn} \& {Gerhard}(2016)}]{Bland_Hawthorn_2016}
{Bland-Hawthorn}, J., \& {Gerhard}, O. 2016, \araa, 54, 529,
  \dodoi{10.1146/annurev-astro-081915-023441}

\bibitem[{{Boggess} {et~al.}(1992){Boggess}, {Mather}, {Weiss}, {Bennett},
  {Cheng}, {Dwek}, {Gulkis}, {Hauser}, {Janssen}, {Kelsall}, {Meyer},
  {Moseley}, {Murdock}, {Shafer}, {Silverberg}, {Smoot}, {Wilkinson}, \&
  {Wright}}]{DIRBE}
{Boggess}, N.~W., {Mather}, J.~C., {Weiss}, R., {et~al.} 1992, \apj, 397, 420,
  \dodoi{10.1086/171797}

\bibitem[{{Burstein} \& {Heiles}(1978)}]{Burstein_Heiles_1978}
{Burstein}, D., \& {Heiles}, C. 1978, \apj, 225, 40, \dodoi{10.1086/156466}

\bibitem[{{Burstein} \& {Heiles}(1982)}]{Burstein_Heiles_1982}
---. 1982, \aj, 87, 1165, \dodoi{10.1086/113199}

\bibitem[{{Cargile} {et~al.}(2020){Cargile}, {Conroy}, {Johnson}, {Ting},
  {Bonaca}, {Dotter}, \& {Speagle}}]{Cargile_2020}
{Cargile}, P.~A., {Conroy}, C., {Johnson}, B.~D., {et~al.} 2020, \apj, 900, 28,
  \dodoi{10.3847/1538-4357/aba43b}

\bibitem[{{Chambers} {et~al.}(2016){Chambers}, {Magnier}, {Metcalfe},
  {Flewelling}, {Huber}, {Waters}, {Denneau}, {Draper}, {Farrow}, {Finkbeiner},
  {Holmberg}, {Koppenhoefer}, {Price}, {Rest}, {Saglia}, {Schlafly}, {Smartt},
  {Sweeney}, {Wainscoat}, {Burgett}, {Chastel}, {Grav}, {Heasley}, {Hodapp},
  {Jedicke}, {Kaiser}, {Kudritzki}, {Luppino}, {Lupton}, {Monet}, {Morgan},
  {Onaka}, {Shiao}, {Stubbs}, {Tonry}, {White}, {Ba{\~n}ados}, {Bell},
  {Bender}, {Bernard}, {Boegner}, {Boffi}, {Botticella}, {Calamida},
  {Casertano}, {Chen}, {Chen}, {Cole}, {Deacon}, {Frenk}, {Fitzsimmons},
  {Gezari}, {Gibbs}, {Goessl}, {Goggia}, {Gourgue}, {Goldman}, {Grant},
  {Grebel}, {Hambly}, {Hasinger}, {Heavens}, {Heckman}, {Henderson}, {Henning},
  {Holman}, {Hopp}, {Ip}, {Isani}, {Jackson}, {Keyes}, {Koekemoer}, {Kotak},
  {Le}, {Liska}, {Long}, {Lucey}, {Liu}, {Martin}, {Masci}, {McLean}, {Mindel},
  {Misra}, {Morganson}, {Murphy}, {Obaika}, {Narayan}, {Nieto-Santisteban},
  {Norberg}, {Peacock}, {Pier}, {Postman}, {Primak}, {Rae}, {Rai}, {Riess},
  {Riffeser}, {Rix}, {R{\"o}ser}, {Russel}, {Rutz}, {Schilbach}, {Schultz},
  {Scolnic}, {Strolger}, {Szalay}, {Seitz}, {Small}, {Smith}, {Soderblom},
  {Taylor}, {Thomson}, {Taylor}, {Thakar}, {Thiel}, {Thilker}, {Unger},
  {Urata}, {Valenti}, {Wagner}, {Walder}, {Walter}, {Watters}, {Werner},
  {Wood-Vasey}, \& {Wyse}}]{Chambers_2016}
{Chambers}, K.~C., {Magnier}, E.~A., {Metcalfe}, N., {et~al.} 2016, arXiv
  e-prints, arXiv:1612.05560, \dodoi{10.48550/arXiv.1612.05560}

\bibitem[{{Chen} {et~al.}(2019){Chen}, {Huang}, {Yuan}, {Wang}, {Fan}, {Xiang},
  {Zhang}, {Tian}, \& {Liu}}]{Chen_2019}
{Chen}, B.~Q., {Huang}, Y., {Yuan}, H.~B., {et~al.} 2019, \mnras, 483, 4277,
  \dodoi{10.1093/mnras/sty3341}

\bibitem[{{Choi} {et~al.}(2016){Choi}, {Dotter}, {Conroy}, {Cantiello},
  {Paxton}, \& {Johnson}}]{Choi_2016}
{Choi}, J., {Dotter}, A., {Conroy}, C., {et~al.} 2016, \apj, 823, 102,
  \dodoi{10.3847/0004-637X/823/2/102}

\bibitem[{{Cutri} {et~al.}(2013){Cutri}, {Wright}, {Conrow}, {Fowler},
  {Eisenhardt}, {Grillmair}, {Kirkpatrick}, {Masci}, {McCallon}, {Wheelock},
  {Fajardo-Acosta}, {Yan}, {Benford}, {Harbut}, {Jarrett}, {Lake}, {Leisawitz},
  {Ressler}, {Stanford}, {Tsai}, {Liu}, {Helou}, {Mainzer}, {Gettings},
  {Gonzalez}, {Hoffman}, {Marsh}, {Padgett}, {Skrutskie}, {Beck}, {Papin}, \&
  {Wittman}}]{Cutri_2013}
{Cutri}, R.~M., {Wright}, E.~L., {Conrow}, T., {et~al.} 2013, {Explanatory
  Supplement to the AllWISE Data Release Products}, Explanatory Supplement to
  the AllWISE Data Release Products, by R. M. Cutri et al.

\bibitem[{{Czekaj} {et~al.}(2014){Czekaj}, {Robin}, {Figueras}, {Luri}, \&
  {Haywood}}]{Czekaj_2014}
{Czekaj}, M.~A., {Robin}, A.~C., {Figueras}, F., {Luri}, X., \& {Haywood}, M.
  2014, \aap, 564, A102, \dodoi{10.1051/0004-6361/201322139}

\bibitem[{{Dharmawardena} {et~al.}(2024){Dharmawardena}, {Bailer-Jones},
  {Fouesneau}, {Foreman-Mackey}, {Coronica}, {Colnaghi}, {M{\"u}ller}, \&
  {Wilson}}]{Dharmawardena_2024}
{Dharmawardena}, T.~E., {Bailer-Jones}, C.~A.~L., {Fouesneau}, M., {et~al.}
  2024, \mnras, 532, 3480, \dodoi{10.1093/mnras/stae1474}

\bibitem[{{Draine}(2003)}]{Draine_2003}
{Draine}, B.~T. 2003, \araa, 41, 241,
  \dodoi{10.1146/annurev.astro.41.011802.094840}

\bibitem[{{Drew} {et~al.}(2005){Drew}, {Greimel}, {Irwin}, {Aungwerojwit},
  {Barlow}, {Corradi}, {Drake}, {G{\"a}nsicke}, {Groot}, {Hales}, {Hopewell},
  {Irwin}, {Knigge}, {Leisy}, {Lennon}, {Mampaso}, {Masheder}, {Matsuura},
  {Morales-Rueda}, {Morris}, {Parker}, {Phillipps}, {Rodriguez-Gil}, {Roelofs},
  {Skillen}, {Sokoloski}, {Steeghs}, {Unruh}, {Viironen}, {Vink}, {Walton},
  {Witham}, {Wright}, {Zijlstra}, \& {Zurita}}]{IPHAS}
{Drew}, J.~E., {Greimel}, R., {Irwin}, M.~J., {et~al.} 2005, \mnras, 362, 753,
  \dodoi{10.1111/j.1365-2966.2005.09330.x}

\bibitem[{{Drimmel} \& {Spergel}(2001)}]{Drimmel_Spergel_2001}
{Drimmel}, R., \& {Spergel}, D.~N. 2001, \apj, 556, 181, \dodoi{10.1086/321556}

\bibitem[{{Edenhofer} {et~al.}(2023){Edenhofer}, {Zucker}, {Frank}, {Saydjari},
  {Speagle}, {Finkbeiner}, \& {En{\ss}lin}}]{Edenhofer_2023}
{Edenhofer}, G., {Zucker}, C., {Frank}, P., {et~al.} 2023, arXiv e-prints,
  arXiv:2308.01295, \dodoi{10.48550/arXiv.2308.01295}

\bibitem[{{Ehlerov{\'a}} \& {Palou{\v{s}}}(2013)}]{Ehlerov_2013}
{Ehlerov{\'a}}, S., \& {Palou{\v{s}}}, J. 2013, \aap, 550, A23,
  \dodoi{10.1051/0004-6361/201220341}

\bibitem[{{Fabricius} {et~al.}(2021){Fabricius}, {Luri}, {Arenou}, {Babusiaux},
  {Helmi}, {Muraveva}, {Reyl{\'e}}, {Spoto}, {Vallenari}, {Antoja}, {Balbinot},
  {Barache}, {Bauchet}, {Bragaglia}, {Busonero}, {Cantat-Gaudin}, {Carrasco},
  {Diakit{\'e}}, {Fabrizio}, {Figueras}, {Garcia-Gutierrez}, {Garofalo},
  {Jordi}, {Kervella}, {Khanna}, {Leclerc}, {Licata}, {Lambert}, {Marrese},
  {Masip}, {Ramos}, {Robichon}, {Robin}, {Romero-G{\'o}mez}, {Rubele}, \&
  {Weiler}}]{Fabricius_2021}
{Fabricius}, C., {Luri}, X., {Arenou}, F., {et~al.} 2021, \aap, 649, A5,
  \dodoi{10.1051/0004-6361/202039834}

\bibitem[{{Finkbeiner} {et~al.}(1999){Finkbeiner}, {Davis}, \&
  {Schlegel}}]{Finkbeiner_1999}
{Finkbeiner}, D.~P., {Davis}, M., \& {Schlegel}, D.~J. 1999, \apj, 524, 867,
  \dodoi{10.1086/307852}

\bibitem[{{Gaia Collaboration}(2022)}]{GaiaDR3Cat}
{Gaia Collaboration}. 2022, {VizieR Online Data Catalog: Gaia DR3 Part 1. Main
  source (Gaia Collaboration, 2022)}, VizieR On-line Data Catalog: I/355.
  Originally published in: Astron. Astrophys., in prep. (2022),
  \dodoi{10.26093/cds/vizier.1355}

\bibitem[{{Gaia Collaboration} {et~al.}(2016){Gaia Collaboration}, {Prusti},
  {de Bruijne}, {Brown}, {Vallenari}, {Babusiaux}, {Bailer-Jones}, {Bastian},
  {Biermann}, {Evans}, {Eyer}, {Jansen}, {Jordi}, {Klioner}, {Lammers},
  {Lindegren}, {Luri}, {Mignard}, {Milligan}, {Panem}, {Poinsignon},
  {Pourbaix}, {Randich}, {Sarri}, {Sartoretti}, {Siddiqui}, {Soubiran},
  {Valette}, {van Leeuwen}, {Walton}, {Aerts}, {Arenou}, {Cropper}, {Drimmel},
  {H{\o}g}, {Katz}, {Lattanzi}, {O'Mullane}, {Grebel}, {Holland}, {Huc},
  {Passot}, {Bramante}, {Cacciari}, {Casta{\~n}eda}, {Chaoul}, {Cheek}, {De
  Angeli}, {Fabricius}, {Guerra}, {Hern{\'a}ndez}, {Jean-Antoine-Piccolo},
  {Masana}, {Messineo}, {Mowlavi}, {Nienartowicz}, {Ord{\'o}{\~n}ez-Blanco},
  {Panuzzo}, {Portell}, {Richards}, {Riello}, {Seabroke}, {Tanga},
  {Th{\'e}venin}, {Torra}, {Els}, {Gracia-Abril}, {Comoretto},
  {Garcia-Reinaldos}, {Lock}, {Mercier}, {Altmann}, {Andrae}, {Astraatmadja},
  {Bellas-Velidis}, {Benson}, {Berthier}, {Blomme}, {Busso}, {Carry},
  {Cellino}, {Clementini}, {Cowell}, {Creevey}, {Cuypers}, {Davidson}, {De
  Ridder}, {de Torres}, {Delchambre}, {Dell'Oro}, {Ducourant}, {Fr{\'e}mat},
  {Garc{\'\i}a-Torres}, {Gosset}, {Halbwachs}, {Hambly}, {Harrison}, {Hauser},
  {Hestroffer}, {Hodgkin}, {Huckle}, {Hutton}, {Jasniewicz}, {Jordan},
  {Kontizas}, {Korn}, {Lanzafame}, {Manteiga}, {Moitinho}, {Muinonen},
  {Osinde}, {Pancino}, {Pauwels}, {Petit}, {Recio-Blanco}, {Robin}, {Sarro},
  {Siopis}, {Smith}, {Smith}, {Sozzetti}, {Thuillot}, {van Reeven}, {Viala},
  {Abbas}, {Abreu Aramburu}, {Accart}, {Aguado}, {Allan}, {Allasia},
  {Altavilla}, {{\'A}lvarez}, {Alves}, {Anderson}, {Andrei}, {Anglada Varela},
  {Antiche}, {Antoja}, {Ant{\'o}n}, {Arcay}, {Atzei}, {Ayache}, {Bach},
  {Baker}, {Balaguer-N{\'u}{\~n}ez}, {Barache}, {Barata}, {Barbier}, {Barblan},
  {Baroni}, {Barrado y Navascu{\'e}s}, {Barros}, {Barstow}, {Becciani},
  {Bellazzini}, {Bellei}, {Bello Garc{\'\i}a}, {Belokurov}, {Bendjoya},
  {Berihuete}, {Bianchi}, {Bienaym{\'e}}, {Billebaud}, {Blagorodnova},
  {Blanco-Cuaresma}, {Boch}, {Bombrun}, {Borrachero}, {Bouquillon}, {Bourda},
  {Bouy}, {Bragaglia}, {Breddels}, {Brouillet}, {Br{\"u}semeister},
  {Bucciarelli}, {Budnik}, {Burgess}, {Burgon}, {Burlacu}, {Busonero}, {Buzzi},
  {Caffau}, {Cambras}, {Campbell}, {Cancelliere}, {Cantat-Gaudin}, {Carlucci},
  {Carrasco}, {Castellani}, {Charlot}, {Charnas}, {Charvet}, {Chassat},
  {Chiavassa}, {Clotet}, {Cocozza}, {Collins}, {Collins}, {Costigan}, {Crifo},
  {Cross}, {Crosta}, {Crowley}, {Dafonte}, {Damerdji}, {Dapergolas}, {David},
  {David}, {De Cat}, {de Felice}, {de Laverny}, {De Luise}, {De March}, {de
  Martino}, {de Souza}, {Debosscher}, {del Pozo}, {Delbo}, {Delgado},
  {Delgado}, {di Marco}, {Di Matteo}, {Diakite}, {Distefano}, {Dolding}, {Dos
  Anjos}, {Drazinos}, {Dur{\'a}n}, {Dzigan}, {Ecale}, {Edvardsson}, {Enke},
  {Erdmann}, {Escolar}, {Espina}, {Evans}, {Eynard Bontemps}, {Fabre},
  {Fabrizio}, {Faigler}, {Falc{\~a}o}, {Farr{\`a}s Casas}, {Faye}, {Federici},
  {Fedorets}, {Fern{\'a}ndez-Hern{\'a}ndez}, {Fernique}, {Fienga}, {Figueras},
  {Filippi}, {Findeisen}, {Fonti}, {Fouesneau}, {Fraile}, {Fraser}, {Fuchs},
  {Furnell}, {Gai}, {Galleti}, {Galluccio}, {Garabato}, {Garc{\'\i}a-Sedano},
  {Gar{\'e}}, {Garofalo}, {Garralda}, {Gavras}, {Gerssen}, {Geyer}, {Gilmore},
  {Girona}, {Giuffrida}, {Gomes}, {Gonz{\'a}lez-Marcos},
  {Gonz{\'a}lez-N{\'u}{\~n}ez}, {Gonz{\'a}lez-Vidal}, {Granvik}, {Guerrier},
  {Guillout}, {Guiraud}, {G{\'u}rpide}, {Guti{\'e}rrez-S{\'a}nchez}, {Guy},
  {Haigron}, {Hatzidimitriou}, {Haywood}, {Heiter}, {Helmi}, {Hobbs},
  {Hofmann}, {Holl}, {Holland}, {Hunt}, {Hypki}, {Icardi}, {Irwin}, {Jevardat
  de Fombelle}, {Jofr{\'e}}, {Jonker}, {Jorissen}, {Julbe}, {Karampelas},
  {Kochoska}, {Kohley}, {Kolenberg}, {Kontizas}, {Koposov}, {Kordopatis},
  {Koubsky}, {Kowalczyk}, {Krone-Martins}, {Kudryashova}, {Kull}, {Bachchan},
  {Lacoste-Seris}, {Lanza}, {Lavigne}, {Le Poncin-Lafitte}, {Lebreton},
  {Lebzelter}, {Leccia}, {Leclerc}, {Lecoeur-Taibi}, {Lemaitre}, {Lenhardt},
  {Leroux}, {Liao}, {Licata}, {Lindstr{\o}m}, {Lister}, {Livanou}, {Lobel},
  {L{\"o}ffler}, {L{\'o}pez}, {Lopez-Lozano}, {Lorenz}, {Loureiro},
  {MacDonald}, {Magalh{\~a}es Fernandes}, {Managau}, {Mann}, {Mantelet},
  {Marchal}, {Marchant}, {Marconi}, {Marie}, {Marinoni}, {Marrese},
  {Marschalk{\'o}}, {Marshall}, {Mart{\'\i}n-Fleitas}, {Martino}, {Mary},
  {Matijevi{\v{c}}}, {Mazeh}, {McMillan}, {Messina}, {Mestre}, {Michalik},
  {Millar}, {Miranda}, {Molina}, {Molinaro}, {Molinaro}, {Moln{\'a}r},
  {Moniez}, {Montegriffo}, {Monteiro}, {Mor}, {Mora}, {Morbidelli}, {Morel},
  {Morgenthaler}, {Morley}, {Morris}, {Mulone}, {Muraveva}, {Musella},
  {Narbonne}, {Nelemans}, {Nicastro}, {Noval}, {Ord{\'e}novic},
  {Ordieres-Mer{\'e}}, {Osborne}, {Pagani}, {Pagano}, {Pailler}, {Palacin},
  {Palaversa}, {Parsons}, {Paulsen}, {Pecoraro}, {Pedrosa}, {Pentik{\"a}inen},
  {Pereira}, {Pichon}, {Piersimoni}, {Pineau}, {Plachy}, {Plum}, {Poujoulet},
  {Pr{\v{s}}a}, {Pulone}, {Ragaini}, {Rago}, {Rambaux}, {Ramos-Lerate},
  {Ranalli}, {Rauw}, {Read}, {Regibo}, {Renk}, {Reyl{\'e}}, {Ribeiro},
  {Rimoldini}, {Ripepi}, {Riva}, {Rixon}, {Roelens}, {Romero-G{\'o}mez},
  {Rowell}, {Royer}, {Rudolph}, {Ruiz-Dern}, {Sadowski}, {Sagrist{\`a}
  Sell{\'e}s}, {Sahlmann}, {Salgado}, {Salguero}, {Sarasso}, {Savietto},
  {Schnorhk}, {Schultheis}, {Sciacca}, {Segol}, {Segovia}, {Segransan},
  {Serpell}, {Shih}, {Smareglia}, {Smart}, {Smith}, {Solano}, {Solitro},
  {Sordo}, {Soria Nieto}, {Souchay}, {Spagna}, {Spoto}, {Stampa}, {Steele},
  {Steidelm{\"u}ller}, {Stephenson}, {Stoev}, {Suess}, {S{\"u}veges}, {Surdej},
  {Szabados}, {Szegedi-Elek}, {Tapiador}, {Taris}, {Tauran}, {Taylor},
  {Teixeira}, {Terrett}, {Tingley}, {Trager}, {Turon}, {Ulla}, {Utrilla},
  {Valentini}, {van Elteren}, {Van Hemelryck}, {van Leeuwen}, {Varadi},
  {Vecchiato}, {Veljanoski}, {Via}, {Vicente}, {Vogt}, {Voss}, {Votruba},
  {Voutsinas}, {Walmsley}, {Weiler}, {Weingrill}, {Werner}, {Wevers},
  {Whitehead}, {Wyrzykowski}, {Yoldas}, {{\v{Z}}erjal}, {Zucker}, {Zurbach},
  {Zwitter}, {Alecu}, {Allen}, {Allende Prieto}, {Amorim},
  {Anglada-Escud{\'e}}, {Arsenijevic}, {Azaz}, {Balm}, {Beck}, {Bernstein},
  {Bigot}, {Bijaoui}, {Blasco}, {Bonfigli}, {Bono}, {Boudreault}, {Bressan},
  {Brown}, {Brunet}, {Bunclark}, {Buonanno}, {Butkevich}, {Carret}, {Carrion},
  {Chemin}, {Ch{\'e}reau}, {Corcione}, {Darmigny}, {de Boer}, {de Teodoro}, {de
  Zeeuw}, {Delle Luche}, {Domingues}, {Dubath}, {Fodor}, {Fr{\'e}zouls},
  {Fries}, {Fustes}, {Fyfe}, {Gallardo}, {Gallegos}, {Gardiol}, {Gebran},
  {Gomboc}, {G{\'o}mez}, {Grux}, {Gueguen}, {Heyrovsky}, {Hoar}, {Iannicola},
  {Isasi Parache}, {Janotto}, {Joliet}, {Jonckheere}, {Keil}, {Kim},
  {Klagyivik}, {Klar}, {Knude}, {Kochukhov}, {Kolka}, {Kos}, {Kutka}, {Lainey},
  {LeBouquin}, {Liu}, {Loreggia}, {Makarov}, {Marseille}, {Martayan},
  {Martinez-Rubi}, {Massart}, {Meynadier}, {Mignot}, {Munari}, {Nguyen},
  {Nordlander}, {Ocvirk}, {O'Flaherty}, {Olias Sanz}, {Ortiz}, {Osorio},
  {Oszkiewicz}, {Ouzounis}, {Palmer}, {Park}, {Pasquato}, {Peltzer}, {Peralta},
  {P{\'e}turaud}, {Pieniluoma}, {Pigozzi}, {Poels}, {Prat}, {Prod'homme},
  {Raison}, {Rebordao}, {Risquez}, {Rocca-Volmerange}, {Rosen}, {Ruiz-Fuertes},
  {Russo}, {Sembay}, {Serraller Vizcaino}, {Short}, {Siebert}, {Silva},
  {Sinachopoulos}, {Slezak}, {Soffel}, {Sosnowska}, {Strai{\v{z}}ys}, {ter
  Linden}, {Terrell}, {Theil}, {Tiede}, {Troisi}, {Tsalmantza}, {Tur},
  {Vaccari}, {Vachier}, {Valles}, {Van Hamme}, {Veltz}, {Virtanen}, {Wallut},
  {Wichmann}, {Wilkinson}, {Ziaeepour}, \& {Zschocke}}]{Gaia_Mission}
{Gaia Collaboration}, {Prusti}, T., {de Bruijne}, J.~H.~J., {et~al.} 2016,
  \aap, 595, A1, \dodoi{10.1051/0004-6361/201629272}

\bibitem[{{Gaia Collaboration} {et~al.}(2022){Gaia Collaboration}, {Vallenari},
  {Brown}, {Prusti}, {de Bruijne}, {Arenou}, {Babusiaux}, {Biermann},
  {Creevey}, {Ducourant}, {Evans}, {Eyer}, {Guerra}, {Hutton}, {Jordi},
  {Klioner}, {Lammers}, {Lindegren}, {Luri}, {Mignard}, {Panem}, {Pourbaix},
  {Randich}, {Sartoretti}, {Soubiran}, {Tanga}, {Walton}, {Bailer-Jones},
  {Bastian}, {Drimmel}, {Jansen}, {Katz}, {Lattanzi}, {van Leeuwen}, {Bakker},
  {Cacciari}, {Casta{\~n}eda}, {De Angeli}, {Fabricius}, {Fouesneau},
  {Fr{\'e}mat}, {Galluccio}, {Guerrier}, {Heiter}, {Masana}, {Messineo},
  {Mowlavi}, {Nicolas}, {Nienartowicz}, {Pailler}, {Panuzzo}, {Riclet}, {Roux},
  {Seabroke}, {Sordo{\o}rcit}, {Th{\'e}venin}, {Gracia-Abril}, {Portell},
  {Teyssier}, {Altmann}, {Andrae}, {Audard}, {Bellas-Velidis}, {Benson},
  {Berthier}, {Blomme}, {Burgess}, {Busonero}, {Busso}, {C{\'a}novas}, {Carry},
  {Cellino}, {Cheek}, {Clementini}, {Damerdji}, {Davidson}, {de Teodoro},
  {Nu{\~n}ez Campos}, {Delchambre}, {Dell'Oro}, {Esquej},
  {Fern{\'a}ndez-Hern{\'a}ndez}, {Fraile}, {Garabato}, {Garc{\'\i}a-Lario},
  {Gosset}, {Haigron}, {Halbwachs}, {Hambly}, {Harrison}, {Hern{\'a}ndez},
  {Hestroffer}, {Hodgkin}, {Holl}, {Jan{\ss}en}, {Jevardat de Fombelle},
  {Jordan}, {Krone-Martins}, {Lanzafame}, {L{\"o}ffler}, {Marchal}, {Marrese},
  {Moitinho}, {Muinonen}, {Osborne}, {Pancino}, {Pauwels}, {Recio-Blanco},
  {Reyl{\'e}}, {Riello}, {Rimoldini}, {Roegiers}, {Rybizki}, {Sarro}, {Siopis},
  {Smith}, {Sozzetti}, {Utrilla}, {van Leeuwen}, {Abbas}, {{\'A}brah{\'a}m},
  {Abreu Aramburu}, {Aerts}, {Aguado}, {Ajaj}, {Aldea-Montero}, {Altavilla},
  {{\'A}lvarez}, {Alves}, {Anders}, {Anderson}, {Anglada Varela}, {Antoja},
  {Baines}, {Baker}, {Balaguer-N{\'u}{\~n}ez}, {Balbinot}, {Balog}, {Barache},
  {Barbato}, {Barros}, {Barstow}, {Bartolom{\'e}}, {Bassilana}, {Bauchet},
  {Becciani}, {Bellazzini}, {Berihuete}, {Bernet}, {Bertone}, {Bianchi},
  {Binnenfeld}, {Blanco-Cuaresma}, {Blazere}, {Boch}, {Bombrun}, {Bossini},
  {Bouquillon}, {Bragaglia}, {Bramante}, {Breedt}, {Bressan}, {Brouillet},
  {Brugaletta}, {Bucciarelli}, {Burlacu}, {Butkevich}, {Buzzi}, {Caffau},
  {Cancelliere}, {Cantat-Gaudin}, {Carballo}, {Carlucci}, {Carnerero},
  {Carrasco}, {Casamiquela}, {Castellani}, {Castro-Ginard}, {Chaoul},
  {Charlot}, {Chemin}, {Chiaramida}, {Chiavassa}, {Chornay}, {Comoretto},
  {Contursi}, {Cooper}, {Cornez}, {Cowell}, {Crifo}, {Cropper}, {Crosta},
  {Crowley}, {Dafonte}, {Dapergolas}, {David}, {David}, {de Laverny}, {De
  Luise}, {De March}, {De Ridder}, {de Souza}, {de Torres}, {del Peloso}, {del
  Pozo}, {Delbo}, {Delgado}, {Delisle}, {Demouchy}, {Dharmawardena}, {Di
  Matteo}, {Diakite}, {Diener}, {Distefano}, {Dolding}, {Edvardsson}, {Enke},
  {Fabre}, {Fabrizio}, {Faigler}, {Fedorets}, {Fernique}, {Fienga}, {Figueras},
  {Fournier}, {Fouron}, {Fragkoudi}, {Gai}, {Garcia-Gutierrez},
  {Garcia-Reinaldos}, {Garc{\'\i}a-Torres}, {Garofalo}, {Gavel}, {Gavras},
  {Gerlach}, {Geyer}, {Giacobbe}, {Gilmore}, {Girona}, {Giuffrida}, {Gomel},
  {Gomez}, {Gonz{\'a}lez-N{\'u}{\~n}ez}, {Gonz{\'a}lez-Santamar{\'\i}a},
  {Gonz{\'a}lez-Vidal}, {Granvik}, {Guillout}, {Guiraud},
  {Guti{\'e}rrez-S{\'a}nchez}, {Guy}, {Hatzidimitriou}, {Hauser}, {Haywood},
  {Helmer}, {Helmi}, {Sarmiento}, {Hidalgo}, {Hilger}, {H{\l}adczuk}, {Hobbs},
  {Holland}, {Huckle}, {Jardine}, {Jasniewicz}, {Jean-Antoine Piccolo},
  {Jim{\'e}nez-Arranz}, {Jorissen}, {Juaristi Campillo}, {Julbe}, {Karbevska},
  {Kervella}, {Khanna}, {Kontizas}, {Kordopatis}, {Korn}, {K{\'o}sp{\'a}l},
  {Kostrzewa-Rutkowska}, {Kruszy{\'n}ska}, {Kun}, {Laizeau}, {Lambert},
  {Lanza}, {Lasne}, {Le Campion}, {Lebreton}, {Lebzelter}, {Leccia}, {Leclerc},
  {Lecoeur-Taibi}, {Liao}, {Licata}, {Lindstr{\o}m}, {Lister}, {Livanou},
  {Lobel}, {Lorca}, {Loup}, {Madrero Pardo}, {Magdaleno Romeo}, {Managau},
  {Mann}, {Manteiga}, {Marchant}, {Marconi}, {Marcos}, {Marcos Santos},
  {Mar{\'\i}n Pina}, {Marinoni}, {Marocco}, {Marshall}, {Polo},
  {Mart{\'\i}n-Fleitas}, {Marton}, {Mary}, {Masip}, {Massari},
  {Mastrobuono-Battisti}, {Mazeh}, {McMillan}, {Messina}, {Michalik}, {Millar},
  {Mints}, {Molina}, {Molinaro}, {Moln{\'a}r}, {Monari}, {Mongui{\'o}},
  {Montegriffo}, {Montero}, {Mor}, {Mora}, {Morbidelli}, {Morel}, {Morris},
  {Muraveva}, {Murphy}, {Musella}, {Nagy}, {Noval}, {Oca{\~n}a}, {Ogden},
  {Ordenovic}, {Osinde}, {Pagani}, {Pagano}, {Palaversa}, {Palicio},
  {Pallas-Quintela}, {Panahi}, {Payne-Wardenaar}, {Pe{\~n}alosa Esteller},
  {Penttil{\"a}}, {Pichon}, {Piersimoni}, {Pineau}, {Plachy}, {Plum}, {Poggio},
  {Pr{\v{s}}a}, {Pulone}, {Racero}, {Ragaini}, {Rainer}, {Raiteri}, {Rambaux},
  {Ramos}, {Ramos-Lerate}, {Re Fiorentin}, {Regibo}, {Richards}, {Rios Diaz},
  {Ripepi}, {Riva}, {Rix}, {Rixon}, {Robichon}, {Robin}, {Robin}, {Roelens},
  {Rogues}, {Rohrbasser}, {Romero-G{\'o}mez}, {Rowell}, {Royer}, {Ruz Mieres},
  {Rybicki}, {Sadowski}, {S{\'a}ez N{\'u}{\~n}ez}, {Sagrist{\`a} Sell{\'e}s},
  {Sahlmann}, {Salguero}, {Samaras}, {Sanchez Gimenez}, {Sanna},
  {Santove{\~n}a}, {Sarasso}, {Schultheis}, {Sciacca}, {Segol}, {Segovia},
  {S{\'e}gransan}, {Semeux}, {Shahaf}, {Siddiqui}, {Siebert}, {Siltala},
  {Silvelo}, {Slezak}, {Slezak}, {Smart}, {Snaith}, {Solano}, {Solitro},
  {Souami}, {Souchay}, {Spagna}, {Spina}, {Spoto}, {Steele},
  {Steidelm{\"u}ller}, {Stephenson}, {S{\"u}veges}, {Surdej}, {Szabados},
  {Szegedi-Elek}, {Taris}, {Taylo}, {Teixeira}, {Tolomei}, {Tonello}, {Torra},
  {Torra}, {Torralba Elipe}, {Trabucchi}, {Tsounis}, {Turon}, {Ulla}, {Unger},
  {Vaillant}, {van Dillen}, {van Reeven}, {Vanel}, {Vecchiato}, {Viala},
  {Vicente}, {Voutsinas}, {Weiler}, {Wevers}, {Wyrzykowski}, {Yoldas}, {Yvard},
  {Zhao}, {Zorec}, {Zucker}, \& {Zwitter}}]{Gaia_DR3}
{Gaia Collaboration}, {Vallenari}, A., {Brown}, A.~G.~A., {et~al.} 2022, arXiv
  e-prints, arXiv:2208.00211, \dodoi{10.48550/arXiv.2208.00211}

\bibitem[{{Gonzalez} {et~al.}(2012){Gonzalez}, {Rejkuba}, {Zoccali}, {Valenti},
  {Minniti}, {Schultheis}, {Tobar}, \& {Chen}}]{Gonzalez_2012}
{Gonzalez}, O.~A., {Rejkuba}, M., {Zoccali}, M., {et~al.} 2012, \aap, 543, A13,
  \dodoi{10.1051/0004-6361/201219222}

\bibitem[{{G{\'o}rski} {et~al.}(2005{\natexlab{a}}){G{\'o}rski}, {Hivon},
  {Banday}, {Wandelt}, {Hansen}, {Reinecke}, \& {Bartelmann}}]{Gorski_2005}
{G{\'o}rski}, K.~M., {Hivon}, E., {Banday}, A.~J., {et~al.} 2005{\natexlab{a}},
  \apj, 622, 759, \dodoi{10.1086/427976}

\bibitem[{{G{\'o}rski} {et~al.}(2005{\natexlab{b}}){G{\'o}rski}, {Hivon},
  {Banday}, {Wandelt}, {Hansen}, {Reinecke}, \& {Bartelmann}}]{healpy_two}
---. 2005{\natexlab{b}}, \apj, 622, 759, \dodoi{10.1086/427976}

\bibitem[{{Green}(2018)}]{dustmaps}
{Green}, G.~M. 2018, The Journal of Open Source Software, 3, 695,
  \dodoi{10.21105/joss.00695}

\bibitem[{{Green} {et~al.}(2019){Green}, {Schlafly}, {Zucker}, {Speagle}, \&
  {Finkbeiner}}]{Green_2019}
{Green}, G.~M., {Schlafly}, E., {Zucker}, C., {Speagle}, J.~S., \&
  {Finkbeiner}, D. 2019, \apj, 887, 93, \dodoi{10.3847/1538-4357/ab5362}

\bibitem[{{Green} {et~al.}(2014){Green}, {Schlafly}, {Finkbeiner}, {Juri{\'c}},
  {Rix}, {Burgett}, {Chambers}, {Draper}, {Flewelling}, {Kudritzki}, {Magnier},
  {Martin}, {Metcalfe}, {Tonry}, {Wainscoat}, \& {Waters}}]{Green_2014}
{Green}, G.~M., {Schlafly}, E.~F., {Finkbeiner}, D.~P., {et~al.} 2014, \apj,
  783, 114, \dodoi{10.1088/0004-637X/783/2/114}

\bibitem[{{Green} {et~al.}(2015){Green}, {Schlafly}, {Finkbeiner}, {Rix},
  {Martin}, {Burgett}, {Draper}, {Flewelling}, {Hodapp}, {Kaiser}, {Kudritzki},
  {Magnier}, {Metcalfe}, {Price}, {Tonry}, \& {Wainscoat}}]{Green_2015}
---. 2015, \apj, 810, 25, \dodoi{10.1088/0004-637X/810/1/25}

\bibitem[{{Green} {et~al.}(2018){Green}, {Schlafly}, {Finkbeiner}, {Rix},
  {Martin}, {Burgett}, {Draper}, {Flewelling}, {Hodapp}, {Kaiser}, {Kudritzki},
  {Magnier}, {Metcalfe}, {Tonry}, {Wainscoat}, \& {Waters}}]{Green_2018}
{Green}, G.~M., {Schlafly}, E.~F., {Finkbeiner}, D., {et~al.} 2018, \mnras,
  478, 651, \dodoi{10.1093/mnras/sty1008}

\bibitem[{{Green} {et~al.}(2021){Green}, {Rix}, {Tschesche}, {Finkbeiner},
  {Zucker}, {Schlafly}, {Rybizki}, {Fouesneau}, {Andrae}, \&
  {Speagle}}]{Green_2021}
{Green}, G.~M., {Rix}, H.-W., {Tschesche}, L., {et~al.} 2021, \apj, 907, 57,
  \dodoi{10.3847/1538-4357/abd1dd}

\bibitem[{{Griffin} {et~al.}(2010){Griffin}, {Abergel}, {Abreu}, {Ade},
  {Andr{\'e}}, {Augueres}, {Babbedge}, {Bae}, {Baillie}, {Baluteau}, {Barlow},
  {Bendo}, {Benielli}, {Bock}, {Bonhomme}, {Brisbin}, {Brockley-Blatt},
  {Caldwell}, {Cara}, {Castro-Rodriguez}, {Cerulli}, {Chanial}, {Chen},
  {Clark}, {Clements}, {Clerc}, {Coker}, {Communal}, {Conversi}, {Cox},
  {Crumb}, {Cunningham}, {Daly}, {Davis}, {de Antoni}, {Delderfield}, {Devin},
  {di Giorgio}, {Didschuns}, {Dohlen}, {Donati}, {Dowell}, {Dowell}, {Duband},
  {Dumaye}, {Emery}, {Ferlet}, {Ferrand}, {Fontignie}, {Fox}, {Franceschini},
  {Frerking}, {Fulton}, {Garcia}, {Gastaud}, {Gear}, {Glenn}, {Goizel},
  {Griffin}, {Grundy}, {Guest}, {Guillemet}, {Hargrave}, {Harwit}, {Hastings},
  {Hatziminaoglou}, {Herman}, {Hinde}, {Hristov}, {Huang}, {Imhof}, {Isaak},
  {Israelsson}, {Ivison}, {Jennings}, {Kiernan}, {King}, {Lange}, {Latter},
  {Laurent}, {Laurent}, {Leeks}, {Lellouch}, {Levenson}, {Li}, {Li},
  {Lilienthal}, {Lim}, {Liu}, {Lu}, {Madden}, {Mainetti}, {Marliani}, {McKay},
  {Mercier}, {Molinari}, {Morris}, {Moseley}, {Mulder}, {Mur}, {Naylor},
  {Nguyen}, {O'Halloran}, {Oliver}, {Olofsson}, {Olofsson}, {Orfei}, {Page},
  {Pain}, {Panuzzo}, {Papageorgiou}, {Parks}, {Parr-Burman}, {Pearce},
  {Pearson}, {P{\'e}rez-Fournon}, {Pinsard}, {Pisano}, {Podosek}, {Pohlen},
  {Polehampton}, {Pouliquen}, {Rigopoulou}, {Rizzo}, {Roseboom}, {Roussel},
  {Rowan-Robinson}, {Rownd}, {Saraceno}, {Sauvage}, {Savage}, {Savini},
  {Sawyer}, {Scharmberg}, {Schmitt}, {Schneider}, {Schulz}, {Schwartz},
  {Shafer}, {Shupe}, {Sibthorpe}, {Sidher}, {Smith}, {Smith}, {Smith},
  {Spencer}, {Stobie}, {Sudiwala}, {Sukhatme}, {Surace}, {Stevens}, {Swinyard},
  {Trichas}, {Tourette}, {Triou}, {Tseng}, {Tucker}, {Turner}, {Vaccari},
  {Valtchanov}, {Vigroux}, {Virique}, {Voellmer}, {Walker}, {Ward}, {Waskett},
  {Weilert}, {Wesson}, {White}, {Whitehouse}, {Wilson}, {Winter}, {Woodcraft},
  {Wright}, {Xu}, {Zavagno}, {Zemcov}, {Zhang}, \& {Zonca}}]{Spire}
{Griffin}, M.~J., {Abergel}, A., {Abreu}, A., {et~al.} 2010, \aap, 518, L3,
  \dodoi{10.1051/0004-6361/201014519}

\bibitem[{{Han}(2017)}]{Han_2017}
{Han}, J.~L. 2017, \araa, 55, 111, \dodoi{10.1146/annurev-astro-091916-055221}

\bibitem[{Harris {et~al.}(2020)Harris, Millman, van~der Walt, Gommers,
  Virtanen, Cournapeau, Wieser, Taylor, Berg, Smith, Kern, Picus, Hoyer, van
  Kerkwijk, Brett, Haldane, del R{'{\i}}o, Wiebe, Peterson,
  G{'{e}}rard-Marchant, Sheppard, Reddy, Weckesser, Abbasi, Gohlke, \&
  Oliphant}]{numpy}
Harris, C.~R., Millman, K.~J., van~der Walt, S.~J., {et~al.} 2020, Nature, 585,
  357, \dodoi{10.1038/s41586-020-2649-2}

\bibitem[{{Indebetouw} {et~al.}(2005){Indebetouw}, {Mathis}, {Babler}, {Meade},
  {Watson}, {Whitney}, {Wolff}, {Wolfire}, {Cohen}, {Bania}, {Benjamin},
  {Clemens}, {Dickey}, {Jackson}, {Kobulnicky}, {Marston}, {Mercer},
  {Stauffer}, {Stolovy}, \& {Churchwell}}]{Indebetouw_2005}
{Indebetouw}, R., {Mathis}, J.~S., {Babler}, B.~L., {et~al.} 2005, \apj, 619,
  931, \dodoi{10.1086/426679}

\bibitem[{{Ivezi{\'c}} {et~al.}(2019){Ivezi{\'c}}, {Kahn}, {Tyson}, {Abel},
  {Acosta}, {Allsman}, {Alonso}, {AlSayyad}, {Anderson}, {Andrew}, {Angel},
  {Angeli}, {Ansari}, {Antilogus}, {Araujo}, {Armstrong}, {Arndt}, {Astier},
  {Aubourg}, {Auza}, {Axelrod}, {Bard}, {Barr}, {Barrau}, {Bartlett}, {Bauer},
  {Bauman}, {Baumont}, {Bechtol}, {Bechtol}, {Becker}, {Becla}, {Beldica},
  {Bellavia}, {Bianco}, {Biswas}, {Blanc}, {Blazek}, {Blandford}, {Bloom},
  {Bogart}, {Bond}, {Booth}, {Borgland}, {Borne}, {Bosch}, {Boutigny},
  {Brackett}, {Bradshaw}, {Brandt}, {Brown}, {Bullock}, {Burchat}, {Burke},
  {Cagnoli}, {Calabrese}, {Callahan}, {Callen}, {Carlin}, {Carlson},
  {Chandrasekharan}, {Charles-Emerson}, {Chesley}, {Cheu}, {Chiang}, {Chiang},
  {Chirino}, {Chow}, {Ciardi}, {Claver}, {Cohen-Tanugi}, {Cockrum}, {Coles},
  {Connolly}, {Cook}, {Cooray}, {Covey}, {Cribbs}, {Cui}, {Cutri}, {Daly},
  {Daniel}, {Daruich}, {Daubard}, {Daues}, {Dawson}, {Delgado}, {Dellapenna},
  {de Peyster}, {de Val-Borro}, {Digel}, {Doherty}, {Dubois},
  {Dubois-Felsmann}, {Durech}, {Economou}, {Eifler}, {Eracleous}, {Emmons},
  {Fausti Neto}, {Ferguson}, {Figueroa}, {Fisher-Levine}, {Focke}, {Foss},
  {Frank}, {Freemon}, {Gangler}, {Gawiser}, {Geary}, {Gee}, {Geha}, {Gessner},
  {Gibson}, {Gilmore}, {Glanzman}, {Glick}, {Goldina}, {Goldstein}, {Goodenow},
  {Graham}, {Gressler}, {Gris}, {Guy}, {Guyonnet}, {Haller}, {Harris},
  {Hascall}, {Haupt}, {Hernandez}, {Herrmann}, {Hileman}, {Hoblitt}, {Hodgson},
  {Hogan}, {Howard}, {Huang}, {Huffer}, {Ingraham}, {Innes}, {Jacoby}, {Jain},
  {Jammes}, {Jee}, {Jenness}, {Jernigan}, {Jevremovi{\'c}}, {Johns}, {Johnson},
  {Johnson}, {Jones}, {Juramy-Gilles}, {Juri{\'c}}, {Kalirai}, {Kallivayalil},
  {Kalmbach}, {Kantor}, {Karst}, {Kasliwal}, {Kelly}, {Kessler}, {Kinnison},
  {Kirkby}, {Knox}, {Kotov}, {Krabbendam}, {Krughoff}, {Kub{\'a}nek},
  {Kuczewski}, {Kulkarni}, {Ku}, {Kurita}, {Lage}, {Lambert}, {Lange},
  {Langton}, {Le Guillou}, {Levine}, {Liang}, {Lim}, {Lintott}, {Long},
  {Lopez}, {Lotz}, {Lupton}, {Lust}, {MacArthur}, {Mahabal}, {Mandelbaum},
  {Markiewicz}, {Marsh}, {Marshall}, {Marshall}, {May}, {McKercher}, {McQueen},
  {Meyers}, {Migliore}, {Miller}, {Mills}, {Miraval}, {Moeyens}, {Moolekamp},
  {Monet}, {Moniez}, {Monkewitz}, {Montgomery}, {Morrison}, {Mueller},
  {Muller}, {Mu{\~n}oz Arancibia}, {Neill}, {Newbry}, {Nief}, {Nomerotski},
  {Nordby}, {O'Connor}, {Oliver}, {Olivier}, {Olsen}, {O'Mullane}, {Ortiz},
  {Osier}, {Owen}, {Pain}, {Palecek}, {Parejko}, {Parsons}, {Pease},
  {Peterson}, {Peterson}, {Petravick}, {Libby Petrick}, {Petry},
  {Pierfederici}, {Pietrowicz}, {Pike}, {Pinto}, {Plante}, {Plate}, {Plutchak},
  {Price}, {Prouza}, {Radeka}, {Rajagopal}, {Rasmussen}, {Regnault}, {Reil},
  {Reiss}, {Reuter}, {Ridgway}, {Riot}, {Ritz}, {Robinson}, {Roby}, {Roodman},
  {Rosing}, {Roucelle}, {Rumore}, {Russo}, {Saha}, {Sassolas}, {Schalk},
  {Schellart}, {Schindler}, {Schmidt}, {Schneider}, {Schneider}, {Schoening},
  {Schumacher}, {Schwamb}, {Sebag}, {Selvy}, {Sembroski}, {Seppala}, {Serio},
  {Serrano}, {Shaw}, {Shipsey}, {Sick}, {Silvestri}, {Slater}, {Smith},
  {Smith}, {Sobhani}, {Soldahl}, {Storrie-Lombardi}, {Stover}, {Strauss},
  {Street}, {Stubbs}, {Sullivan}, {Sweeney}, {Swinbank}, {Szalay}, {Takacs},
  {Tether}, {Thaler}, {Thayer}, {Thomas}, {Thornton}, {Thukral}, {Tice},
  {Trilling}, {Turri}, {Van Berg}, {Vanden Berk}, {Vetter}, {Virieux},
  {Vucina}, {Wahl}, {Walkowicz}, {Walsh}, {Walter}, {Wang}, {Wang}, {Warner},
  {Wiecha}, {Willman}, {Winters}, {Wittman}, {Wolff}, {Wood-Vasey}, {Wu},
  {Xin}, {Yoachim}, \& {Zhan}}]{Ivezic_2019}
{Ivezi{\'c}}, {\v{Z}}., {Kahn}, S.~M., {Tyson}, J.~A., {et~al.} 2019, \apj,
  873, 111, \dodoi{10.3847/1538-4357/ab042c}

\bibitem[{{Juric}(2012)}]{Juric_2012}
{Juric}, M. 2012, {LSD: Large Survey Database framework}, Astrophysics Source
  Code Library, record ascl:1209.003.
\newblock \doeprint{1209.003}

\bibitem[{{Knollm{\"u}ller} \& {En{\ss}lin}(2019)}]{Knollmuller_2019}
{Knollm{\"u}ller}, J., \& {En{\ss}lin}, T.~A. 2019, arXiv e-prints,
  arXiv:1901.11033, \dodoi{10.48550/arXiv.1901.11033}

\bibitem[{{Kroupa}(2001)}]{Kroupa_2001}
{Kroupa}, P. 2001, \mnras, 322, 231, \dodoi{10.1046/j.1365-8711.2001.04022.x}

\bibitem[{{Kuhn} {et~al.}(2021){Kuhn}, {Benjamin}, {Zucker}, {Krone-Martins},
  {de Souza}, {Castro-Ginard}, {Ishida}, {Povich}, \&
  {Hillenbrand}}]{Kuhn_2021}
{Kuhn}, M.~A., {Benjamin}, R.~A., {Zucker}, C., {et~al.} 2021, \aap, 651, L10,
  \dodoi{10.1051/0004-6361/202141198}

\bibitem[{{Lallement} {et~al.}(2019){Lallement}, {Babusiaux}, {Vergely},
  {Katz}, {Arenou}, {Valette}, {Hottier}, \& {Capitanio}}]{Lallement_2019}
{Lallement}, R., {Babusiaux}, C., {Vergely}, J.~L., {et~al.} 2019, \aap, 625,
  A135, \dodoi{10.1051/0004-6361/201834695}

\bibitem[{{Lallement} {et~al.}(2022){Lallement}, {Vergely}, {Babusiaux}, \&
  {Cox}}]{Lallement_2022}
{Lallement}, R., {Vergely}, J.~L., {Babusiaux}, C., \& {Cox}, N.~L.~J. 2022,
  \aap, 661, A147, \dodoi{10.1051/0004-6361/202142846}

\bibitem[{{Lallement} {et~al.}(2018){Lallement}, {Capitanio}, {Ruiz-Dern},
  {Danielski}, {Babusiaux}, {Vergely}, {Elyajouri}, {Arenou}, \&
  {Leclerc}}]{Lallement_2018}
{Lallement}, R., {Capitanio}, L., {Ruiz-Dern}, L., {et~al.} 2018, \aap, 616,
  A132, \dodoi{10.1051/0004-6361/201832832}

\bibitem[{{Lang}(2014)}]{Lang_2014}
{Lang}, D. 2014, \aj, 147, 108, \dodoi{10.1088/0004-6256/147/5/108}

\bibitem[{{Lee} {et~al.}(2022){Lee}, {Whitmore}, {Thilker}, {Deger}, {Larson},
  {Ubeda}, {Anand}, {Boquien}, {Chandar}, {Dale}, {Emsellem}, {Leroy},
  {Rosolowsky}, {Schinnerer}, {Schmidt}, {Lilly}, {Turner}, {Van Dyk}, {White},
  {Barnes}, {Belfiore}, {Bigiel}, {Blanc}, {Cao}, {Chevance}, {Congiu},
  {Egorov}, {Glover}, {Grasha}, {Groves}, {Henshaw}, {Hughes}, {Klessen},
  {Koch}, {Kreckel}, {Kruijssen}, {Liu}, {Lopez}, {Mayker}, {Meidt}, {Murphy},
  {Pan}, {Pety}, {Querejeta}, {Razza}, {Saito}, {S{\'a}nchez-Bl{\'a}zquez},
  {Santoro}, {Sardone}, {Scheuermann}, {Schruba}, {Sun}, {Usero}, {Watkins}, \&
  {Williams}}]{Lee_2022}
{Lee}, J.~C., {Whitmore}, B.~C., {Thilker}, D.~A., {et~al.} 2022, \apjs, 258,
  10, \dodoi{10.3847/1538-4365/ac1fe5}

\bibitem[{{Leike} \& {En{\ss}lin}(2019)}]{Leike_2019}
{Leike}, R.~H., \& {En{\ss}lin}, T.~A. 2019, \aap, 631, A32,
  \dodoi{10.1051/0004-6361/201935093}

\bibitem[{{Leike} {et~al.}(2020){Leike}, {Glatzle}, \&
  {En{\ss}lin}}]{Leike_2020}
{Leike}, R.~H., {Glatzle}, M., \& {En{\ss}lin}, T.~A. 2020, \aap, 639, A138,
  \dodoi{10.1051/0004-6361/202038169}

\bibitem[{{Lenz} {et~al.}(2017){Lenz}, {Hensley}, \& {Dor{\'e}}}]{Lenz_2017}
{Lenz}, D., {Hensley}, B.~S., \& {Dor{\'e}}, O. 2017, \apj, 846, 38,
  \dodoi{10.3847/1538-4357/aa84af}

\bibitem[{{Lindegren} {et~al.}(2021){Lindegren}, {Bastian}, {Biermann},
  {Bombrun}, {de Torres}, {Gerlach}, {Geyer}, {Hern{\'a}ndez}, {Hilger},
  {Hobbs}, {Klioner}, {Lammers}, {McMillan}, {Ramos-Lerate},
  {Steidelm{\"u}ller}, {Stephenson}, \& {van Leeuwen}}]{Lindegren_2021_ZP}
{Lindegren}, L., {Bastian}, U., {Biermann}, M., {et~al.} 2021, \aap, 649, A4,
  \dodoi{10.1051/0004-6361/202039653}

\bibitem[{{Lindegren, L.} {et~al.}(2021){Lindegren, L.}, {Klioner, S. A.},
  {Hern\'andez, J.}, {Bombrun, A.}, {Ramos-Lerate, M.}, {Steidelm\"uller, H.},
  {Bastian, U.}, {Biermann, M.}, {de Torres, A.}, {Gerlach, E.}, {Geyer, R.},
  {Hilger, T.}, {Hobbs, D.}, {Lammers, U.}, {McMillan, P. J.}, {Stephenson, C.
  A.}, {Casta\~neda, J.}, {Davidson, M.}, {Fabricius, C.}, {Gracia-Abril, G.},
  {Portell, J.}, {Rowell, N.}, {Teyssier, D.}, {Torra, F.}, {Bartolom\'e, S.},
  {Clotet, M.}, {Garralda, N.}, {Gonz\'alez-Vidal, J. J.}, {Torra, J.}, {Abbas,
  U.}, {Altmann, M.}, {Anglada Varela, E.}, {Balaguer-N\'u\~nez, L.}, {Balog,
  Z.}, {Barache, C.}, {Becciani, U.}, {Bernet, M.}, {Bertone, S.}, {Bianchi,
  L.}, {Bouquillon, S.}, {Brown, A. G. A.}, {Bucciarelli, B.}, {Busonero, D.},
  {Butkevich, A. G.}, {Buzzi, R.}, {Cancelliere, R.}, {Carlucci, T.}, {Charlot,
  P.}, {Cioni, M.-R. L.}, {Crosta, M.}, {Crowley, C.}, {del Peloso, E. F.},
  {del Pozo, E.}, {Drimmel, R.}, {Esquej, P.}, {Fienga, A.}, {Fraile, E.},
  {Gai, M.}, {Garcia-Reinaldos, M.}, {Guerra, R.}, {Hambly, N. C.}, {Hauser,
  M.}, {Jan\ss{}en, K.}, {Jordan, S.}, {Kostrzewa-Rutkowska, Z.}, {Lattanzi, M.
  G.}, {Liao, S.}, {Licata, E.}, {Lister, T. A.}, {L\"offler, W.}, {Marchant,
  J. M.}, {Masip, A.}, {Mignard, F.}, {Mints, A.}, {Molina, D.}, {Mora, A.},
  {Morbidelli, R.}, {Murphy, C. P.}, {Pagani, C.}, {Panuzzo, P.}, {Pe\~nalosa
  Esteller, X.}, {Poggio, E.}, {Re Fiorentin, P.}, {Riva, A.}, {Sagrist\`a
  Sell\'es, A.}, {Sanchez Gimenez, V.}, {Sarasso, M.}, {Sciacca, E.},
  {Siddiqui, H. I.}, {Smart, R. L.}, {Souami, D.}, {Spagna, A.}, {Steele, I.
  A.}, {Taris, F.}, {Utrilla, E.}, {van Reeven, W.}, \& {Vecchiato,
  A.}}]{Gaia_EDR3_Astrometry}
{Lindegren, L.}, {Klioner, S. A.}, {Hern\'andez, J.}, {et~al.} 2021, A\&A, 649,
  A2, \dodoi{10.1051/0004-6361/202039709}

\bibitem[{{LSST}(2024)}]{lsst_recommendations_2024}
{LSST}. 2024, Survey Cadence Optimization Committee’s Phase 3
  Recommendations, \url{https://pstn-056.lsst.io/}

\bibitem[{{Marshall} {et~al.}(2025){Marshall}, {Montillaud}, {Cambresy}, \&
  {Cornu}}]{Marshall_2024}
{Marshall}, D., {Montillaud}, J., {Cambresy}, L., \& {Cornu}, D. 2025, {A New
  Dust Map of the Milky Way I : Principal Features}

\bibitem[{{Marshall} {et~al.}(2006){Marshall}, {Robin}, {Reyl{\'e}},
  {Schultheis}, \& {Picaud}}]{Marshall_2006}
{Marshall}, D.~J., {Robin}, A.~C., {Reyl{\'e}}, C., {Schultheis}, M., \&
  {Picaud}, S. 2006, \aap, 453, 635, \dodoi{10.1051/0004-6361:20053842}

\bibitem[{{Meisner} \& {Finkbeiner}(2015)}]{Meisner_Finkbeiner_2015}
{Meisner}, A.~M., \& {Finkbeiner}, D.~P. 2015, \apj, 798, 88,
  \dodoi{10.1088/0004-637X/798/2/88}

\bibitem[{{Minniti} {et~al.}(2010){Minniti}, {Lucas}, {Emerson}, {Saito},
  {Hempel}, {Pietrukowicz}, {Ahumada}, {Alonso}, {Alonso-Garcia}, {Arias},
  {Bandyopadhyay}, {Barb{\'a}}, {Barbuy}, {Bedin}, {Bica}, {Borissova},
  {Bronfman}, {Carraro}, {Catelan}, {Clari{\'a}}, {Cross}, {de Grijs},
  {D{\'e}k{\'a}ny}, {Drew}, {Fari{\~n}a}, {Feinstein}, {Fern{\'a}ndez
  Laj{\'u}s}, {Gamen}, {Geisler}, {Gieren}, {Goldman}, {Gonzalez}, {Gunthardt},
  {Gurovich}, {Hambly}, {Irwin}, {Ivanov}, {Jord{\'a}n}, {Kerins}, {Kinemuchi},
  {Kurtev}, {L{\'o}pez-Corredoira}, {Maccarone}, {Masetti}, {Merlo},
  {Messineo}, {Mirabel}, {Monaco}, {Morelli}, {Padilla}, {Palma}, {Parisi},
  {Pignata}, {Rejkuba}, {Roman-Lopes}, {Sale}, {Schreiber}, {Schr{\"o}der},
  {Smith}, {}, {Soto}, {Tamura}, {Tappert}, {Thompson}, {Toledo}, {Zoccali}, \&
  {Pietrzynski}}]{Minniti_2010}
{Minniti}, D., {Lucas}, P.~W., {Emerson}, J.~P., {et~al.} 2010, \na, 15, 433,
  \dodoi{10.1016/j.newast.2009.12.002}

\bibitem[{{Molinari} {et~al.}(2016){Molinari}, {Schisano}, {Elia},
  {Pestalozzi}, {Traficante}, {Pezzuto}, {Swinyard}, {Noriega-Crespo}, {Bally},
  {Moore}, {Plume}, {Zavagno}, {di Giorgio}, {Liu}, {Pilbratt}, {Mottram},
  {Russeil}, {Piazzo}, {Veneziani}, {Benedettini}, {Calzoletti}, {Faustini},
  {Natoli}, {Piacentini}, {Merello}, {Palmese}, {Del Grande}, {Polychroni},
  {Rygl}, {Polenta}, {Barlow}, {Bernard}, {Martin}, {Testi}, {Ali},
  {Andr{\'e}}, {Beltr{\'a}n}, {Billot}, {Carey}, {Cesaroni}, {Compi{\`e}gne},
  {Eden}, {Fukui}, {Garcia-Lario}, {Hoare}, {Huang}, {Joncas}, {Lim}, {Lord},
  {Martinavarro-Armengol}, {Motte}, {Paladini}, {Paradis}, {Peretto},
  {Robitaille}, {Schilke}, {Schneider}, {Schulz}, {Sibthorpe}, {Strafella},
  {Thompson}, {Umana}, {Ward-Thompson}, \& {Wyrowski}}]{Molinari_2016}
{Molinari}, S., {Schisano}, E., {Elia}, D., {et~al.} 2016, \aap, 591, A149,
  \dodoi{10.1051/0004-6361/201526380}

\bibitem[{{NASA}(2024)}]{nasa_wfi_technical_2024}
{NASA}. 2024, Wide Field Imager (WFI) Technical Information,
  \url{https://roman.gsfc.nasa.gov/science/WFI_technical.html}

\bibitem[{{Nataf} {et~al.}(2013){Nataf}, {Gould}, {Fouqu{\'e}}, {Gonzalez},
  {Johnson}, {Skowron}, {Udalski}, {Szyma{\'n}ski}, {Kubiak},
  {Pietrzy{\'n}ski}, {Soszy{\'n}ski}, {Ulaczyk}, {Wyrzykowski}, \&
  {Poleski}}]{Nataf_2013}
{Nataf}, D.~M., {Gould}, A., {Fouqu{\'e}}, P., {et~al.} 2013, \apj, 769, 88,
  \dodoi{10.1088/0004-637X/769/2/88}

\bibitem[{{Neugebauer} {et~al.}(1984){Neugebauer}, {Habing}, {van Duinen},
  {Aumann}, {Baud}, {Beichman}, {Beintema}, {Boggess}, {Clegg}, {de Jong},
  {Emerson}, {Gautier}, {Gillett}, {Harris}, {Hauser}, {Houck}, {Jennings},
  {Low}, {Marsden}, {Miley}, {Olnon}, {Pottasch}, {Raimond}, {Rowan-Robinson},
  {Soifer}, {Walker}, {Wesselius}, \& {Young}}]{IRAS}
{Neugebauer}, G., {Habing}, H.~J., {van Duinen}, R., {et~al.} 1984, \apjl, 278,
  L1, \dodoi{10.1086/184209}

\bibitem[{{Paladini} {et~al.}(2023){Paladini}, {Zucker}, {Benjamin}, {Nataf},
  {Minniti}, {Zasowski}, {Peek}, {Carey}, {Allen}, {Alonso-Garcia}, {Alves},
  {Anders}, {Athanassoula}, {Beers}, {Bird}, {Bland-Hwathorn}, {Brown},
  {Buder}, {Casagrande}, {Casey}, {Cassisi}, {Catelan}, {Chary}, {Chene},
  {Ciardi}, {Comeron}, {Cohen}, {Dame}, {Drimmel}, {Fernandez Trincado},
  {Finkbeiner}, {Geisler}, {Gennaro}, {Goodman}, {Green}, {Hajdu}, {Henderson},
  {Hora}, {Ivanov}, {Kirkpatrick}, {Kobayashi}, {Kuhn}, {Kunder}, {Lu},
  {Lucas}, {Majaess}, {Megeath}, {Meisner}, {Molinari}, {Mroz}, {Ness},
  {Neumayer}, {Nogueras-Lara}, {Noriega-Crespo}, {Poleski}, {Rix}, {Rebull},
  {Reggiani}, {Rejkuba}, {Saito}, {Schoenrich}, {Saydjari}, {Schisano},
  {Schlafly}, {Schlaufman}, {Smith}, {Speagle}, {Wisz}, {Wyse}, \&
  {Zakamska}}]{Paladini_2023}
{Paladini}, R., {Zucker}, C., {Benjamin}, R., {et~al.} 2023, arXiv e-prints,
  arXiv:2307.07642, \dodoi{10.48550/arXiv.2307.07642}

\bibitem[{{Planck Collaboration} {et~al.}(2014){Planck Collaboration},
  {Abergel}, {Ade}, {Aghanim}, {Alves}, {Aniano}, {Armitage-Caplan}, {Arnaud},
  {Ashdown}, {Atrio-Barandela}, {Aumont}, {Baccigalupi}, {Banday}, {Barreiro},
  {Bartlett}, {Battaner}, {Benabed}, {Beno{\^\i}t}, {Benoit-L{\'e}vy},
  {Bernard}, {Bersanelli}, {Bielewicz}, {Bobin}, {Bock}, {Bonaldi}, {Bond},
  {Borrill}, {Bouchet}, {Boulanger}, {Bridges}, {Bucher}, {Burigana}, {Butler},
  {Cardoso}, {Catalano}, {Chamballu}, {Chary}, {Chiang}, {Chiang},
  {Christensen}, {Church}, {Clemens}, {Clements}, {Colombi}, {Colombo},
  {Combet}, {Couchot}, {Coulais}, {Crill}, {Curto}, {Cuttaia}, {Danese},
  {Davies}, {Davis}, {de Bernardis}, {de Rosa}, {de Zotti}, {Delabrouille},
  {Delouis}, {D{\'e}sert}, {Dickinson}, {Diego}, {Dole}, {Donzelli},
  {Dor{\'e}}, {Douspis}, {Draine}, {Dupac}, {Efstathiou}, {En{\ss}lin},
  {Eriksen}, {Falgarone}, {Finelli}, {Forni}, {Frailis}, {Fraisse},
  {Franceschi}, {Galeotta}, {Ganga}, {Ghosh}, {Giard}, {Giardino},
  {Giraud-H{\'e}raud}, {Gonz{\'a}lez-Nuevo}, {G{\'o}rski}, {Gratton},
  {Gregorio}, {Grenier}, {Gruppuso}, {Guillet}, {Hansen}, {Hanson}, {Harrison},
  {Helou}, {Henrot-Versill{\'e}}, {Hern{\'a}ndez-Monteagudo}, {Herranz},
  {Hildebrandt}, {Hivon}, {Hobson}, {Holmes}, {Hornstrup}, {Hovest},
  {Huffenberger}, {Jaffe}, {Jaffe}, {Jewell}, {Joncas}, {Jones}, {Juvela},
  {Keih{\"a}nen}, {Keskitalo}, {Kisner}, {Knoche}, {Knox}, {Kunz},
  {Kurki-Suonio}, {Lagache}, {L{\"a}hteenm{\"a}ki}, {Lamarre}, {Lasenby},
  {Laureijs}, {Lawrence}, {Leonardi}, {Le{\'o}n-Tavares}, {Lesgourgues},
  {Levrier}, {Liguori}, {Lilje}, {Linden-V{\o}rnle}, {L{\'o}pez-Caniego},
  {Lubin}, {Mac{\'\i}as-P{\'e}rez}, {Maffei}, {Maino}, {Mandolesi}, {Maris},
  {Marshall}, {Martin}, {Mart{\'\i}nez-Gonz{\'a}lez}, {Masi}, {Massardi},
  {Matarrese}, {Matthai}, {Mazzotta}, {McGehee}, {Melchiorri}, {Mendes},
  {Mennella}, {Migliaccio}, {Mitra}, {Miville-Desch{\^e}nes}, {Moneti},
  {Montier}, {Morgante}, {Mortlock}, {Munshi}, {Murphy}, {Naselsky}, {Nati},
  {Natoli}, {Netterfield}, {N{\o}rgaard-Nielsen}, {Noviello}, {Novikov},
  {Novikov}, {Osborne}, {Oxborrow}, {Paci}, {Pagano}, {Pajot}, {Paladini},
  {Paoletti}, {Pasian}, {Patanchon}, {Perdereau}, {Perotto}, {Perrotta},
  {Piacentini}, {Piat}, {Pierpaoli}, {Pietrobon}, {Plaszczynski},
  {Pointecouteau}, {Polenta}, {Ponthieu}, {Popa}, {Poutanen}, {Pratt},
  {Pr{\'e}zeau}, {Prunet}, {Puget}, {Rachen}, {Reach}, {Rebolo}, {Reinecke},
  {Remazeilles}, {Renault}, {Ricciardi}, {Riller}, {Ristorcelli}, {Rocha},
  {Rosset}, {Roudier}, {Rowan-Robinson}, {Rubi{\~n}o-Mart{\'\i}n}, {Rusholme},
  {Sandri}, {Santos}, {Savini}, {Scott}, {Seiffert}, {Shellard}, {Spencer},
  {Starck}, {Stolyarov}, {Stompor}, {Sudiwala}, {Sunyaev}, {Sureau}, {Sutton},
  {Suur-Uski}, {Sygnet}, {Tauber}, {Tavagnacco}, {Terenzi}, {Toffolatti},
  {Tomasi}, {Tristram}, {Tucci}, {Tuovinen}, {T{\"u}rler}, {Umana},
  {Valenziano}, {Valiviita}, {Van Tent}, {Verstraete}, {Vielva}, {Villa},
  {Vittorio}, {Wade}, {Wandelt}, {Welikala}, {Ysard}, {Yvon}, {Zacchei}, \&
  {Zonca}}]{Planck_2014}
{Planck Collaboration}, {Abergel}, A., {Ade}, P.~A.~R., {et~al.} 2014, \aap,
  571, A11, \dodoi{10.1051/0004-6361/201323195}

\bibitem[{{Planck Collaboration} {et~al.}(2016){Planck Collaboration}, {Adam},
  {Ade}, {Aghanim}, {Alves}, {Arnaud}, {Ashdown}, {Aumont}, {Baccigalupi},
  {Banday}, {Barreiro}, {Bartlett}, {Bartolo}, {Battaner}, {Benabed},
  {Beno{\^\i}t}, {Benoit-L{\'e}vy}, {Bernard}, {Bersanelli}, {Bielewicz},
  {Bock}, {Bonaldi}, {Bonavera}, {Bond}, {Borrill}, {Bouchet}, {Boulanger},
  {Bucher}, {Burigana}, {Butler}, {Calabrese}, {Cardoso}, {Catalano},
  {Challinor}, {Chamballu}, {Chary}, {Chiang}, {Christensen}, {Clements},
  {Colombi}, {Colombo}, {Combet}, {Couchot}, {Coulais}, {Crill}, {Curto},
  {Cuttaia}, {Danese}, {Davies}, {Davis}, {de Bernardis}, {de Rosa}, {de
  Zotti}, {Delabrouille}, {D{\'e}sert}, {Dickinson}, {Diego}, {Dole},
  {Donzelli}, {Dor{\'e}}, {Douspis}, {Ducout}, {Dupac}, {Efstathiou}, {Elsner},
  {En{\ss}lin}, {Eriksen}, {Falgarone}, {Fergusson}, {Finelli}, {Forni},
  {Frailis}, {Fraisse}, {Franceschi}, {Frejsel}, {Galeotta}, {Galli}, {Ganga},
  {Ghosh}, {Giard}, {Giraud-H{\'e}raud}, {Gjerl{\o}w}, {Gonz{\'a}lez-Nuevo},
  {G{\'o}rski}, {Gratton}, {Gregorio}, {Gruppuso}, {Gudmundsson}, {Hansen},
  {Hanson}, {Harrison}, {Helou}, {Henrot-Versill{\'e}},
  {Hern{\'a}ndez-Monteagudo}, {Herranz}, {Hildebrandt}, {Hivon}, {Hobson},
  {Holmes}, {Hornstrup}, {Hovest}, {Huffenberger}, {Hurier}, {Jaffe}, {Jaffe},
  {Jones}, {Juvela}, {Keih{\"a}nen}, {Keskitalo}, {Kisner}, {Kneissl},
  {Knoche}, {Kunz}, {Kurki-Suonio}, {Lagache}, {L{\"a}hteenm{\"a}ki},
  {Lamarre}, {Lasenby}, {Lattanzi}, {Lawrence}, {Le Jeune}, {Leahy},
  {Leonardi}, {Lesgourgues}, {Levrier}, {Liguori}, {Lilje}, {Linden-V{\o}rnle},
  {L{\'o}pez-Caniego}, {Lubin}, {Mac{\'\i}as-P{\'e}rez}, {Maggio}, {Maino},
  {Mandolesi}, {Mangilli}, {Maris}, {Marshall}, {Martin},
  {Mart{\'\i}nez-Gonz{\'a}lez}, {Masi}, {Matarrese}, {McGehee}, {Meinhold},
  {Melchiorri}, {Mendes}, {Mennella}, {Migliaccio}, {Mitra},
  {Miville-Desch{\^e}nes}, {Moneti}, {Montier}, {Morgante}, {Mortlock}, {Moss},
  {Munshi}, {Murphy}, {Naselsky}, {Nati}, {Natoli}, {Netterfield},
  {N{\o}rgaard-Nielsen}, {Noviello}, {Novikov}, {Novikov}, {Orlando},
  {Oxborrow}, {Paci}, {Pagano}, {Pajot}, {Paladini}, {Paoletti}, {Partridge},
  {Pasian}, {Patanchon}, {Pearson}, {Perdereau}, {Perotto}, {Perrotta},
  {Pettorino}, {Piacentini}, {Piat}, {Pierpaoli}, {Pietrobon}, {Plaszczynski},
  {Pointecouteau}, {Polenta}, {Pratt}, {Pr{\'e}zeau}, {Prunet}, {Puget},
  {Rachen}, {Reach}, {Rebolo}, {Reinecke}, {Remazeilles}, {Renault}, {Renzi},
  {Ristorcelli}, {Rocha}, {Rosset}, {Rossetti}, {Roudier},
  {Rubi{\~n}o-Mart{\'\i}n}, {Rusholme}, {Sandri}, {Santos}, {Savelainen},
  {Savini}, {Scott}, {Seiffert}, {Shellard}, {Spencer}, {Stolyarov}, {Stompor},
  {Strong}, {Sudiwala}, {Sunyaev}, {Sutton}, {Suur-Uski}, {Sygnet}, {Tauber},
  {Terenzi}, {Toffolatti}, {Tomasi}, {Tristram}, {Tucci}, {Tuovinen}, {Umana},
  {Valenziano}, {Valiviita}, {Van Tent}, {Vielva}, {Villa}, {Wade}, {Wandelt},
  {Wehus}, {Wilkinson}, {Yvon}, {Zacchei}, \& {Zonca}}]{Planck_2016}
{Planck Collaboration}, {Adam}, R., {Ade}, P.~A.~R., {et~al.} 2016, \aap, 594,
  A10, \dodoi{10.1051/0004-6361/201525967}

\bibitem[{{Reid} {et~al.}(2019){Reid}, {Menten}, {Brunthaler}, {Zheng}, {Dame},
  {Xu}, {Li}, {Sakai}, {Wu}, {Immer}, {Zhang}, {Sanna}, {Moscadelli}, {Rygl},
  {Bartkiewicz}, {Hu}, {Quiroga-Nu{\~n}ez}, \& {van Langevelde}}]{Reid_2019}
{Reid}, M.~J., {Menten}, K.~M., {Brunthaler}, A., {et~al.} 2019, \apj, 885,
  131, \dodoi{10.3847/1538-4357/ab4a11}

\bibitem[{{Rezaei Kh.} {et~al.}(2024){Rezaei Kh.}, {Beuther}, {Benjamin},
  {Eilers}, {Henning}, {Jim{\'e}nez-Donaire}, \&
  {Miville-Desch{\^e}nes}}]{Rezaei_2024}
{Rezaei Kh.}, S., {Beuther}, H., {Benjamin}, R.~A., {et~al.} 2024, arXiv
  e-prints, arXiv:2405.09634, \dodoi{10.48550/arXiv.2405.09634}

\bibitem[{{Riello} {et~al.}(2021){Riello}, {De Angeli}, {Evans}, {Montegriffo},
  {Carrasco}, {Busso}, {Palaversa}, {Burgess}, {Diener}, {Davidson}, {Rowell},
  {Fabricius}, {Jordi}, {Bellazzini}, {Pancino}, {Harrison}, {Cacciari}, {van
  Leeuwen}, {Hambly}, {Hodgkin}, {Osborne}, {Altavilla}, {Barstow}, {Brown},
  {Castellani}, {Cowell}, {De Luise}, {Gilmore}, {Giuffrida}, {Hidalgo},
  {Holland}, {Marinoni}, {Pagani}, {Piersimoni}, {Pulone}, {Ragaini}, {Rainer},
  {Richards}, {Sanna}, {Walton}, {Weiler}, \& {Yoldas}}]{Riello_2021}
{Riello}, M., {De Angeli}, F., {Evans}, D.~W., {et~al.} 2021, \aap, 649, A3,
  \dodoi{10.1051/0004-6361/202039587}

\bibitem[{{Robin} {et~al.}(2003){Robin}, {Reyl{\'e}}, {Derri{\`e}re}, \&
  {Picaud}}]{Robin_2003}
{Robin}, A.~C., {Reyl{\'e}}, C., {Derri{\`e}re}, S., \& {Picaud}, S. 2003,
  \aap, 409, 523, \dodoi{10.1051/0004-6361:20031117}

\bibitem[{{Robitaille} {et~al.}(2019){Robitaille}, {Beaumont}, {Qian},
  {Borkin}, \& {Goodman}}]{glueviz}
{Robitaille}, T., {Beaumont}, C., {Qian}, P., {Borkin}, M., \& {Goodman}, A.
  2019, {glueviz v0.15.2: multidimensional data exploration}, 0.15.2,  Zenodo,
  \dodoi{10.5281/zenodo.3385920}

\bibitem[{{Rybizki} {et~al.}(2018){Rybizki}, {Demleitner}, {Fouesneau},
  {Bailer-Jones}, {Rix}, \& {Andrae}}]{Rybizki_2018}
{Rybizki}, J., {Demleitner}, M., {Fouesneau}, M., {et~al.} 2018, \pasp, 130,
  074101, \dodoi{10.1088/1538-3873/aabd70}

\bibitem[{{Saito} {et~al.}(2024){Saito}, {Hempel}, {Alonso-Garc{\'\i}a},
  {Lucas}, {Minniti}, {Alonso}, {Baravalle}, {Borissova}, {Caceres},
  {Chen{\'e}}, {Cross}, {Duplancic}, {Garro}, {G{\'o}mez}, {Ivanov}, {Kurtev},
  {Luna}, {Majaess}, {Navarro}, {Pullen}, {Rejkuba}, {Sanders}, {Smith},
  {Albino}, {Alonso}, {Am{\^o}res}, {Angeloni}, {Arias}, {Arnaboldi}, {Barbuy},
  {Bayo}, {Beamin}, {Bedin}, {Bellini}, {Benjamin}, {Bica}, {Bonatto}, {Botan},
  {Braga}, {Brown}, {Cabral}, {Camargo}, {Caratti o Garatti}, {Carballo-Bello},
  {Catelan}, {Chavero}, {Chijani}, {Clari{\'a}}, {Coldwell}, {Pe{\~n}a},
  {Ramos}, {Corral-Santana}, {Cort{\'e}s}, {Cort{\'e}s-Contreras}, {Cruz},
  {Daza-Perilla}, {Debattista}, {Dias}, {Donoso}, {D'Souza}, {Emerson},
  {Federle}, {Fermiano}, {Fernandez}, {Fern{\'a}ndez-Trincado}, {Ferreira},
  {Lopes}, {Firpo}, {Flores-Quintana}, {Fraga}, {Froebrich}, {Galdeano},
  {Gavignaud}, {Geisler}, {Gerhard}, {Gieren}, {Gonzalez}, {Gramajo}, {Gran},
  {Granitto}, {Griggio}, {Guo}, {Gurovich}, {Hilker}, {Jones}, {Kammers},
  {Kuhn}, {Kumar}, {Kundu}, {Lares}, {Libralato}, {Lima}, {Maccarone},
  {Cort{\'e}s}, {Martin}, {Masetti}, {Matsunaga}, {Mauro}, {McDonald},
  {Mej{\'\i}as}, {Mesa}, {Milla-Castro}, {Minniti}, {Bidin}, {Montenegro},
  {Morris}, {Motta}, {Navarete}, {Molina}, {Nikzat}, {Castell{\'o}n}, {Obasi},
  {Ortigoza-Urdaneta}, {Palma}, {Parisi}, {Ram{\'\i}rez}, {Pereyra}, {Perez},
  {Petralia}, {Pichel}, {Pignata}, {Alegr{\'\i}a}, {Rojas}, {Rojas},
  {Roman-Lopes}, {Rovero}, {Saroon}, {Schmidt}, {Schr{\"o}der}, {Schultheis},
  {Sgr{\'o}}, {Solano}, {Soto}, {Stecklum}, {Steeghs}, {Tamura}, {Tissera},
  {Valcarce}, {Valotto}, {Vasquez}, {Villalon}, {Villanova}, {C{\'a}diz},
  {Bacigalupo}, {Zijlstra}, \& {Zoccali}}]{Saito_2024}
{Saito}, R.~K., {Hempel}, M., {Alonso-Garc{\'\i}a}, J., {et~al.} 2024, \aap,
  689, A148, \dodoi{10.1051/0004-6361/202450584}

\bibitem[{{Sale} {et~al.}(2014){Sale}, {Drew}, {Barentsen}, {Farnhill},
  {Raddi}, {Barlow}, {Eisl{\"o}ffel}, {Vink}, {Rodr{\'\i}guez-Gil}, \&
  {Wright}}]{Sale_2014}
{Sale}, S.~E., {Drew}, J.~E., {Barentsen}, G., {et~al.} 2014, \mnras, 443,
  2907, \dodoi{10.1093/mnras/stu1090}

\bibitem[{{Sanderson} {et~al.}(2024){Sanderson}, {Hickox}, {Hirata}, {Holman},
  {Lu}, \& {Villar}}]{Sanderson_2024}
{Sanderson}, R.~E., {Hickox}, R., {Hirata}, C.~M., {et~al.} 2024, arXiv
  e-prints, arXiv:2404.14342, \dodoi{10.48550/arXiv.2404.14342}

\bibitem[{{Saydjari} \& {Finkbeiner}(2022)}]{Saydjari_2022}
{Saydjari}, A.~K., \& {Finkbeiner}, D.~P. 2022, \apj, 933, 155,
  \dodoi{10.3847/1538-4357/ac6875}

\bibitem[{{Saydjari} {et~al.}(2023{\natexlab{a}}){Saydjari}, {M. Uzsoy},
  {Zucker}, {Peek}, \& {Finkbeiner}}]{Saydjari_2023_DIBs}
{Saydjari}, A.~K., {M. Uzsoy}, A.~S., {Zucker}, C., {Peek}, J.~E.~G., \&
  {Finkbeiner}, D.~P. 2023{\natexlab{a}}, \apj, 954, 141,
  \dodoi{10.3847/1538-4357/acd454}

\bibitem[{{Saydjari} {et~al.}(2023{\natexlab{b}}){Saydjari}, {Schlafly},
  {Lang}, {Meisner}, {Green}, {Zucker}, {Zelko}, {Speagle}, {Daylan}, {Lee},
  {Valdes}, {Schlegel}, \& {Finkbeiner}}]{Saydjari_2023}
{Saydjari}, A.~K., {Schlafly}, E.~F., {Lang}, D., {et~al.} 2023{\natexlab{b}},
  \apjs, 264, 28, \dodoi{10.3847/1538-4365/aca594}

\bibitem[{{Schechter} {et~al.}(1993){Schechter}, {Mateo}, \&
  {Saha}}]{DoPhot_1993}
{Schechter}, P.~L., {Mateo}, M., \& {Saha}, A. 1993, \pasp, 105, 1342,
  \dodoi{10.1086/133316}

\bibitem[{{Schlafly}(2021)}]{Schlafly_2021}
{Schlafly}, E.~F. 2021, {crowdsource: Crowded field photometry pipeline},
  Astrophysics Source Code Library, record ascl:2106.004.
\newblock \doeprint{2106.004}

\bibitem[{{Schlafly} \& {Finkbeiner}(2011)}]{Schlafly_2011}
{Schlafly}, E.~F., \& {Finkbeiner}, D.~P. 2011, \apj, 737, 103,
  \dodoi{10.1088/0004-637X/737/2/103}

\bibitem[{{Schlafly} {et~al.}(2019){Schlafly}, {Meisner}, \&
  {Green}}]{Schlafly_2019}
{Schlafly}, E.~F., {Meisner}, A.~M., \& {Green}, G.~M. 2019, \apjs, 240, 30,
  \dodoi{10.3847/1538-4365/aafbea}

\bibitem[{{Schlafly} {et~al.}(2016){Schlafly}, {Meisner}, {Stutz},
  {Kainulainen}, {Peek}, {Tchernyshyov}, {Rix}, {Finkbeiner}, {Covey}, {Green},
  {Bell}, {Burgett}, {Chambers}, {Draper}, {Flewelling}, {Hodapp}, {Kaiser},
  {Magnier}, {Martin}, {Metcalfe}, {Wainscoat}, \& {Waters}}]{Schlafly_2016}
{Schlafly}, E.~F., {Meisner}, A.~M., {Stutz}, A.~M., {et~al.} 2016, \apj, 821,
  78, \dodoi{10.3847/0004-637X/821/2/78}

\bibitem[{{Schlafly} {et~al.}(2018){Schlafly}, {Green}, {Lang}, {Daylan},
  {Finkbeiner}, {Lee}, {Meisner}, {Schlegel}, \& {Valdes}}]{Schlafly_2018}
{Schlafly}, E.~F., {Green}, G.~M., {Lang}, D., {et~al.} 2018, \apjs, 234, 39,
  \dodoi{10.3847/1538-4365/aaa3e2}

\bibitem[{{Schlegel} {et~al.}(1998){Schlegel}, {Finkbeiner}, \& {Davis}}]{SFD}
{Schlegel}, D.~J., {Finkbeiner}, D.~P., \& {Davis}, M. 1998, \apj, 500, 525,
  \dodoi{10.1086/305772}

\bibitem[{{Sellgren}(1984)}]{Sellgren_1984}
{Sellgren}, K. 1984, \apj, 277, 623, \dodoi{10.1086/161733}

\bibitem[{{Skrutskie} {et~al.}(2006){Skrutskie}, {Cutri}, {Stiening},
  {Weinberg}, {Schneider}, {Carpenter}, {Beichman}, {Capps}, {Chester},
  {Elias}, {Huchra}, {Liebert}, {Lonsdale}, {Monet}, {Price}, {Seitzer},
  {Jarrett}, {Kirkpatrick}, {Gizis}, {Howard}, {Evans}, {Fowler}, {Fullmer},
  {Hurt}, {Light}, {Kopan}, {Marsh}, {McCallon}, {Tam}, {Van Dyk}, \&
  {Wheelock}}]{Skrutskie_2006}
{Skrutskie}, M.~F., {Cutri}, R.~M., {Stiening}, R., {et~al.} 2006, \aj, 131,
  1163, \dodoi{10.1086/498708}

\bibitem[{{Speagle}(2019)}]{Speagle_2019}
{Speagle}, J.~S. 2019, arXiv e-prints, arXiv:1909.12313,
  \dodoi{10.48550/arXiv.1909.12313}

\bibitem[{{Speagle} {et~al.}(2024){Speagle}, {Zucker}, {Bonaca}, {Cargile},
  {Johnson}, {Beane}, {Conroy}, {Finkbeiner}, {Green}, {Kamdar}, {Naidu},
  {Rix}, {Schlafly}, {Dotter}, {Eadie}, {Eisenstein}, {Goodman}, {Han},
  {Saydjari}, {Ting}, \& {Zelko}}]{Speagle_2023b}
{Speagle}, J.~S., {Zucker}, C., {Bonaca}, A., {et~al.} 2024, \apj, 970, 121,
  \dodoi{10.3847/1538-4357/ad2b62}

\bibitem[{{Speagle} {et~al.}(2025){Speagle}, {Zucker}, {Bonaca}, {Cargile},
  {Johnson}, {Beane}, {Conroy}, {Finkbeiner}, {Green}, {Kamdar}, {Naidu},
  {Rix}, {Schlafly}, {Dotter}, {Eadie}, {Eisenstein}, {Goodman}, {Han},
  {Saydjari}, {Ting}, \& {Zelko}}]{Speagle_2023a}
---. 2025, {Deriving Stellar Properties, Distances, and Reddenings using
  Photometry and Astrometry with BRUTUS}

\bibitem[{{Stetson}(1987)}]{DaoPHOT_1987}
{Stetson}, P.~B. 1987, \pasp, 99, 191, \dodoi{10.1086/131977}

\bibitem[{{Surot} {et~al.}(2020){Surot}, {Valenti}, {Gonzalez}, {Zoccali},
  {S{\"o}kmen}, {Hidalgo}, \& {Minniti}}]{Surot_2020}
{Surot}, F., {Valenti}, E., {Gonzalez}, O.~A., {et~al.} 2020, \aap, 644, A140,
  \dodoi{10.1051/0004-6361/202038346}

\bibitem[{{Tchernyshyov} {et~al.}(2018){Tchernyshyov}, {Peek}, \&
  {Zasowski}}]{KTII}
{Tchernyshyov}, K., {Peek}, J.~E.~G., \& {Zasowski}, G. 2018, \aj, 156, 248,
  \dodoi{10.3847/1538-3881/aae68d}

\bibitem[{{Vergely} {et~al.}(2022){Vergely}, {Lallement}, \&
  {Cox}}]{Vergely_2022}
{Vergely}, J.~L., {Lallement}, R., \& {Cox}, N.~L.~J. 2022, \aap, 664, A174,
  \dodoi{10.1051/0004-6361/202243319}

\bibitem[{Virtanen {et~al.}(2020)Virtanen, Gommers, Oliphant, Haberland, Reddy,
  Cournapeau, Burovski, Peterson, Weckesser, Bright, {van der Walt}, Brett,
  Wilson, Millman, Mayorov, Nelson, Jones, Kern, Larson, Carey, Polat, Feng,
  Moore, {VanderPlas}, Laxalde, Perktold, Cimrman, Henriksen, Quintero, Harris,
  Archibald, Ribeiro, Pedregosa, {van Mulbregt}, \& {SciPy 1.0
  Contributors}}]{scipy}
Virtanen, P., Gommers, R., Oliphant, T.~E., {et~al.} 2020, Nature Methods, 17,
  261, \dodoi{10.1038/s41592-019-0686-2}

\bibitem[{{Wang} \& {Chen}(2019)}]{Wang_2019}
{Wang}, S., \& {Chen}, X. 2019, \apj, 877, 116,
  \dodoi{10.3847/1538-4357/ab1c61}

\bibitem[{{Weingartner} \& {Draine}(2001)}]{Weingartner_2001}
{Weingartner}, J.~C., \& {Draine}, B.~T. 2001, \apjs, 134, 263,
  \dodoi{10.1086/320852}

\bibitem[{{Xue} {et~al.}(2015){Xue}, {Rix}, {Ma}, {Morrison}, {Bovy}, {Sesar},
  \& {Janesh}}]{Xue_2015}
{Xue}, X.-X., {Rix}, H.-W., {Ma}, Z., {et~al.} 2015, \apj, 809, 144,
  \dodoi{10.1088/0004-637X/809/2/144}

\bibitem[{{Zhang} \& {Kainulainen}(2019)}]{Zhang_2019}
{Zhang}, M., \& {Kainulainen}, J. 2019, \aap, 632, A85,
  \dodoi{10.1051/0004-6361/201935513}

\bibitem[{{Zhang} \& {Kainulainen}(2022)}]{Zhang_Kainulainen_2022}
---. 2022, \mnras, 517, 5180, \dodoi{10.1093/mnras/stac3012}

\bibitem[{{Zhang} \& {Green}(2024)}]{Zhang_2024}
{Zhang}, X., \& {Green}, G. 2024, arXiv e-prints, arXiv:2407.14594,
  \dodoi{10.48550/arXiv.2407.14594}

\bibitem[{{Zhang} {et~al.}(2023){Zhang}, {Green}, \& {Rix}}]{Zhang_2023}
{Zhang}, X., {Green}, G.~M., \& {Rix}, H.-W. 2023, \mnras, 524, 1855,
  \dodoi{10.1093/mnras/stad1941}

\bibitem[{Zonca {et~al.}(2019)Zonca, Singer, Lenz, Reinecke, Rosset, Hivon, \&
  Gorski}]{healpy_one}
Zonca, A., Singer, L., Lenz, D., {et~al.} 2019, Journal of Open Source
  Software, 4, 1298, \dodoi{10.21105/joss.01298}

\bibitem[{{Zucker} {et~al.}(2019){Zucker}, {Speagle}, {Schlafly}, {Green},
  {Finkbeiner}, {Goodman}, \& {Alves}}]{Zucker_Speagle_2019}
{Zucker}, C., {Speagle}, J.~S., {Schlafly}, E.~F., {et~al.} 2019, \apj, 879,
  125, \dodoi{10.3847/1538-4357/ab2388}

\bibitem[{{Zucker} {et~al.}(2021){Zucker}, {Goodman}, {Alves}, {Bialy}, {Koch},
  {Speagle}, {Foley}, {Finkbeiner}, {Leike}, {En{\ss}lin}, {Peek}, \&
  {Edenhofer}}]{Zucker_2021}
{Zucker}, C., {Goodman}, A., {Alves}, J., {et~al.} 2021, \apj, 919, 35,
  \dodoi{10.3847/1538-4357/ac1f96}

\end{thebibliography}
\bibliographystyle{aasjournal}

\appendix

\section{Artifacts} \label{sec:artifacts}
Here we highlight several artifacts in our 3D dust map that potential users should be cognizant of. Many of these artifacts preferentially manifest in certain parts of the sky and/or over a certain range in distances. Likewise, some artifacts may only be visible if the map is stretched in a particularly extreme way. A few of these artifacts are captured by our accompanying bitmask, including for example, the known limitations of the map outside our minimum and maximum reliable distance range. 

\begin{figure*}[ht!]
\begin{center}
\includegraphics[width=1.\textwidth]{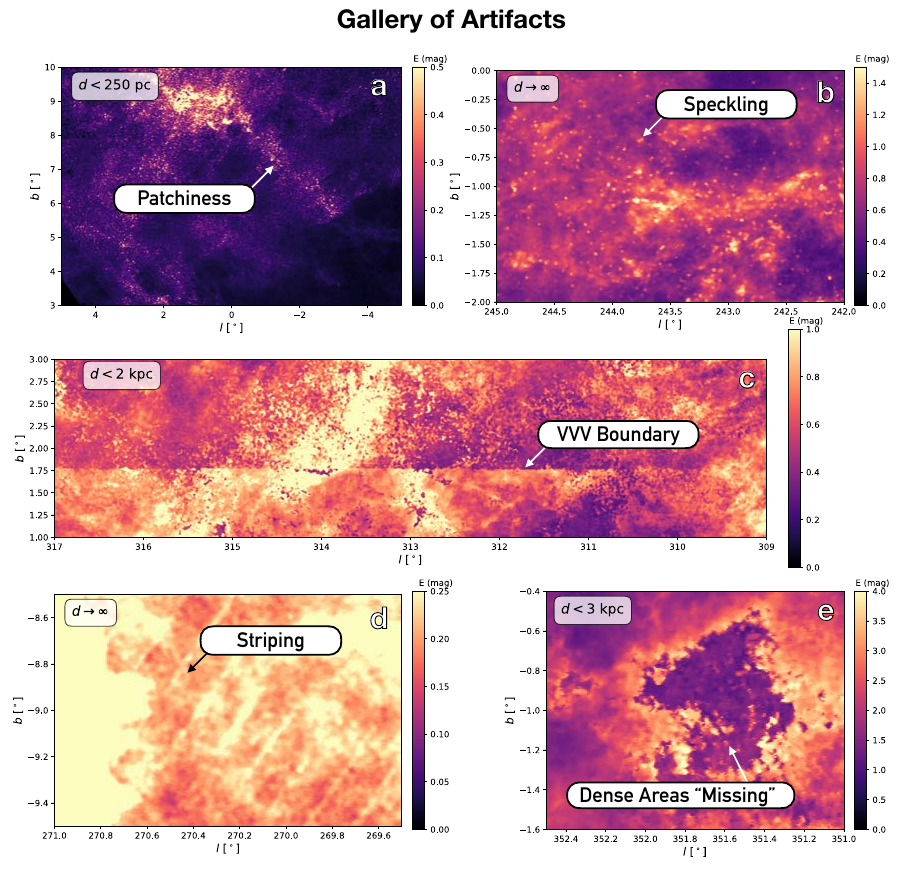}
\end{center}
\caption{Gallery of known artifacts in the DECaPS 3D dust map. Each panel highlights a different artifact, where the stretch and the distance integration range (summarized in the top left corner of each panel) were chosen to highlight each artifact in its most extreme form. See \S \ref{sec:artifacts} for a description of the artifacts.} 
\label{fig:artifacts}
\end{figure*}

In Figure \ref{fig:artifacts}, we show a gallery of known artifacts, where each panel highlights a different artifact. In panel A, we show the limitations of our map at very close distances towards a local molecular cloud, using the Pipe nebula \citep[$d=152$ pc;][]{Zucker_2021} as an example. We integrate our map out to $d=200$ pc (beyond the nominal distance of the Pipe nebula) to show that while we recover the gross morphology of the nebula, the map is very patchy at close distances. The patchiness stems from a combination of our M-dwarf cut (removing stars foreground to the cloud; see \S \ref{subsec:filtering}), our high-angular resolution (see \S \ref{subsec:pixelization}), and our lack of a spatial regularization scheme to correlate distance slices across neighboring pixels. This artifact is typically captured by our minimum reliable distance bitmask. 

In panel B we show a speckling that appears in our map over a small fraction of the sky when integrated out to large distances. As an example, we show a piece of the Herschel filament originally seen in Figure \ref{fig:herschel_comp} but instead of integrating out to $d=7 \; \rm kpc$ (where no artifacts are seen), we integrate out to the edge of the map. We are unable to explain the root cause of this artifact, but note that it is typically captured by our maximum reliable distance bitmask. 

In Panel C, we highlight the artifact associated with the VVV boundary. We show an $8^\circ$ longitude strip of the Galactic plane integrated out to $d=2\;$ kpc, where the VVV boundary is clearly visible at $b=+1.75^\circ$. This artifact is also visible over a broader distance range, manifesting between $d=1-5 \; \rm kpc$. This artifacts stems from the fact that not only do we detect more highly reddened stars with VVV compared to 2MASS, but the inclusion of VVV photometry alongside 2MASS photometry (in regions of overlap) can modify the underlying distance-extinction posterior, which can cause clouds to be placed at slightly distances above and below the boundary. 

In Panel D, we show a striping artifact that appears at the edges of the DECaPS map (near $b\pm 8-10^\circ$) when integrated out to $d \rightarrow \infty$. This artifact stems from chip gaps in the underlying DECaPS survey, causing stars within the gaps to have fewer photometric detections, which again modifies the underlying distance-extinction posteriors, and propagates to the inferred amount of reddening within the gaps. 

Finally, in Panel E, we show an artifact that manifests in all dense regions across the plane, where the cores of dense clouds seem to be ``missing" in the sense that they they appear to be in ``shadow" (at much lower extinction) than their envelopes. As an example, we show a cloud envelope (integrated out to $d=3$ kpc) at $E_{B-V} \approx 3-4$ mag, where the core of the cloud lies at lower extinction ($E_{B-V} \approx 1$ mag). This artifact stems from the fact that the cloud is so dense that it is fully extinguishing, rather than simply reddening, the light from stars background to the clouds, so the core of the cloud is undetectable in 3D dust maps. 

\begin{figure*}[ht!]
\begin{center}
\includegraphics[width=1.\textwidth]{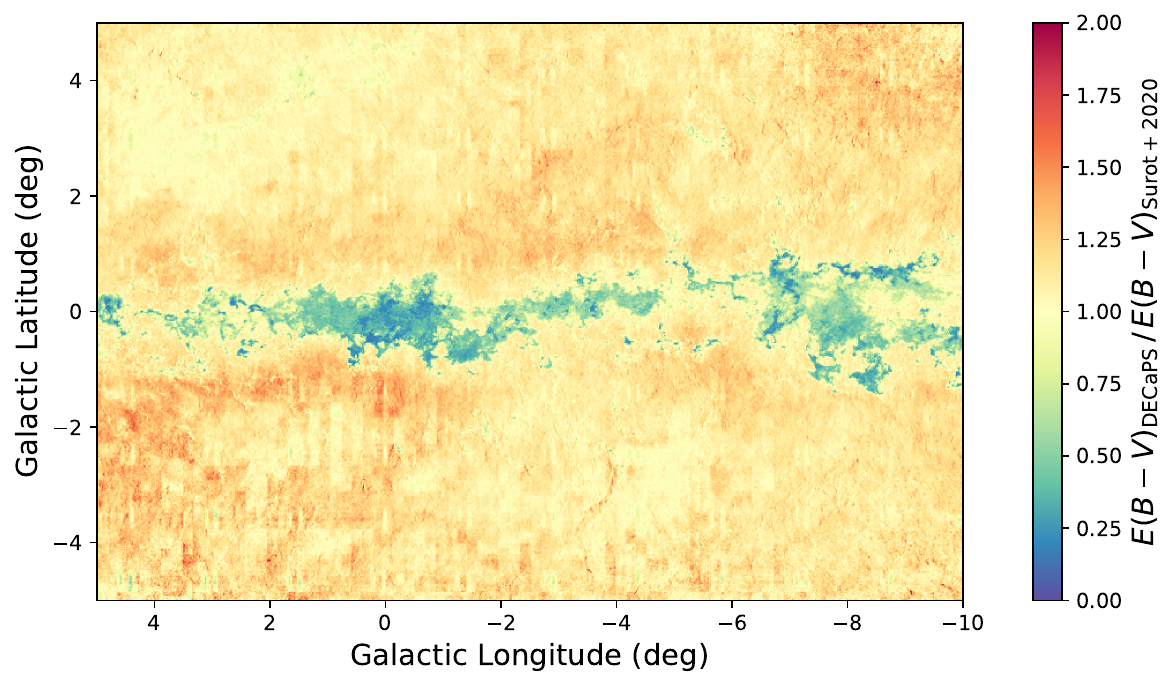}
\end{center}
\caption{Ratio of the reddening of the DECaPS map (integrated out to 8.5 kpc) over the reddening from the 2D \citet{Surot_2020} dust map. We find good agreement except for in the midplane, where we underpredict the reddening with respect to 2D maps based solely on infrared photometry.} 
\label{fig:surot}
\end{figure*}

In Figure \ref{fig:surot}, we better illustrate the DECaPS map inability to recover the total extinction in very dense regions. Specifically, we show the ratio of the reddening of our projected DECaPS map (integrated out to 8.5 kpc, roughly the distance of the Galactic center) over the reddening obtained from the 2D infrared star-based dust map of \citet{Surot_2020}. We convert the \citet{Surot_2020} map, in units of $E_{J-K_s}$ to $E_{B-V}$ assuming $E_{J-K_s}=0.57 \times E_{B-V}$ based on our reddening vector in Table \ref{tab:extcurve}. Unlike our DECaPS map, \citet{Surot_2020} leverages VVV infrared photometry towards RC and RGB stars and does not require a detection in the optical. Therefore, the \citet{Surot_2020} map is able to detect more stars in dense regions and recover more of the total extinction. However outside $|b| \lesssim 0.5^\circ$, we generally find strong agreement with the \citet{Surot_2020} 2D dust map. 

\section{Computational Details} \label{sec:compute_deets}
The generation of this 3D dust map took significant computational resources.  In total, we utilized roughly 3.5 million CPU hours on the FASRC Cannon cluster, supported by Harvard's FAS Division of Science Research Computing. The per-star inference took $\approx$ 887K CPU hours, primarily with 9 GB per process. Across all 793 million stars, we averaged roughly 4 s/star.  The line-of-sight inference took $\approx$ 2.6 million CPU hours, ranging from $4-8$ GB per process. Across all 51 million pixels, we averaged roughly 182 s per pixel. Infilling the 3D dust map took $\approx 320$ CPU hours, with $\approx 90$ GB per process.

\section{Inferred $R_V$ Distribution}
\label{sec:rv}

\begin{figure*}[ht!]
\begin{center}
\includegraphics[width=0.65\textwidth]{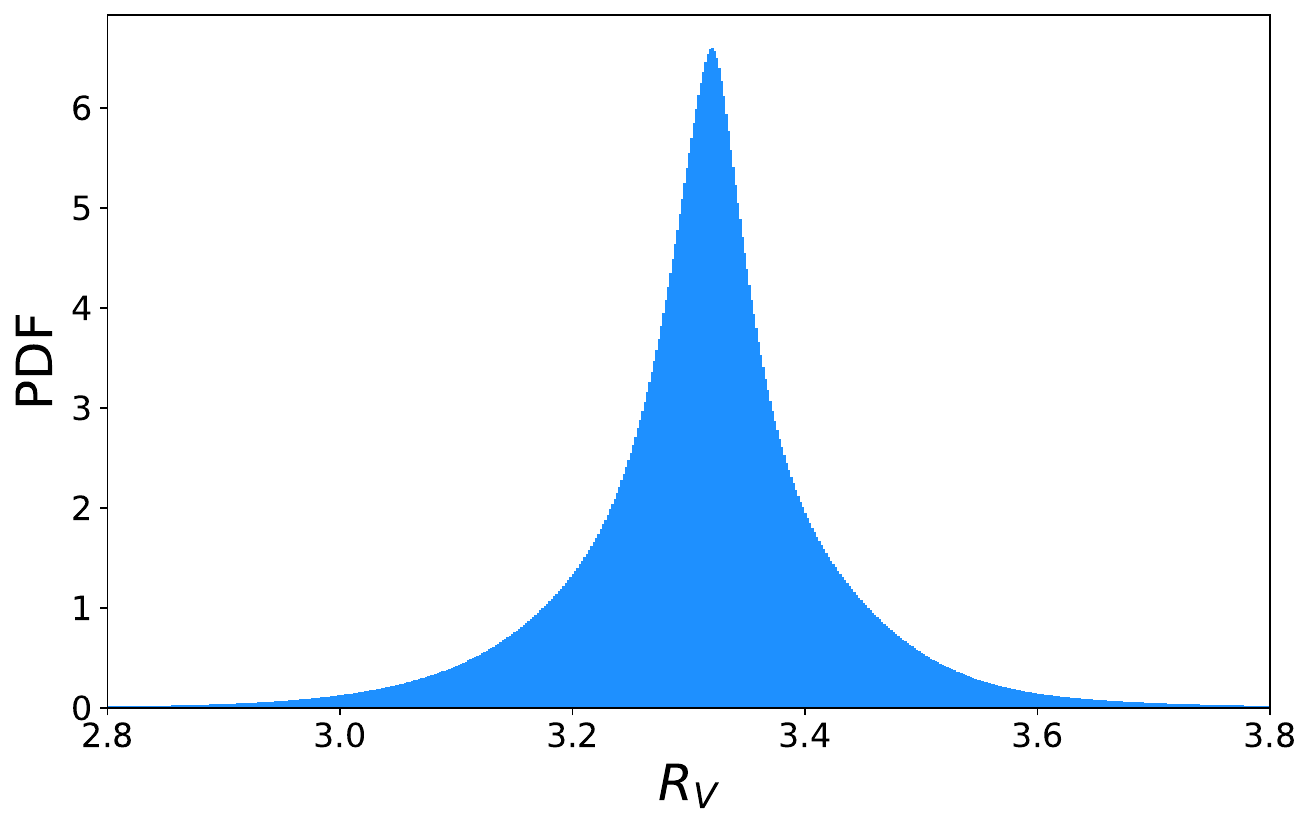}
\caption{Normalized PDF of the distribution of  $R_V$ inferred per star for the high-quality stellar sample (709 million stars). For each star, we draw a random sample. The distribution closely follows our adopted prior, limiting potential implications of our work for understanding the variation in the extinction curve across the Milky Way.} 
\label{fig:rv_dist}
\end{center}
\end{figure*}

Recall that we infer the total-to-selective extinction ratio, $R_V$, toward every star as part of the methodology described in \S \ref{subsec:perstar_inference}. In Figure \ref{fig:rv_dist}, we show the distribution of $R_V$, tabulated across the full sample of 709 stars used in the generation of the 3D dust map. For each star, we draw a random $R_V$ sample. As apparent in Figure \ref{fig:rv_dist}, we caution that $R_V$ is not strongly constrained by our data and closely aligns with the adopted mean $R_V$ of $\mu_{R_V}=3.32$ and standard deviation of $\sigma_{R_V}=0.18$ of our prior. 

\section{Predicting Gaia $G$-band Magnitudes}
\label{sec:gaia_G}

\begin{figure*}[ht!]
\begin{center}
\includegraphics[width=0.65\textwidth]{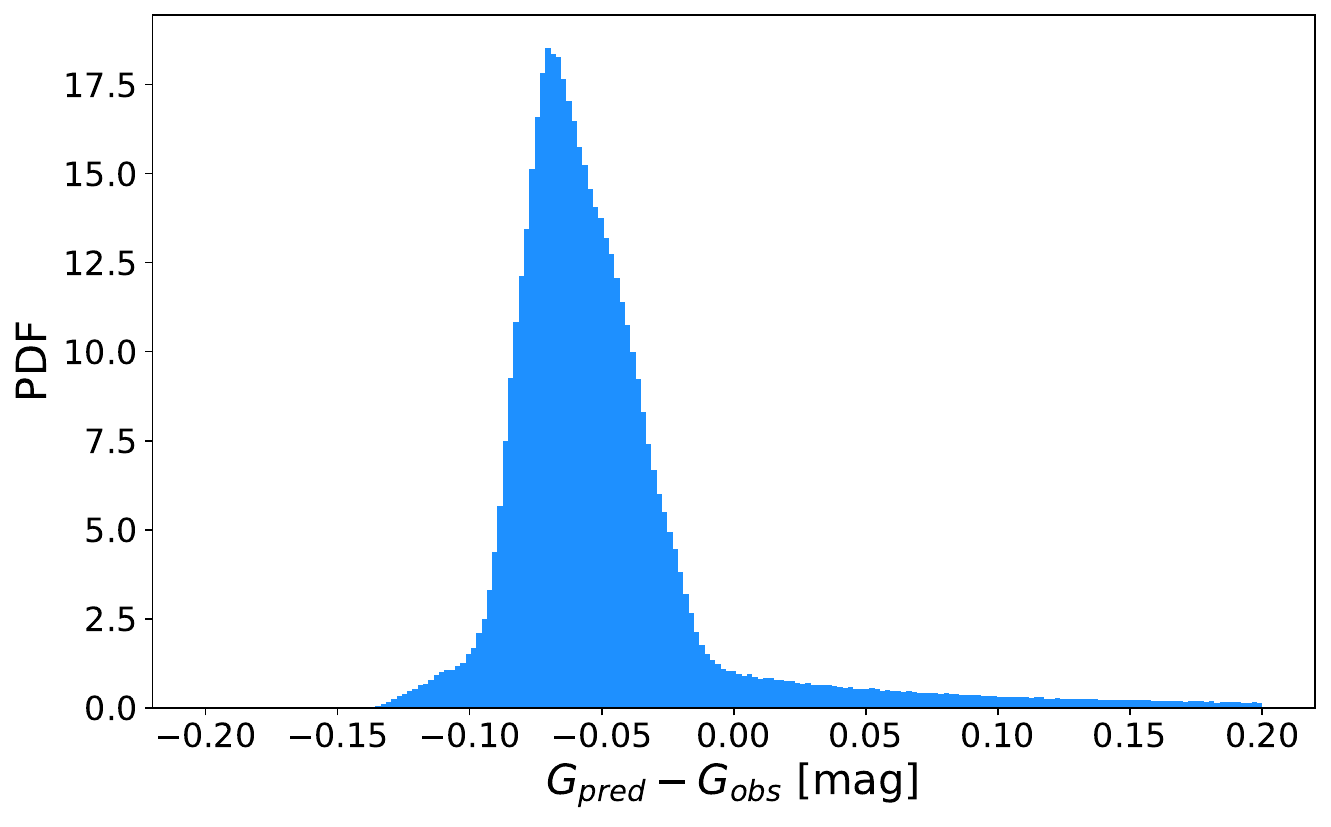}
\caption{Offset between the predicted and observed Gaia $G$ band magnitudes, computed as an average per $N_{\rm side}=1024$ over the full DECaPS2 footprint for stars with a Gaia detection. While we find good agreement overall, we tend to systematically predict slightly smaller (brighter) Gaia G magnitudes than observed, at the level of $\approx 0.05$ mag} 
\label{fig:gaia_G}
\end{center}
\end{figure*}

While we do not incorporate Gaia photometry into our stellar modeling pipeline described in \S \ref{subsec:perstar_inference}, the Gaia photometry provides an additional means to verify the fidelity of our inferred stellar parameters. For each star with a Gaia detection (roughly half the sample see; see Figure \ref{fig:coverage_hist}), we predict the Gaia $G$ band magnitudes based on the inferred stellar type constrained using the photometry from other wide-field surveys (see \S \ref{sec:data}). In Figure \ref{fig:gaia_G} we show the difference between the predicted and observed Gaia G band magnitude, averaged in $N_{\rm side}=1024$ pixels over the full DECaPS2 footprint. We overall find good agreement, but underpredict the observed Gaia G band magnitudes by $\approx 0.05$ mag, favoring a slightly brighter Gaia G band detection. There are a number of potential causes for this discrepancy. For example, we adopt a different reddening vector to produce the predicted G band magnitudes because the Gaia passbands are not modeled in \citet{Schlafly_2016}. Moreover, the intrinsic Gaia colors in our current model grid are based on the Gaia DR2 photometric system, but we compare to magnitudes reported in the Gaia DR3 photometric system in Figure \ref{fig:gaia_G} \citep{Riello_2021}.  We defer a detailed investigation of this small systematic offset to future work. 

\section{Minimum and Maximum Reliable Distance}
\label{sec:reliable}

To compute an estimate of the variation in the minimum and maximum reliable distance of the 3D dust map across the sky, we utilize the set of 2D binned posteriors on distance and reddening (\S \ref{subsec:perstar_setup}) used in the line-of-sight fits (\S \ref{subsec:bayestarlos}). The reliability of our inference is directly tied to how the stars are distributed along the line of sight toward a given patch of sky --- if there is an insufficient number of stars constraining the line-of-sight fit at a given distance, we cannot adequately constrain the amount of reddening at that distance. To quantify this effect for each $N_{side} = 8192$ pixel, we sum the set of binned posteriors on distance and reddening (of shape $N_{\rm stars} \times N_{E_{\rm bins}} \times N_{\mu_{\rm bins}}$) over the $N_{\rm stars}$ and $N_{E_{\rm bins}}$ axes to obtain the probability distribution in distance modulus along that line of sight. We then compute the cumulative probability distribution in distance modulus stepping both forward (starting at $\mu = 4$ mag) and backward in distance (starting at $\mu = 19$ mag) to characterize the minimum and maximum reliable distance modulus, respectively. 

We define the minimum and maximum reliable distance modulus as the first distance bin (computed forward in distance for the minimum and backward in distance for the maximum) to exceed some probability threshold. We tested a range of probability thresholds ranging from $p=0.025$ (cumulative probability of a small fraction of a single star) to $p=10$ (cumulative probability of ten stars) over a range of testbeds, including those shown in the gallery of artifacts in Figure \ref{fig:artifacts}. While the selection of a probability threshold is subjective, our goal was to select thresholds that minimized the known artifacts across the testbeds, including striping, speckling, patchiness, and undersampling of the reddening in very dense regions. We find an optimal value of $p=0.5$ for the minimum reliable distance modulus and $p=5$ for the maximum reliable distance modulus. In many, but not all, lines of sight, we find that the $p=0.5$ minimum cumulative probability threshold roughly corresponds to the distance of the first foreground star. 

\begin{figure*}[ht!]
\begin{center}
\includegraphics[width=1.\textwidth]{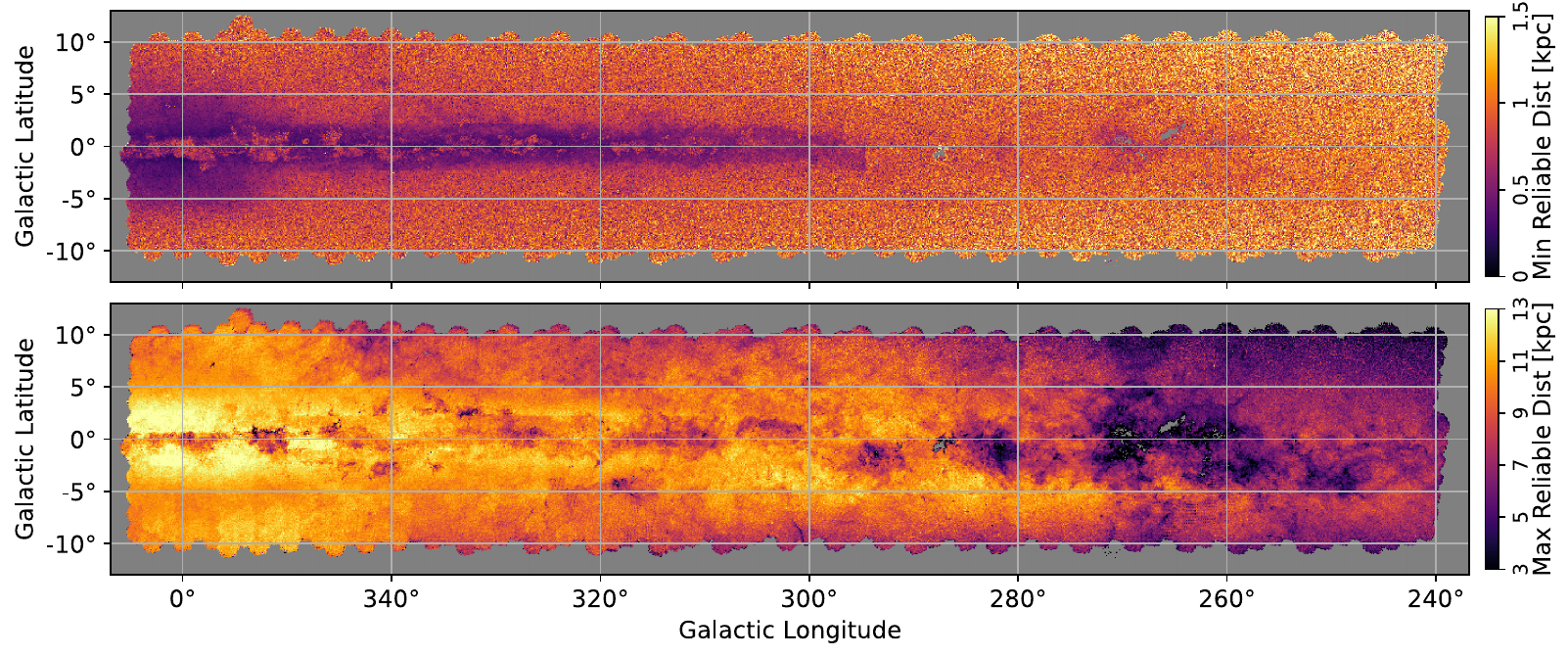}
\end{center}
\caption{\textit{Top:} Plane-of-sky variation in the minimum reliable distance, computed for each $N_{side} = 8192$ pixel as described in \S \ref{sec:reliable}. \textit{Bottom:} Plane-of-sky variation in the maximum reliable distance.} 
\label{fig:minmax_pos}
\end{figure*}

\begin{figure*}[ht!]
\begin{center}
\includegraphics[width=1.\textwidth]{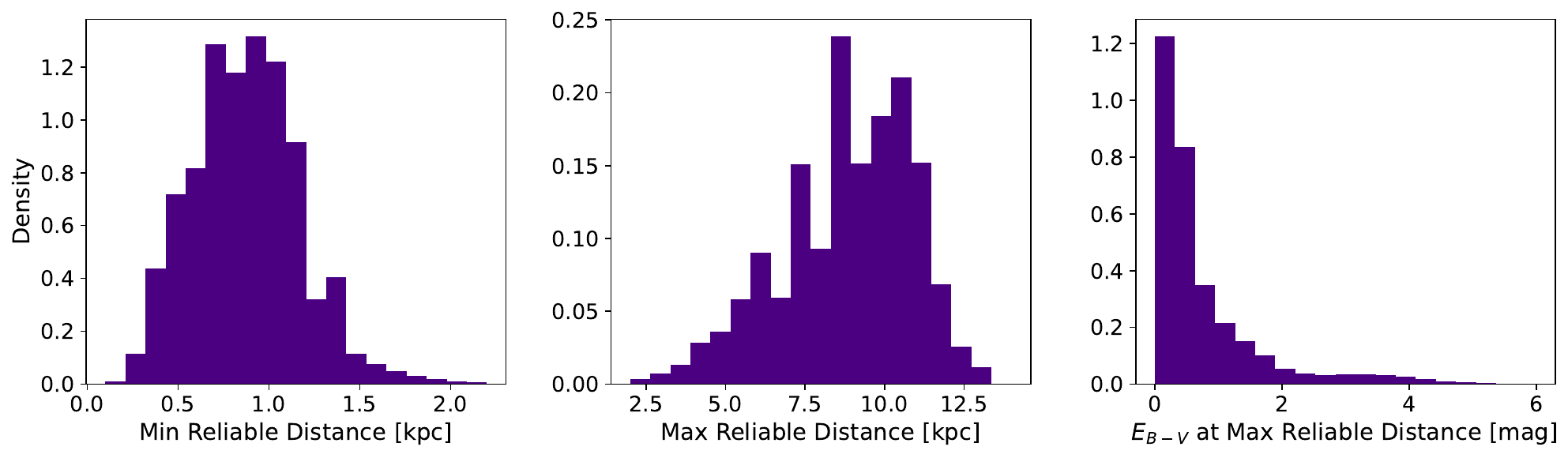}
\end{center}
\caption{Normalized PDFs of the distribution of the minimum reliable distance (left), maximum reliable distance (middle) and reddening at the maximum reliable distance (right), computed over the 51 million pixels in the DECaPS footprint.} 
\label{fig:minmax_histograms}
\end{figure*}

In Figure \ref{fig:minmax_pos}, we show the spatial variation in the minimum (top) and maximum (bottom) reliable distance across our footprint. In Figure \ref{fig:minmax_histograms}, we show normalized histograms of the minimum reliable distance (left), maximum reliable distance (middle), and correspondng reddening at the maximum reliable distance (right, computed from the mean map) over the full footprint. We find a median and $1 \sigma$ spread in the minimum reliable distance modulus of $\mu_{\rm min} = 9.6^{+0.8}_{-0.9}$ ($d_{\rm min} = 0.8^{+0.3}_{-0.3}$ kpc). We find a median and $1 \sigma$ spread in the maximum reliable distance modulus of $\mu_{\rm max} = 14.9^{+0.4}_{-0.8}$ ($d_{\rm max} = 9.4^{+1.8}_{-2.8}$ kpc). Since the maximum reliable reddening --- defined as the reddening of the map at the maximum reliable distance --- is a function of both the sensitivity limit of the map and of natural variation in reddening over the plane (i.e. higher latitudes will have less reddening), we take $2 \sigma$ above the median as the adopted maximum reliable extinction (spread spanning $+1\sigma$ to $+3\sigma$ above the median) finding, $E_{B-V_{\rm max}} = 3.7^{+1.5}_{-2.4}$ mag ($A_{V_{\rm max}} = 12.3^{+5.0}_{-7.9}$ mag) for the mean 3D dust map (see \S \ref{sec:dustresults}).

\end{document}